\documentclass[12pt]{article}
\pdfoutput=1
\usepackage[centertags]{amsmath}
\allowdisplaybreaks[1]
\usepackage{array,multirow}
\numberwithin{equation}{section}
\usepackage{amssymb}
\usepackage{float}
\usepackage{graphicx}
\usepackage{color}
\usepackage{xcolor}
\usepackage{relsize}
\usepackage{verbatim}
\usepackage{mathtools}
\usepackage{amsmath,amsfonts,amssymb,amsthm, bm}

\usepackage{epsfig}

\def\Or[#1]{{\text{O}}\left({#1}\right)}
\def\dotl[#1,#2]{\left\langle #1, #2 \right\rangle}
\def\dotlb[#1,#2]{[ #1, #2 ]}
\def\dotp[#1,#2]{(#1) \cdot (#2)}
\def\aff[#1,#2]{\hat{#1}(#2)}
\def\n4sym{{\cal N}=4 SYM}
\def\>{\rangle}
\def\<{\langle}
\def\weight[#1,#2,#3]{\{(#1),#2,#3\}}
\def\ads[#1]{$\text{AdS}_{#1}$}

\newcommand{\ba}{\begin{eqnarray}}
\newcommand{\ea}{\end{eqnarray}}
\newcommand{\be}{\begin{eqnarray}}
\newcommand{\ee}{\end{eqnarray}}
\newcommand{\bq}{\begin{equation}}
\newcommand{\eq}{\end{equation}}
\newcommand{\benn}{\begin{equation*}}
\newcommand{\eenn}{\end{equation*}}
\newcommand{\bi}{\begin{itemize}}  
\newcommand{\ei}{\end{itemize}}

\newcommand{\ca}{{\cal A}}
\newcommand{\cb}{{\cal B}}

\newcommand{\CC}{{\cal C}}
\newcommand{\cc}{{\cal C}}

\newcommand{\cd}{{\cal D}}

\newcommand{\cf}{{\cal F}}

\newcommand{\CL}{{\cal L}}
\newcommand{\cl}{{\cal L}}
\newcommand{\CJ}{{\cal J}}
\newcommand{\cj}{{\cal J}}

\newcommand{\CO}{{\cal O}}
\newcommand{\co}{{\cal O}}

\newcommand{\cp}{{\cal P}}

\newcommand{\cv}{{\cal V}}
\newcommand{\nn}{\nonumber}

\newcommand\oo\infty
\newcommand\s\sigma
\newcommand\de\delta
\newcommand\De\Delta

\newcommand\f\phi
\newcommand\g\gamma
\newcommand\x\times

\usepackage{jheppub}

\makeatletter
\def\@fpheader{\vspace{-.1cm}}
\makeatother

\title{AdS$_3$ Reconstruction with  General Gravitational Dressings}

\author{Hongbin Chen,} 
\author{Jared Kaplan,} 
\author{Utkarsh Sharma} 
\affiliation{Department of Physics and Astronomy,  Johns Hopkins University, \\
Charles Street, Baltimore, MD 21218, USA} 

\abstract{ The gauge redundancy of quantum gravity makes the definition of local  operators ambiguous, as they depend on the choice of gauge or on a `gravitational dressing' analogous to a choice of Wilson line attachments.  Recent work identified exact AdS$_3$  proto-fields by fixing to a Fefferman-Graham gauge.   Here we extend that work and define proto-fields with general gravitational dressing. We first study bulk fields charged under a $U(1)$ Chern-Simons gauge theory  as an illustrative warm-up, and then generalize the results to gravity.  As an application, we compute a gravitational loop  correction to the bulk-boundary correlator in the background of a  black hole microstate, and then verify this calculation using a newly adapted recursion relation.   Branch points at the Euclidean horizon are present in the $1/c$ corrections to semiclassical correlators. } 

\arxivnumber{}
 
\begin{document} 
      
\maketitle
\flushbottom

\section{Introduction and Summary} 

A complete description of AdS/CFT  requires an exact prescription for bulk reconstruction, which would ideally provide a quantitative guide to its own limitations.   This problem may decomposed into two (overlapping) sub-problems:
\begin{itemize}
\item Reconstruction of interacting bulk fields from dual boundary `CFT' operators in the absence of AdS gravity.  It's easy to solve this problem for free bulk fields and generalized free theory (GFT) duals, and it can also be addressed order-by-order in bulk perturbation theory \cite{Kabat:2011rz, Kabat:2012av, Kabat:2013wga, Kabat:2016zzr}.  This problem is very similar \cite{Paulos:2016fap} to the question of how to relate the operators in a CFT which ends at a boundary to the BCFT operators living on that boundary.  
\item Bulk reconstruction in the presence of gravity.  This problem is qualitatively different, because we do not expect local bulk operators to be uniquely defined -- they must be associated with a `gravitational Wilson line' or `gravitational dressing'.  These complications arise because of the gauge redundancy of bulk diffeomorphisms and the universality of the gravitational force.
\end{itemize}
Both gravitational and non-gravitational interactions seem to require  bulk field operators $\Phi(X)$ to include mixtures of infinitely many CFT operators.

In AdS$_3$ the purely gravitational component of bulk reconstruction  can be more precisely  specified by taking advantage of the relation between bulk gravity and Virasoro symmetry.    This makes it possible to solve one aspect of reconstruction exactly.  Prior work \cite{Anand:2017dav}  defined bulk operators by first fixing to Fefferman-Graham gauge, thereby assuming a specific and arbitrarily chosen gravitational dressing.  The purpose of this paper is to define bulk proto-fields  with much more general gravitational dressings, or equivalently, by defining the bulk field in a more general gauge. 

\subsubsection*{Bulk Operators from Symmetry}

The AdS/CFT dictionary specifies that
\be
\lim_{y \to 0} y^{-2h}\Phi(y, z, \bar z) = \CO(z, \bar z)
\ee
for a bulk scalar field $\Phi(Y)$ and a dual boundary CFT primary $\CO$.  However, at finite $y$ the bulk field operator $\Phi(Y)$ will include an infinite sum of contributions from  other primaries.  Ultra-schematically, we may write \cite{Kabat:2011rz}
\be
\nn
\Phi &=& \CO  +  \sqrt{G_N} [T \CO]  +  G_N [T T \CO]  + \cdots 
\nn \\ && +  g [\CO_i \CO_j] +  g^2 [\CO_i \CO_j \CO_k] + \cdots
\ee
to indicate that $\Phi$ includes a mixture of multi-trace operators made from the stress-tensor $T$ and $\CO$, as well as multi-trace operators made from other primaries $\CO_i$, with perturbative coefficients that can be computed \cite{Kabat:2011rz, Kabat:2012av, Kabat:2013wga, Kabat:2016zzr} when such a description applies.  

\begin{figure}[t]
\begin{center}
\includegraphics[width=0.45\textwidth]{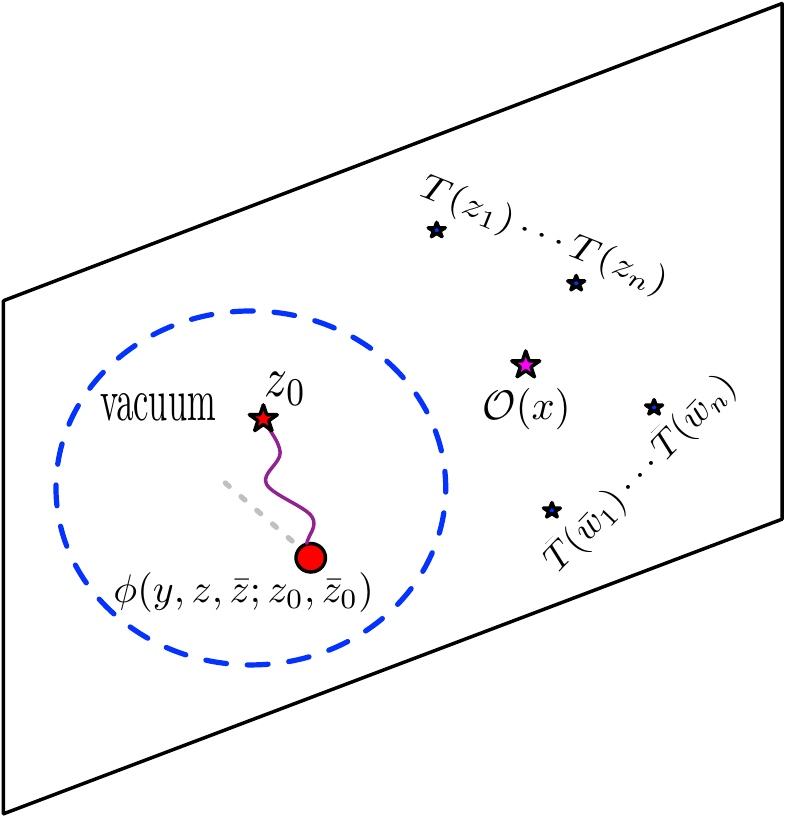}
\caption{This figure suggests many aspects of our reconstruction strategy.  We reconstruct a bulk operator $\phi$ connected by a Wilson line or `gravitational dressing' that attaches to the boundary at  $z_0$.  Correlators of $\phi$ with stress tensors will only have singularities when the stress tensors approach $z_0$.   When $\phi$ acts on the vacuum, it creates a state that we expand in radial quantization, and so when we define $\phi$ we assume it is surrounded by empty space.  As our methods are ultimately based on symmetry, we only compute the proto-field $\phi$ as a linear combination of a  CFT$_2$ primary $\CO$ and its  Virasoro descendants.   All of these statements have analogs for the $U(1)$ charged $\phi$ we discuss as a warm-up in section \ref{sec:BulkCSOperators}, with $T \to J$. }
\label{fig:RadialQuantizationPicturePhi}
\end{center}
\end{figure}

We will be studying the terms in $\Phi$ involving $\CO$ and any number of stress tensors, as these are determined by the Virasoro symmetry\footnote{There has been much recent work on AdS$_3$ reconstruction \cite{Anand:2017dav, Maxfield:2017rkn, Nakayama:2016xvw, Chen:2017dnl, daCunha:2016crm, Guica:2015zpf, Guica:2016pid, Castro:2018srf, Das:2018ojl, Chen:2018qzm}.  Our results \cite{Anand:2017dav} differ from the proposal \cite{Lewkowycz:2016ukf}, which produces a field that does not  seem to satisfy the interacting bulk equation of motion in a known gauge when expanded perturbatively in $1/c$ (e.g., compare $\< \phi \CO T \>$ correlators to the results of appendix D.4 of \cite{Anand:2017dav}).   } in AdS$_3$.  Just as conformal symmetry dictates that CFT correlators  must be decomposable as a sum of conformal blocks, bulk scalar fields $\Phi$ can  be written as a sum of bulk proto-field operators $\phi$ that are fixed by symmetry.

The proto-fields $\phi$ also have another interpretation, as sources or sinks for one-particle states in a first-quantized worldline action description.  Correlators of  protofields $\< \phi(X_1) \phi(X_2) \cdots \>$ with other CFT operators will match to all orders in perturbation theory with the propagation of a particle from $X_1$ to $X_2$ in the gravitational background created by these other CFT operators, including the effect of gravitational loops on the propagation.  But the proto-field correlators do not include non-gravitational interactions, or mixings with multi-trace operators induced by gravity.

\subsection*{`Dressings' and Correlators with Symmetry Currents}

Charged operators in gauge theories and local bulk operators in quantum gravity are not gauge-invariant.  This means that their definition is ambiguous, and  we need to supply  more information to fully specify them.  This additional information may be a Wilson line, a specific choice of gauge, or a `gravitational dressing' (by this term we roughly mean `gravitational Wilson line').  We discuss the relation between these ideas in section \ref{sec:ChargeWilsonLinesGaugeFixing}.  

The necessity and ambiguity of these dressings has a simple interpretation in the CFT.  If we are to write a bulk proto-field $\phi(X)$ as a CFT operator, then the charge and energy in $\phi$ must be visible to the charge $Q$ and spacetime symmetries $D, P_\mu, K_\nu, M_{\mu \nu}$ in the CFT.  These quantities can be computed by integrals of $J_\mu(x)$ or $T_{\mu \nu}(x)$ over Cauchy surfaces on the boundary \cite{Ginsparg}, but the specific spacetime distribution of current and energy-momentum associated with $\phi(X)$ is somewhat arbitrary.  This explains  the ambiguity in $\phi(X)$, and also suggests how it can be fixed -- the gauge and gravitational dressings are specified by the form of $\phi(X)$ correlators with $J_\mu(x)$ or $T_{\mu \nu}(x)$.  

To make sense of this logic, it must be possible to distinguish the energy-momentum in $\phi(X)$ from that of other sources in any state or correlator.  We accomplish this by assuming that $\phi(X)$ is surrounded by vacuum, so that we can define $\phi(X)$ in a series expansion\footnote{Another common approach \cite{Hamilton:2006az} defines bulk fields by integrating local CFT operators over a region.  This procedure may have equivalent issues when other local CFT operators are present in the region of integration and OPE singularities are encountered.} in the bulk coordinate \cite{Paulos:2016fap}, with local CFT operators as coefficients.  In this way we can use radial quantization to define $\phi(X)$.  

We will specify general gravitational dressings in two equivalent ways.  In section \ref{sec:DressingfromDiffeo} we use a trick:  starting with the proto-field defined through Fefferman-Graham \cite{Anand:2017dav}, we use a diffeomorphism to bend the gravitational dressing.  Whereas in section \ref{sec:GeneralGravitationalDressing} we take a more abstract route, and simply construct an operator $\phi(X; x_0)$ at a point $X$ in the bulk, but where $T(x)$ on the boundary detects $\phi$'s associated stress-energy at a general point $x_0$. 

\subsubsection*{Summary of Results}

Our main result is a simple formula for a proto-field with general dressing
\begin{equation}
\phi\left(u,x,\bar{x}; x_0 , \bar x_0 \right)=\sum_{n=0}^{\infty}\sum_{m,\bar{m}=0}^{\infty} \frac{(-1)^n u^{2h+2n}}{n! (2h)_n} \frac{\left(x-x_0\right)^{m}\left(\bar{x}-\bar{x}_0\right)^{\bar{m}}}{m!\bar{m}!}\mathcal{L}_{-n-m}\bar{\mathcal{L}}_{-n-\bar{m}}\mathcal{O}(x_0,\bar{x}_0)
\end{equation}
The interpretation of this operator is discussed in section \ref{sec:GravitationalReconstruction}, but roughly speaking, the proto-field is located $(u,x, \bar x)$ in AdS$_3$, with its associated energy-momentum localized at $(x_0, \bar x_0)$ on the boundary.  The $\CL_{-N}$ are polynomials in the Virasoro generators determined by the bulk primary condition \cite{Anand:2017dav}, with coefficients that are rational functions in the central charge $c$ and holomorphic dimension $h$ of $\CO$.  We verify that this result has the expected correlators with stress tensors $T(x)$.  Our formula can be integrated against a positive, normalized distribution 
$\rho$ via
\be
\phi[\rho](X) \equiv \int d^2x_0  \, \rho(x_0,\bar x_0) \phi(X; x_0,\bar x_0)
\ee
to obtain a very general\footnote{It's not entirely clear what operators the full space of gravitational dressings should include, but by letting $\rho$ depend on $X$ we can parameterize a large space of possibilities.  An average over $x_0$ is not equivalent to averaging over different exponents in a Wilson line \cite{Dirac:1955uv, Donnelly:2015hta}, since the average of an exponential is not the exponential of an average.  Wilson lines should be path-ordered,  so averaging over complete Wilson lines should be the more generally valid approach.  } gravitational dressing for the proto-field.

We also show in section \ref{sec:ZRR} that correlators of $\phi(X; x_0, \bar x_0)$ can be computed by a further adaptation \cite{Chen:2017dnl, Chen:2018qzm} of Zamolodchikov's recursion relations \cite{ZamolodchikovRecursion, Zamolodchikovq}.  Then in section \ref{sec:HeavyLightCorrelator} we analytically calculate the $1/c$ correction to the heavy-light, bulk-boundary propagator on the cylinder using a recent quantization \cite{Cotler:2018zff} of AdS$_3$ gravity.  We demonstrate that our analytic result matches that of the recursion relation.  We also observe that as expected \cite{Fitzpatrick:2015dlt, Fitzpatrick:2016ive}, the analytic $1/c$ correction to the correlator is not periodic in Euclidean time \cite{Chen:2018qzm}, and so it has a branch cut singularity at the Euclidean horizon. This is surprising from the point of view of perturbation theory in a fixed black hole background.

The outline of the paper is as follows.  In section \ref{sec:BulkCSOperators}  we provide a detailed discussion of bulk reconstruction  for fields charged under a $U(1)$ Chern-Simons field.  This serves as a warm-up where many of the  ideas can be more straightforwardly illustrated.  Then in section \ref{sec:GravitationalReconstruction} we turn to gravity, where many of our results are analogous to the simpler $U(1)$ setting.  In section \ref{sec:ZRR} we adapt a recursion relation to compute correlators of $\phi$ with general dressing. In section \ref{sec:HeavyLightCorrelator} we explain some rather technical calculations, including the recursion relation in a specific configuration and an analytic computation of the one-loop gravitational correction (i.e., order $1/c$) to a $\< \CO_H \CO_H \CO_L \phi_L \>$ correlator.  We provide a brief discussion in section \ref{sec:Discussion}. Many technical results are relegated to the appendices.

\section{Bulk Proto-Fields with $U(1)$ Chern-Simons Charge}
\label{sec:BulkCSOperators}

This section will serve as a warm-up in preparation for our eventual discussion of bulk gravity, where most of these ingredients will have a direct analog.

\subsection{Charged Fields, Wilson Lines, and Gauge Fixing}
\label{sec:ChargeWilsonLinesGaugeFixing}

Consider a bulk field $\varphi(X)$  charged under a $U(1)$ gauge symmetry.  It transforms as 
\be
\varphi(X) \to e^{i q \Lambda(X)} \varphi(X)
\ee
under the gauge redundancy, so it cannot be regarded as a physical observable.  We can remedy this problem in two equivalent ways -- by fixing the gauge, or by attaching $\varphi(X)$ to a Wilson line.

The latter approach has the clear advantage that it makes the gauge-invariant nature of  our  observable manifest.  Given a Wilson line 
\be
W_{\CC}(\infty, X) = e^{i q \int_{\CC} d x^\mu A_\mu } 
\ee
running from $X$ to infinity, we can form a non-local operator
\be
\phi(X) = W_{\CC}(\infty, X) \varphi(X)
\ee
Since gauge transformations do not act at infinity, $\phi$ will be a gauge-invariant observable.  However, this means that $\varphi(X)$ itself was highly ambiguous, since $\phi$ now depends on the path of the Wilson line.  Note that once we define a gauge-invariant $\phi$ in this way, we can compute observables involving it in any convenient gauge, and we will obtain the same results.

The other (fixing the gauge) approach will be easier to discuss when we generalize to quantum gravity.  However, it's less flexible and can lead to confusing terminology.  In this approach we simply fix a gauge, for example by setting some component of the gauge field $A_y = 0$, and then compute observables involving $\phi(X)$ in this gauge.  The results will then be well-defined observables.  Note that if $A_y = 0$ then the Wilson line in the $\hat y$ direction $W_{\hat y} = 1$ identically, so in this case the underlying gauge invariant observable will  be $\phi = W_{\hat y} \varphi(X)$.  But in general it may not be clear how to compute with our observable in other gauges. And it may seem confusing to refer to an observable defined in a specific gauge as gauge-invariant (though this is in fact true).

Let us develop these ideas in the context of a scalar field $\phi$ in AdS$_3$ charged under a $U(1)$ Chern-Simons theory with level $k$.  The scalar will be dual to a CFT$_2$ primary operator $\CO$ with conformal dimension $h$ and charge $q$, and the gauge field to a holomorphic conserved current $J(z)$.  We will work in Euclidean space with a fixed metric
\be
ds^2 = \frac{dy^2 + dz d \bar z}{y^2}
\ee
and in this section we will not include dynamical gravity.  

We will be viewing $\phi(y, z, \bar z)$ through the lens of radial quantization, as discussed previously in  \cite{Anand:2017dav} and pictured in figure \ref{fig:RadialQuantizationPicturePhi} (the figure denotes the gravitational case, but with $T \to J$ it also applies to the present discussion).  In the CFT $\phi$ will be a non-local operator, but only because it will be written as an infinite sum of local operators, each a coefficient in the near-boundary or small $y$ expansion of $\phi$.  If we turn off both gravity and the Chern-Simons interaction, then $\phi$ is determined by symmetry to be \cite{Nakayama:2015mva}
\be
\label{eq:FreeBulkFieldyExpansion}
\phi_0(y, z, \bar z) = \sum_{n=0}^{\infty} (-1)^n  \frac{y^{2h+2n}}{n! (2h)_n} L_{-1}^n \bar L_{-1}^n \CO(z, \bar z)
\ee
This follows from the form of the vacuum bulk-boundary propagator
\be
\label{eq:FreeBulkBoundary}
\< \phi_0(y, z, \bar z ) \CO^\dag(w, \bar w) \> = \left( \frac{y}{y^2 + (z-w) (\bar {z}-\bar{w})} \right)^{2h}
\ee
if we expand in $y$ and identify the coefficients with global descendants of $\CO$.

Now let us define a gauge-invariant charged scalar field.  As discussed above, we can do this by simply attaching $\varphi$ to a Wilson line that ends on the boundary at $y=0$.  A very simple choice takes the Wilson line to run in the $\hat y$ direction, so that
\be
\phi(y, z , \bar z) = e^{iq \int_0^y dy' A_y(y', z, \bar z)} \varphi(y, z , \bar z)
\ee
This operator also has a very simple definition via gauge fixing -- it is simply $\phi$ defined in the gauge $A_y = 0$.  This makes it clear that the correlator $\< \phi(y, 0, 0 ) \CO^\dag(w, \bar w) \>$ should be equal to the expression (\ref{eq:FreeBulkBoundary}).

Correlators involving the boundary current will be non-trivial.  Computing in perturbation theory gives
\be
\< J(z_1) \CO^\dag(w, \bar w)  \phi(y, 0, 0) \> = \frac{q w}{z_1 (z_1-w)} \left( \frac{y}{y^2 + w \bar w} \right)^{2h},
\ee
where $\co^\dagger$ has charge $-q$. This reduces to $y^{2h} \< J \CO^\dag \CO \>$ in the limit of small $y$, as expected based on the AdS/CFT dictionary.

If we attach $\phi$ to the boundary with a more general Wilson line, then we will have
\be
\phi(y, z, \bar z; z_0, \bar z) = e^{iq\int_{z_0}^X dY^\mu A_\mu(Y)}  \phi(y, z , \bar z)
\ee
The Wilson line takes some general path from $z_0$ on the boundary to the location of $\phi$ in the bulk, yet our notation does not include information about the path.  Since  Chern-Simons theory is topological, our $\phi$ will only depend on this path if other charges or Wilson lines entangle with it.  However, we are invoking radial quantization to define $\phi$, meaning that we will be assuming that there aren't any matter fields or Wilson lines near $\phi$, or between it and the boundary.   Thus our results for the operator $\phi$ will be independent of the choice of path, except through the location of $z_0$.

Correlators of this more general bulk field can still be computed in $A_y=0$ gauge.  Since the vacuum equations of motion set $F_{y\mu} = 0$, in this gauge $A_z$ is independent of $y$.  Since on the boundary $A_z(0, z, \bar z)  = \frac{1}{k}J(z)$, this identification must hold for all $y$, so\footnote{We could also just choose to have the Wilson line run along the boundary from $z_0$ to $z$.}
\be
\phi(y, z, \bar z; z_0, \bar z) = \left. e^{i \frac{q}{k} \int_{z_0}^z J(z') dz'}  \phi(y, z , \bar z) \right|_{A_y = 0}
\ee
This formula requires some regularization to remain consistent with $\CO$ correlators and the dictionary $\CO = \lim_{y \to 0} \left[ y^{-2h}  \phi \right]$.  If we expand to first order in $q$ we find
\begin{smaller}
\begin{align}
\< J(z_1) \CO^\dag(w,\bar w) \phi(y, 0, 0; z_0, 0) \>  &\approx  \< J(z_1) \CO^\dag(w,\bar w) \phi(y, 0, 0) \> + i\frac{q}{k} \int_0^{z_0} \< J(z_1) \CO^\dag(w,\bar w) J(z') \phi(y, 0, \ 0) \>
\nn  \\
&\approx   \frac{q (z_0-z)}{(z_1 - z_0) (z_1 - z) }   \< \CO^\dag \phi \>
\end{align}
\end{smaller}
This formula has a nice interpretation, as the singularities in $z_1$ indicate the presence of charge $\pm q$ at $z_0$ and $z$.  But higher order corrections involving many $J$ will produce divergent integrals.  And even the simpler correlator $\< \CO^\dag \phi \>$ also requires regularization.  

In the next sections we will see how to avoid regularization by defining $\phi$ using symmetry when its Wilson line attachments are simple.  Then we will extend our results to include general Wilson lines by leveraging the singularity structure of  $\phi$ correlators.

\subsection{A Bulk Primary Condition from Symmetry}
\label{sec:BulkPrimaryfromSymmetry}

We can take another approach, and constrain the bulk field $\phi$ using symmetry.  If we can determine how to extend CFT symmetries into the bulk, then we can use their action on a charged bulk field to determine how to write it as a sum of CFT operators.  This approach will provide an exact definition, without needing to regulate Wilson lines, and it will also generalize more directly to gravity.

The CFT current $J(z)$ can be expanded in modes
\begin{equation}
J\left(z\right)=\sum_{n=-\infty}^{\infty}\frac{J_{n}}{ z^{n+1}}.  
\end{equation}
The global conformal generators have an algebra with $J_n$ with commutation relations
\begin{align}
\left[L_{m},L_{n}\right] & =\left(m-n\right)L_{m+n},\nonumber \\
\left[L_{m},J_{n}\right] & =-nJ_{n+m},\\
\left[J_{m},J_{n}\right] & =mk\delta_{n+m,0},\nonumber 
\end{align}
where the subscripts of the $L$ generators runs from $-1$ to $1$, and the subscript of the $J$ generators run from $-\infty$ to $\infty$.  The current acts on local primary operators via
\be
[J_n, \CO(z)] = q z^n \CO(z)
\ee
which can be derived from the $J(x) \CO(z)$ OPE.  This means that a finite transformation $e^{i \delta J_n}$ will rephase $\CO(z) \to e^{i q \delta z^n} \CO(z)$. 

Now we would like to understand how to extend these symmetries so that they act on bulk fields.  This requires either a careful specification of the gauge invariant operators, or a choice of gauge.  We will take the latter route and choose $A_y=0$.  We can still transform $A_z \to A_z + \partial_z \lambda(z)$ while preserving this gauge fixing condition.  But this  is a global (rather than gauge) symmetry transformation, since it acts non-trivially on fields at the boundary $y=0$. 

In the bulk, we expect that a charged field should transform as $\phi \to e^{i \lambda} \phi$.   Since $\lambda$ cannot depend on $y$, this transforms $\phi(X) \to e^{i \lambda(z)} \phi(X)$.  So in $A_y=0$ gauge, the $J_n$  act on $\phi$ in the same way that they act on $\CO$, giving
\be
\left. [J_n, \phi(y,z, \bar z)] = q z^n \phi(y,z, \bar z) \right|_{A_y = 0}
\ee
where we have indicated explicitly that this only holds in $A_y = 0$ gauge.  This further implies a \emph{bulk primary condition}
\be\label{eq:ChargedBulkPrimaryCondition}
[J_n, \phi(y, 0, 0) ] = 0 \ \ \  \mathrm{for} \  \ \  n \geq 1
\ee
for the bulk field $\phi$.  This condition is the $U(1)$ Chern-Simons version of the gravitational bulk primary condition originally derived in \cite{Anand:2017dav}.  Along with the requirement that $\phi$ has the correct bulk-boundary propagator in vacuum (\ref{eq:FreeBulkBoundary}), this bulk primary condition uniquely determines $\phi(y, z, \bar z)$ as an expansion in $y$.  Furthermore, it is an exact result, and does not require a small coupling expansion.

Notice that a gauge-invariant bulk operator $\phi(y,z, \bar z)$ attached to the boundary by a Wilson line in the $\hat y$-direction must transform in the same way.  This simply follows from the fact that $\phi = \varphi$ identically if the latter is defined in the gauge $A_y = 0$.

We can write a formal solution to the bulk primary conditions as
\be
\label{eq:StandardChargedBulkPrimaryField}
\phi(y,z,\bar z) = \sum_{n=0}^\infty
\frac{(-1)^n}{n! (2h)_n}  y^{2 h+2 n} \CJ_{-n} \bar{L}_{-1}^{n} \CO \left( z,\bar z \right)
\ee
where $\CJ_{-n} \CO$ is defined as an $n$th level descendant of $\CO$ satisfying the bulk primary condition (\ref{eq:ChargedBulkPrimaryCondition}). In appendix  \ref{app:ExactSolutions}, we solve
the the bulk primary condition exactly for the first several $\cj_{-n}$s.
As in section 3.2.2 of \cite{Anand:2017dav}, it can be shown that $\cj_{-n}$ can be
written formally in terms of quasi-primaries as
\begin{equation}
\cj_{-n}\co=L_{-1}^{n}\co+n!\left(2h\right)_{n}\sum_{j=1}^{n}\sum_{i}\frac{L_{-1}^{n-j}\co_{h+j}^{\left(i\right)}}{\left|L_{-1}^{n-j}\co_{h+j}^{\left(i\right)}\right|^{2}}\label{eq:CJnInTermsOfQuasiPrimariesSec}
\end{equation}
where the $\co_{h+j}^{\left(i\right)}$ represent the $i$th quasi-primary
at level $j$ (so they satisfy $L_{1}\co_{h+j}^{\left(i\right)}=0$) and the denominator is the norm of the corresponding operator. In writing down this equation, we've used the fact that the quasi-primaries can be chosen to be orthogonal to each
other. This will be a very useful property for some of the discussions
in the following sections. As a concrete example, $\cj_{-1}$ is given by
\begin{equation}
\mathcal{J}_{-1}\mathcal{O}=L_{-1}\mathcal{O}+\frac{q^{2}}{2hk-q^{2}}\left(L_{-1}-\frac{2h}{q}J_{-1}\right)\mathcal{O}.
\end{equation}
where the second term is a quasi-primary satisfying $L_{1}\left(L_{-1}-\frac{2h}{q}J_{-1}\right)\mathcal{O}=0$.

In appendix (\ref{app:LargekSolution}), we solve the bulk primary condition in the
large $k$ limit for the all-order $\frac{1}{k}$ terms in $\cj_{-n}$.
As shown in both appendix (\ref{app:ExactSolutions}) and (\ref{app:LargekSolution}), in the $k\to\infty$, we have
$\cj_{-n}=L_{-1}^{n}+O(1/k)$, so our expansion in $y$ reduces to
that of equation (\ref{eq:FreeBulkFieldyExpansion}).
Note that as in \cite{Anand:2017dav}, $\phi$ is
a non-local operator in the CFT due to the infinite sum in its definition.

Since the bulk proto-field has been defined as an expansion in descendants of a local CFT primary, we will often informally discuss the `OPE' of the current $J$ with $\phi$.
Because of the bulk primary condition above, the singular term in the OPE of $J\left(z_{1}\right)$ and $\phi\left(y,z,z\right)$ is very similar to the $J\CO$ OPE (which is simply $J(z_1)\co(z,\bar{z})\sim \frac{q\co(z,\bar{z})}{z_1-z}+\cdots$):
\begin{equation}\label{eq:JPhiOPE}
J\left(z_{1}\right)\phi\left(y,z,\bar{z}\right)\sim\frac{q\phi\left(y,z,\bar{z}\right)}{z_{1}-z}+\cdots
\end{equation}
where we have used $[J_{0}, \phi] = q\phi$, since the descendant operators in $\phi$ all have the same charge $q$.

Using these CFT definitions of the bulk charged scalar $\phi$, we
can compute various correlation functions, such as
$\left\langle \phi\co^{\dagger}J\cdots J\right\rangle $ and $\left\langle \phi^\dag \phi\right\rangle $. We can first verify that $\phi$ given in (\ref{eq:StandardChargedBulkPrimaryField}) indeed gives the correct bulk-boundary propagator.
Using (\ref{eq:CJnInTermsOfQuasiPrimariesSec}), one can see $\langle\phi\co^{\dagger}\rangle$ is given by 
\begin{equation}
	\left\langle \phi\left(y,z,z\right)\co^{\dagger}\left(w,\bar{w}\right)\right\rangle =\left(\frac{y}{y^{2}+(z-w)(\bar{z}-\bar{w})}\right)^{2h}.
\end{equation}
simply because the quasi-primary terms in $\cj_{-n}\co$ do not contribute to this two-point function and the calculation reduces to that of $\langle\phi_0\co^\dagger\rangle$ with $\phi_0$ given by (\ref{eq:FreeBulkFieldyExpansion}). The bulk-boundary three-point function $\langle JO^{\dagger}\phi\rangle$
can be computed simply using the OPE of $J\co^{\dagger}$ and $J\phi$
(equation (\ref{eq:JPhiOPE})), and the result is given by 
\begin{equation}
\left\langle J\left(z_{1}\right)\co^{\dagger}\left(w,\bar{w}\right)\phi\left(y,z,z\right)\right\rangle =q\left(\frac{1}{z_{1}-z}-\frac{1}{z_{1}-w}\right)\left\langle \phi\left(y,z,z\right)\co^{\dagger}\left(w,\bar w\right)\right\rangle \label{eq:JOPhiBoundary}.
\end{equation}
Correlation functions of the form $\left\langle \phi\co^{\dagger}J\cdots J\right\rangle $
can then be computed recursively using OPEs used above and the $JJ$ OPE.

For the bulk two-point function $\left\langle \phi^{\dagger}\phi\right\rangle$, we compute
up to order $1/k$ using the perturbative approximation to $\phi$ that we derived in appendix (\ref{app:LargekSolution}), which corresponds to the
one photon-loop correction to the bulk propagator. The details of the calculation
are given in appendix \ref{app:CFTphiphiOneLoop}. The result for $\left\langle \phi^{\dagger}\left(y_{1},z_{1},\bar{z}_{1}\right)\phi\left(y_{2},\bar{z}_{2},\bar{z}_{2}\right)\right\rangle $
is given by
\begin{equation}
\left\langle \phi^{\dagger}\phi\right\rangle =\frac{\rho^{h}}{1-\rho}\left[1-\frac{q^{2}}{k}\left(\frac{\rho_{2}^{2}F_{1}(1,2h+1;2(h+1);\rho)}{2h+1}+\frac{\rho}{2h}-\log(1-\rho)\right)\right]+O\left(\frac{1}{k^{2}}\right)\label{eq:PhiPhiOneLopp}
\end{equation}
where $\rho=\left(\frac{\xi}{1+\sqrt{1-\xi^{2}}}\right)^{2}$ with
$\xi=\frac{2y_{1}y_{2}}{y_{1}^{2}+y_{2}^{2}+z_{12}\bar{z}_{12}}$.

In appendix \ref{app:CFTCalculationBulkCorrelationFunctions} we also
use the bulk Witten diagrams to compute $\left\langle J\co^{\dagger}\phi\right\rangle $
and $\left\langle \phi^{\dagger}\phi\right\rangle $, and the results
exactly match equations (\ref{eq:JOPhiBoundary}) and (\ref{eq:PhiPhiOneLopp}).
This provides a non-trivial check of our definition of a charged bulk  scalar field.

\subsection{Singularities in $J(z)$ and the AdS Equations of Motion}
\label{sec:SingularitiesGaussLaw}

\begin{figure}[th!]
\begin{center}
\includegraphics[width=0.25\textwidth]{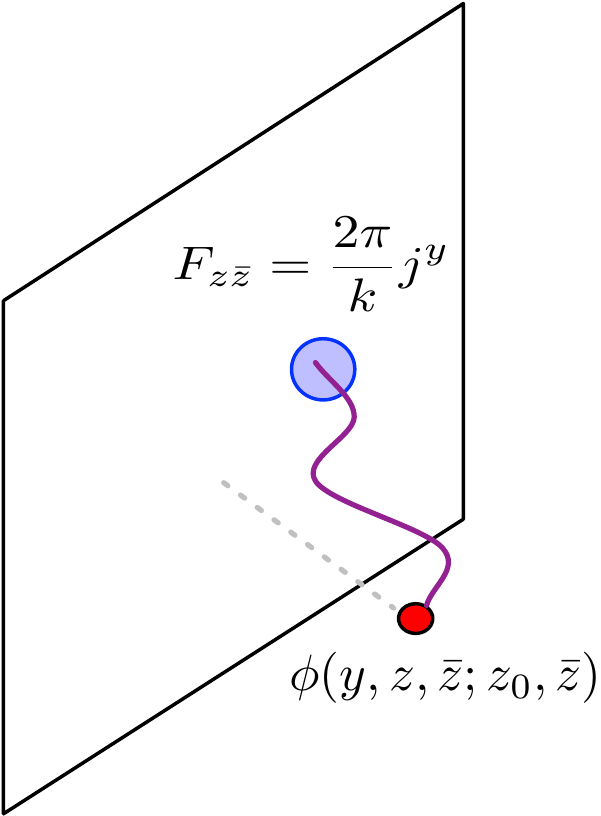}
\caption{This figure indicates the relationship between singularities in the current $J(z)$, the bulk equations of motion, and  Wilson lines.  Analogous statements hold for gravity and connect singularities in $T(z)$ to the gravitational dressing.    }
\label{fig:BulkEoMandSingularities}
\end{center}
\end{figure}

The singularity structure of the $J(z)$ correlators in equation (\ref{eq:JOPhiBoundary}) follow from the bulk equations of motion.  So these singularities indicate the placement of Wilson lines attaching bulk charges to the boundary, and vice versa.  Let us briefly explain these statements, which are illustrated in figure \ref{fig:BulkEoMandSingularities}.

In the $U(1)$ Chern-Simons theory, the equations of motion are
\be
\epsilon^{abc} F_{bc}(X) = \frac{2 \pi}{k} j^a(X)
\ee
for the bulk matter charge $j^a(X)$ and bulk field strength $F_{ab}$.  In the presence of a Wilson line with components in the $\hat y$ direction, $j^y$ will receive a delta function contribution localized to the Wilson line.  This means that $F_{z \bar z}$ must include a delta function.  
In $A_y = 0$ gauge, we can identify $A_z(y, z, \bar z) = \frac{1}{k} J(z)$, so 
\be
\partial_{\bar z} J(z) = 2 \pi q  \, \delta^2(z, \bar z)
\ee
or equivalently
\be
\oint dz J(z) = 2 \pi i q
\ee
To satisfy this constraint,  the current must have a simple pole $J(z) = \frac{q}{z} + \cdots$, where the ellipsis denotes less singular terms as $z \to 0$.  So the singularity structure of correlators with $J(z)$ follows directly from the bulk equations of motion.  Conversely, a singularity in $J(z)$ in correlators with other operators indicates the presence of charge.  

This means that we can use the singularity structure of correlators with $J(z)$ or $T(z)$ in the case of gravity to help to define a bulk field $\phi$ with a more general Wilson line attachment or `gravitational dressing'.   Similar observations also hold in higher dimensions, and may be useful for bulk reconstruction more generally.

\subsection{Charged Bulk Operators and General Wilson Lines}
\label{sec:GeneralWilsonLines}

In section \ref{sec:BulkPrimaryfromSymmetry} we constructed a charged bulk scalar field by fixing to  the  gauge $A_y=0$.  In so doing we defined a gauge-invariant bulk field $\phi(y, z, \bar z)$ connected to the boundary by a Wilson line in the $\hat y$ direction.  In this section, we are going to construct an exact gauge-invariant bulk field whose associated Wilson line attaches to an arbitrary point $z_0$ on the boundary.  

There are several ways to approach the construction of this general $\phi(y, z, \bar z; z_0 , \bar z)$.  The most immediate one was already discussed in section \ref{sec:ChargeWilsonLinesGaugeFixing}, namely including an explicit Wilson line.  An issue with this approach is that it requires a regulator for divergences in intermediate calculations, and this makes it difficult to define $\phi$ non-perturbatively.  Another important limitation is that it's challenging to work with Wilson lines for bulk diffeomorphisms, as would be necessary when we turn to gravity.

Instead of inserting an explicit Wilson line, we can define $\phi$ using the singularity structure of current correlators.  Correlators involving $J(z_1)$ and the field $\phi(y, z, \bar z)$ defined in $A_y=0$ gauge have singularities as $z_1 \to z$.  These singularities represent the charge of the bulk $\phi$ on the boundary, as emphasized in section \ref{sec:SingularitiesGaussLaw}.  If a Wilson line connects $\phi$  to the boundary at $z_0$, then instead we expect that correlators involving $J(z_1) \phi$ will have singularities at $z_0$.

Thus we need a way to move the singularities in $z_1$ for all correlators involving $J(z_1)$ and $\phi$.  In fact, we already have many of the technical tools that we need.  The level $n$ descendants $\CJ_{-n} \CO$ defined in section \ref{sec:BulkPrimaryfromSymmetry}  were constructed so that they would not have any additional singularities in the $J(z_1)\CJ_{-n} \CO$ OPE beyond those already present in $J(z_1) \CO$.   We can use these $\CJ_{-n}$ to move an operator without moving the singularities associated with its charge (we develop this idea in more detail in appendix \ref{app:MirageTranslations}).    The point is that since $\CJ_{-n} = L_{-1}^n + \cdots$ as shown in equation (\ref{eq:CJnInTermsOfQuasiPrimariesSec}), we have that
\be
\sum_{n=0}^{\infty}\frac{x^n}{n!} \CJ_{-n}  = e^{x L_{-1}} + \cdots
\ee
where the ellipsis denotes terms proportional to powers of the charge $q$. So this expression acts like a combination of a translation and a $U(1)$ symmetry transformation.  This means that the non-local operator
\begin{equation}\label{eq:ChargedGeneralizedTranslation}
\tilde{\co}\left(z, \bar z;z_{0}, \bar z\right)=\sum_{n=0}^{\infty}\frac{\left(z-z_{0}\right)^{n}}{n!}\cj_{-n} \co\left(z_{0},\bar z\right).
\end{equation}
behaves like a kind of mirage.  It is defined so that $J\left(z_1\right)\tilde{\co}\left(z, \bar z;z_{0}, \bar z\right)$ only has singularities in $z_1 - z_{0}$, while correlators of $\tilde{\co}\left(z, \bar z;z_{0}, \bar z\right)$ with $\CO^{\dagger}\left(w,\bar w\right)$ will instead have singularities when $w-z$ vanishes\footnote{Since only the $L_{-1}^{n}\co$ terms in $\cj_{-n}$ contribute in
$\left\langle \tilde{\co}\left(z,\bar{z};z_{0},\bar{z}\right)\co^{\dagger}\left(w,\bar{w}\right)\right\rangle $,
the $\cj_{-n}$ in (\ref{eq:ChargedGeneralizedTranslation}) become a translation operator, and one can see
that we have $\left\langle \tilde{\co}\left(z,\bar{z};z_{0},\bar{z}\right)\co^{\dagger}\left(w,\bar{w}\right)\right\rangle =\left\langle \co\left(z,\bar{z}\right)\co^{\dagger}\left(w,\bar{w}\right)\right\rangle $.
}.  Explicitly, by using the OPEs of $\cj\co^\dagger$ and $\cj\tilde\co$ and properties of $\cj_{-n}$, we have
\be
\< J(z_1) \CO^\dag(w,\bar w) \tilde \CO(z, \bar z; z_0,\bar z) \> = \frac{q (z_0 - w)}{(z_1 - z_0)(z_1-w)} \frac{1}{(z-w)^{2h} (\bar z- \bar w)^{2 \bar h}}
\ee
so $\tilde \CO(z, \bar z; z_0,\bar z) $ behaves as though it is in two places at once.

We can generalize the above idea to obtain a bulk proto-field $\phi\left(y_{1},z,\bar{z} ;z_{0},\bar{z}\right)$ at $\left(y ,z,\bar{z}\right)$ with a Wilson line landing at $\left(z_{0},\bar{z} \right)$ on the boundary. This leads to a proposal for a charged bulk field with a more general Wilson line attachment
\begin{equation}
\label{eq:GeneralChargedBulkField}
\phi \left(y, z, \bar{z} ; z_{0}, \bar{z}\right)=\sum_{n=0}^{\infty} \sum_{m=0}^{\infty} (-1)^n  \frac{y^{2 h+2 n}}{n!(2h)_n} \frac{(z-z_0)^{m}}{m !} \mathcal{J}_{-n-m} \bar{L}_{-1}^{n} \mathcal{O}\left(z_0, \bar{z}\right),
\end{equation}
In fact, this form for $\phi$ is uniquely fixed by demanding that:
\begin{enumerate}
\item $\< \phi \CO^\dag \>$ takes the vacuum form given in equation (\ref{eq:FreeBulkBoundary})
\item Correlators $\< \CO^\dag(w,\bar w)  J(z_1) \cdots J(z_n)  \phi(y, z, \bar z; z_0, \bar z) \>$ only have simple poles in the $z_i$, which can only occur when $z_i \to z_0$ or $z_i \to w$.
\end{enumerate}
Note that if $z=z_{0}$, then only the $m=0$ terms contribute to equation (\ref{eq:GeneralChargedBulkField}), and $\phi$ reduces to the bulk field of equation (\ref{eq:StandardChargedBulkPrimaryField}).  To obtain $\CO(z,\bar z)$ as we take $y \to 0$, we need to simultaneously send $z_0 \to z$; otherwise we obtain the non-local operator $\tilde \CO(z,\bar z;z_0,\bar{z})$ of equation  (\ref{eq:ChargedGeneralizedTranslation}).

We can verify that $\phi\left(y,z,\bar{z};z_{0},\bar{z}\right)$ is
really the desired operator by computing $\left\langle \co^{\dagger}\phi\right\rangle $
and $\left\langle J\co^{\dagger}\phi\right\rangle $ using the properties
of $\cj_{-n}$s. For the bulk-boundary propagator $\left\langle \co^{\dagger}\left(w,\bar{w}\right)\phi\left(y,z,\bar{z};z_{0},\bar{z}\right)\right\rangle $,
since the quasi-primaries terms in $\cj_{-n}\co$ (equation (\ref{eq:CJnInTermsOfQuasiPrimariesSec})) will not contribute,
we can simply replace $\cj_{-n}\co$ with $L_{-1}^{n}\co$. But then the
sum over $m$ becomes exactly a translation operator, and the $\phi$
in (\ref{eq:GeneralChargedBulkField}) becomes the free-field $\phi_{0}$ of (\ref{eq:FreeBulkFieldyExpansion}), and we have
\begin{equation}
\left\langle \co^{\dagger}\left(w,\bar{w}\right)\phi\left(y,z,\bar{z};z_{0},\bar{z}\right)\right\rangle =\left\langle \co^{\dagger}\left(w,\bar{w}\right)\phi_{0}\left(y,z,\bar{z}\right)\right\rangle =\left(\frac{y}{y^{2}+\left(z-w\right)\left(\bar{z}-\bar{w}\right)}\right)^{2h},
\end{equation}
as expected. For $\left\langle J\co^{\dagger}\phi\right\rangle $, we can use the
OPEs $J\left(z_{1}\right)\co^{\dagger}\left(w,\bar{w}\right)\sim\frac{-q\co^{\dagger}\left(w,\bar{w}\right)}{z_{1}-w}+\cdots$
and $J\left(z_{1}\right)\phi\left(y,z,\bar{z};z_{0},\bar{z}\right)\sim\frac{q}{z_{1}-z_{0}}\phi\left(y,z,\bar{z};z_{0},\bar{z}\right)+\cdots$,
where the ellipses denotes non-singular terms. $\left\langle J\co^{\dagger}\phi\right\rangle $
can then be computed by only including the singular terms in both
OPEs, and we get 
\begin{equation}
\left\langle J\left(z_{1}\right)\co^{\dagger}\left(w,\bar{w}\right)\phi\left(y,z,\bar{z};z_{0},\bar{z}\right)\right\rangle =q\frac{z_{0}-w}{\left(z_{1}-z_{0}\right)\left(z_{1}-w\right)}\left(\frac{y}{y^{2}+\left(z-w\right)\left(\bar{z}-\bar{w}\right)}\right)^{2h}.\label{eq:JOphiGeneralDress}
\end{equation}
Note that here the singularities are at $z_1=w$ and $z_1=z_0$, while in equation (\ref{eq:JOPhiBoundary}), the singularities were at  $z_1=w$ and $z_1=z$.

Before concluding this section, we should emphasize that our methods are insensitive to the trajectory that the Wilson line takes from $(y, z, \bar z)$ in the bulk to the boundary point $z_0$.  This is possible because the bulk theory is topological, and we have assumed that  $\phi$ is surrounded by a region of vacuum, so that it's possible to define $\phi$ in radial quantization.  Our gravitational $\phi$ will be defined in the same way.

\subsection{More General Bulk Operators from Sums of Wilson Lines}
\label{sec:Smearingz0}

In prior sections we have developed a formalism for exactly defining and evaluating an operator $\phi(X; z_0)$, where a Wilson line connects $X\equiv(y,z,\bar{z})$ to  a point $z_0$ on the boundary.  But there are a host of other choices for gauge-invariant bulk operators.  We can explore this space by defining a new bulk operator
\be
\label{eq:Smearingz0}
\phi[\rho](X) \equiv \int d^2z_0  \, \rho(z_0,\bar z_0) \phi(X; z_0,\bar{z})
\ee
where $\rho(z_0,\bar z_0) > 0$ and $\int d^2z_0 \rho(z_0,\bar z_0) =1$.  The properties of $\phi[\rho]$ are inherited from $\phi(X; z_0)$, which is simply the special case where $\rho(x) = \delta(x - z_0)$.  Note that if we like, we can let $\rho$ depends on $X$ so that $\rho$ varies as we move $\phi$ to different locations $X$ in the bulk.

 Operators like $\phi[\rho]$ make it possible to study bulk fields with far more general `dressings', which include superpositions of Wilson lines.  For example, we might define a bulk field with spherically symmetric dressing. Let the bulk field be located at the point $ (y,z,\bar z) $ in Poincare patch and $ (t_E,r,\theta) $ in global coordinates\footnote{The Poincare patch metric $ds^{2}=\frac{dy^{2}+dzd\bar{z}}{y^{2}}$
and the global AdS$_{3}$ metric $ds^{2}=\left(r^{2}+1\right)dt_{E}^{2}+\frac{dr^{2}}{r^{2}+1}+r^{2}d\theta^{2}$
are related by 
\begin{equation}
y=\frac{e^{t_{E}}}{\sqrt{r^{2}+1}},\quad z=\frac{re^{t_{E}+i\theta}}{\sqrt{r^{2}+1}},\quad\bar{z}=\frac{re^{t_{E}-i\theta}}{\sqrt{r^{2}+1}}.
\end{equation}}, and let the point of attachment of the Wilson line to the boundary be $z_0$. We will average with the measure $ \frac{d\theta}{2\pi}$ over the spatial circle on the boundary at fixed $t_E$. In Poincare patch this means integrating with the measure $\frac{dz_0}{2\pi iz_0}$ along the contour $|z_0|=|z|\sqrt{1+\frac{y^2}{z\bar z}} $. The integration contour ensures that $ z_0 $ is on the same time slice as the bulk field. In summary, the integration measure is
\begin{align}
\rho(z_0, \bar z_0)=\frac{1}{2\pi iz_0}\delta\left(|z_0|-|z|\sqrt{1+\frac{y^2}{z\bar z}}\right)
\end{align} 
Note that $ \int d^2z_0 \rho(z_0, \bar z_0)=1 $. The integration of $ \langle J\mathcal{O}^\dagger \phi\rangle  $ (equation \ref{eq:JOphiGeneralDress}) then gives
\begin{align}
\frac{\langle J(z_1)\mathcal{O}^\dagger(w,\bar w) \phi[\rho](y,z,\bar z)\rangle}{\langle\phi(y,z,\bar z)\co(w,\bar w)\rangle} &= q\left(\frac{\Theta(|z_1|-|z_0|)}{z_1}-\frac{1}{z_1-w} \right)
\end{align}
where $ |z_0| $ denotes $|z|\sqrt{1+\frac{y^2}{z\bar z}}  $ and $ \Theta $ is the Heaviside step function. Since $ J_0=\frac{1}{2\pi i}\oint dz_1 J(z_1) $, we can further contour integrate the above expression with respect to $ z_1 $ to obtain the total charge of the state. The result is given by
\begin{align}
q\left(\Theta(|z_1|-|z_0|)-\Theta(|z_1|-|w|) \right)
\end{align} 
The interpretaton of this is clear: A state created by $ \phi $ or $ \mathcal{O}^\dagger $ carries charges $ q $ and $ -q $ respectively, and the charge content of the state at the time of insertion of $ J $ depends upon the insertions that occur before that  time.  This result should be equivalent to Coulomb gauge \cite{Dirac:1955uv} to lowest order in perturbation theory, but will not agree more generally, because the exponential of an average  is not equal to the average of an exponential.

\section{Gravitational Proto-Fields with General Dressing}
\label{sec:GravitationalReconstruction}

There are no local gauge invariant operators in gravity.  To understand this, observe that diffeomorphism gauge redundancies act on local scalar fields via
\be
\varphi(x) \to \varphi(x) + \xi^\mu(x) \nabla_\mu \varphi(x)
\ee
This is similar to $U(1)$ gauge theory, where a charged field transforms as $\varphi \to \varphi + i q \Lambda(x) \varphi$, insofar as in both cases, local fields by themselves are not gauge-invariant.  As we discussed in  detail in section \ref{sec:BulkCSOperators}, we can form gauge invariant quasi-local charged operators in a $U(1)$ theory using Wilson line attachments.  

Matters are not so simple in the case of gravity, because it is the dependence of $\varphi$ on the bulk coordinates themselves that renders $\varphi$ gauge non-invariant. 
Lacking a simple and general notion of a gravitational Wilson line, we will discuss  diffeomorphism gauge-invariant operators as `gravitationally dressed' local operators.   In this section, we consider gravitational  dressings\footnote{As discussed in section \ref{sec:ChargeWilsonLinesGaugeFixing}, one way to define dressed, gauge-invariant operators is to define the operators after fixing to a specific gauge.  We will use this method and others in this section.  } that associate local bulk operators with specific points on the boundary.  Such dressings are natural analogs of Wilson lines joining a field in the bulk to a point on the boundary.

We will use the notation $\phi(y,z,\bar{z}; z_0, \bar z_0)$ to denote a diffeomorphism invariant bulk proto-field at $(y,z,\bar{z})$ in AdS$_3$ that has a gravitational line dressing landing on the boundary at the point $(z_0, \bar z_0)$. The simplified notation $\phi(y,z,\bar{z})$ is used when the  boundary point of the gravitational dressing is at $(z,\bar{z})$. We sometimes refer to the path that the associated gravitational dressing takes, but strictly speaking our results are not associated with any particular path.  In our formalism all paths are equivalent because we assume that the bulk field can be connected to the boundary through the bulk vacuum.

 Our typical setup is pictured in figure \ref{fig:RadialQuantizationPicturePhi}.  Some of our analysis will be analogous to the simpler and conceptually clearer $U(1)$ Chern-Simons case discussed in section \ref{sec:BulkCSOperators}.  However, we will provide more details about the relationship between diffeomorphisms and gravitational dressing, and explain how some simple correlation functions can be computed.

\subsection{Review of Protofields in Fefferman-Graham Gauge}
\label{sec:ProtofieldFGGauge}

In recent work \cite{Anand:2017dav, Chen:2017dnl, Chen:2018qzm}, an exact gravitational  proto-field was defined with a very specific gravitational dressing.  
The dressing was determined implicitly, by fixing to a Fefferman-Graham or Banados gauge \cite{Banados:1998gg} where the metric takes the form
\begin{equation}
\label{eq:FGMetrichat}
d\hat{s}^{2}=\frac{dy^{2}+dzd\bar{z}}{y^{2}}-\frac{6T(z)}{c}dz^{2}-\frac{6\bar{T}(\bar{z})}{c}d\bar{z}^{2}+y^{2}\frac{36T(z)\bar{T}(\bar{z})}{c^2}dzd\bar{z}.
\end{equation}
Away from  sources of bulk energy, $d\hat{s}^{2}$ may be viewed as an operator whose VEV corresponds to the semiclassical metric.  That is, for a CFT state $|\psi\rangle$, the semiclassical metric will be given by $ds_{|\psi\rangle}^2=\langle\psi|d\hat{s}|\psi\rangle$, which is the RHS of the above equation with $T(z)\rightarrow\langle\psi|T(z)|\psi\rangle$ and $\bar{T}(z)\rightarrow\langle\psi|\bar{T}(z)|\psi\rangle$. In this subsection, we are considering the case where the CFT is living on a flat Euclidean plane with coordinates $(z,\bar z)$.
So for the CFT vacuum $|0\rangle$,  we have $\langle0|T(z)|0\rangle=0$ and $\langle0|\bar{T}(z)|0\rangle=0$, and the bulk metric is the Poincare metric $ds_{|0\rangle}^2=\frac{dy^2+dzd\bar{z}}{y^2}$.


Once we gauge fix,  Virasoro symmetry transformations extend to a unique set of vector fields in the bulk.  Demanding that $\phi(y,z, \bar z)$ transforms as a scalar field under the corresponding infinitesimal diffeomorphisms then implies that it must satisfy the  bulk primary conditions \cite{Anand:2017dav}
\begin{equation}
\label{eq:StandardBulkPrimaryConditions}
	[L_n, \phi(y, 0,0)] = 0, \quad  [\bar{L}_n, \phi(y, 0,0)] = 0, \quad n\ge2,
\end{equation}
along with the condition that in the vacuum $ds_{|0\rangle}^2=\frac{dy^2+dzd\bar{z}}{y^2}$, the bulk-boundary propagator is
\begin{equation}\label{eq:BulkBounaryVacuumPoincare}
	\langle\mathcal{O}(z, \bar{z}) \phi(y, 0,0)\rangle =\frac{y^{2 h_{L}}}{\left(y^{2}+z \bar{z}\right)^{2 h_{L}}}.
\end{equation}
These conditions uniquely determine $\phi(y,0,0)$ as a CFT operator defined by its series expansion in the radial $y$ coordinate:
\begin{equation}\label{eq:OriginalBulkProtoField}
	\phi(y, 0,0)=y^{2 h_{L}} \sum_{N=0}^{\infty} \frac{(-1)^{N} y^{2 N}}{N !\left(2 h_{L}\right)_{N}} \mathcal{L}_{-N} \bar{\mathcal{L}}_{-N} \mathcal{O}(0).
\end{equation}
This is a bulk proto-field in the metric of equation (\ref{eq:FGMetrichat}).
The $\mathcal{L}_{-N}$ are polynomials in the Virasoro generators
at level $N$, with coefficients that are rational functions of the
dimension $h$ of the scalar operator $\CO$ and of the central charge
$c$. They are obtained by solving the bulk-primary conditions (\ref{eq:StandardBulkPrimaryConditions})
exactly (no large $c$ expansion is required)\footnote{As a concrete example 
\begin{equation}
\mathcal{L}_{-2}=\frac{(2h+1)(c+8h)}{(2h+1)c+2h(8h-5)}\left(L_{-1}^{2}-\frac{12h}{c+8h}L_{-2}\right).
\end{equation}
}. And as shown in section 3.2.2 of \cite{Anand:2017dav},
they can be written formally in terms of quasi-primaries as 
\begin{equation}\label{eq:LNInQuasiPrimary}
\mathcal{L}_{-N}\mathcal{O}=L_{-1}^{N}\co+N!(2h)_{N}\sum_{n=2}^{N}\sum_{i}\frac{L_{-1}^{N-n}\mathcal{O}_{h+n}^{(i)}}{\left|L_{-1}^{N-n}\mathcal{O}_{h+n}^{(i)}\right|^{2}}
\end{equation}
where $\mathcal{O}_{h+n}^{(i)}$ is the $i$th quasi-primary\footnote{In prior work \cite{Anand:2017dav, Chen:2017dnl, Chen:2018qzm} the notation $\cl_{-n}^{\text{quasi},i}\co$ has been used,
which is  $\co_{h+n}^{\left(i\right)}$ here.} at level
$n$ and the denominator is the norm of the corresponding operator. In writing this equation, we have used the fact that the quasi-primaries can be chosen so that they are orthogonal to each other. This will be a very useful property of $\cl_{-N}\co$ in later discussions.

The correlation function $\langle\phi\CO T\rangle$ was computed and it was found that \cite{Anand:2017dav}
\begin{equation}\label{eq:phiOT1708}
	\frac{\left\langle\phi(y, 0,0) \mathcal{O}(z, \bar{z}) T\left(z_{1}\right)\right\rangle}{\langle\phi(y, 0,0) \mathcal{O}(z, \bar{z})\rangle}=\frac{h z^{2}}{z_{1}^{3}\left(z_{1}-z\right)^{2}}\left(z_{1}+\frac{2 y^{2}\left(z_{1}-z\right)}{y^{2}+z \bar{z}}\right) .
\end{equation}
Notice that this correlator has a cubic singularity when the position $z_1$ of the energy-momentum tensor $T$  approaches the point $(0,0)$. Roughly speaking, this singularity is the boundary imprint of the energy-momentum of the bulk field.  It demonstrates the presence of a gravitational dressing connecting $\phi$ to the boundary point $(0,0)$.  The boundary energy-momentum tensor $T(z_1)$ can detect this dressing via the cubic singularity, just as the current $J(z_1)$ registered the charge of a bulk field by a similar, but lower-order, singularity, as discussed in section \ref{sec:SingularitiesGaussLaw}.  On a more formal level, contour integrals $\oint z_1^{2+n} T(z_1)$ surrounding the singularity pick out the action of conformal generators $L_{n}$ on the bulk field $\phi$.

 Using our more general notation $\phi(y,z,\bar z;z_0,\bar z_0)$, the proto-field $\phi(y,0,0)$ given in equation (\ref{eq:OriginalBulkProtoField}) is  the gauge-invariant operator $\phi(y,0,0; 0,0)$ evaluated in the Fefferman-Graham gauge. The gravitational dressing  is in the $\hat y$ direction. In the next subsection, we are going to construct bulk proto-fields that have  more general  dressings.

\subsection{Natural Gravitational Dressings from Diffeomorphisms}
\label{sec:DressingfromDiffeo}

\begin{figure}[t]
\begin{center}
\includegraphics[width=0.85\textwidth]{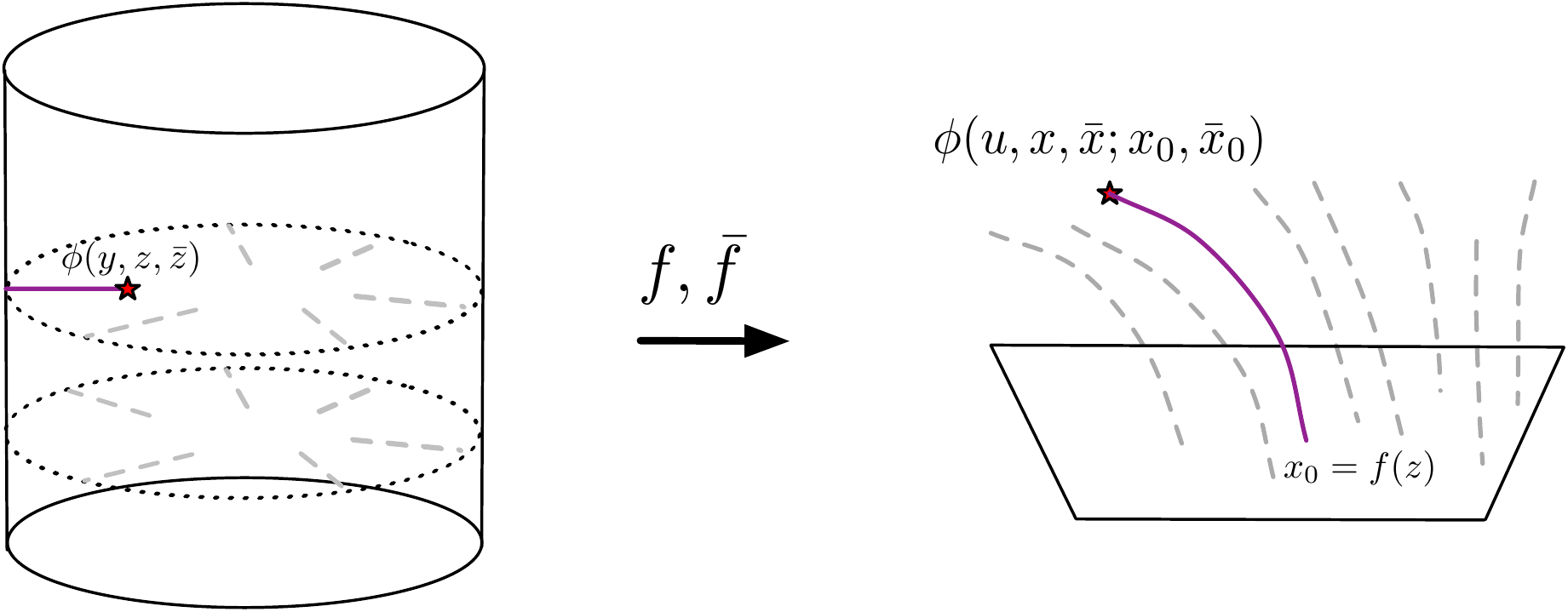}
\caption{ This figure suggests the use of a diffeomorphism to generate a natural gravitational dressing for a CFT living on a curved space, such as the cylinder.  A curved dressing in the $(u, x, \bar x)$ coordinates, with the CFT living on a plane, is mapped using equation (\ref{eq:VacuumAdSDiffeomorphism}) to a radial dressing in global AdS and BTZ black hole backgrounds. }
\label{fig:BendingGravitationalDressing}
\end{center}
\end{figure}

In many situations it is convenient to place the CFT on a two-dimensional surface other than the  plane. For example, in the presence of a BTZ black hole it's more natural to place the CFT on a cylinder, as will be discussed in section \ref{sec:HeavyLightCorrelator}.  In this section we will demonstrate how to define scalar proto-fields with a correspondingly natural dressing, greatly generalizing the results of section \ref{sec:ProtofieldFGGauge}.  The  ideas are illustrated in figure \ref{fig:BendingGravitationalDressing}.

General vacuum AdS$_3$ metrics with different boundary geometries can be obtained from the CFT on a plane
\begin{equation}\label{eq:uxxbPoincare}
	d s^{2}=\frac{d u^{2}+d x d \bar{x}}{u^{2}}
\end{equation}
by the following coordinate transformation \cite{Roberts:2012aq}
\begin{align} \label{eq:VacuumAdSDiffeomorphism}
u &=y \frac{4\left(f^{\prime}(z) \bar{f}^{\prime}(\bar{z})\right)^{\frac{3}{2}}}{4 f^{\prime}(z) \bar{f}^{\prime}(\bar{z})+y^{2} f^{\prime \prime}(z) \bar{f}^{\prime \prime}(\bar{z})} \nn\\ 
x &=f(z)-\frac{2 y^{2}\left(f^{\prime}(z)\right)^{2} \bar{f}^{\prime \prime}(\bar{z})}{4 f^{\prime}(z) \bar{f}^{\prime}(\bar{z})+y^{2} f^{\prime \prime}({z}) \bar f^{\prime \prime}(\bar{z})} 
\\ 
\bar{x} &=\bar{f}(\bar{z})-\frac{2 y^{2}\left(\bar{f}^{\prime}(\bar{z})\right)^{2} f^{\prime \prime}(z)}{4 f^{\prime}(z) \bar{f}^{\prime}(\bar{z})+y^{2} f^{\prime \prime}(z) \bar{f}^{\prime \prime}(\bar{z})}. \nn\end{align}
The resulting  vacuum AdS$_3$ metric is then given by 
\begin{equation}\label{eq:GeneralVacuumMetrics}
	d s^{2}=\frac{d y^{2}+d z d \bar{z}}{y^{2}}-\frac{1}{2} S_{f}(z) d z^{2}-\frac{1}{2} \bar{S}_{\bar{f}}(\bar{z}) d \bar{z}^{2}+y^{2} \frac{S_{f}(z) \bar{S}_{\bar{f}}(\bar{z})}{4} d z d \bar{z}.
\end{equation}
where $S_f(z)\equiv\frac{f^{\prime\prime\prime}(z)}{f^{\prime}(z)}-\frac{3}{2}\left(\frac{f^{\prime\prime}(z)}{f^{\prime}(z)}\right)^{2}$
is the Schwarzian derivative.

In this section, we would like to obtain a formula for a bulk proto-field $\phi(y,z,\bar z)$ with a gravitational dressing in the $\hat{y}$ direction. The difference between this section and section \ref{sec:ProtofieldFGGauge} is that in this section, the CFT is living on a different 2d surface. This may seem confusing, since the  boundary here is also described by $(z,\bar z)$, but here  the energy-momentum tensor  has a non-vanishing VEV given by $\langle0|T(z)|0\rangle=\frac{c}{12}S_f$.  For example, if $f(z)=e^{z}$ and $\bar f(\bar z)=e^{\bar z}$ (see Section \ref{sec:PhiOTGlobalAdS} for more details about this example), then ($z,\bar z$) are naturally coordinates on a cylinder and $S_f=-\frac{1}{2}$ (i.e. $\langle0|T(z)|0\rangle=-\frac{c}{24}$, which is the expectation value of the energy-momentum tensor on the cylinder). In other words, the metric dual to the CFT vacuum $|0\rangle$ here is given by (\ref{eq:GeneralVacuumMetrics}) with a $S_f \neq 0$, whereas in section \ref{sec:ProtofieldFGGauge} the metric dual to the CFT vacuum is given by the Poincare metric $ds^2=\frac{dy^2+dzd\bar z}{y^2}$ with planar boundary.

To obtain a bulk proto-field  operator $\phi(y,z,\bar z)$ with a gravitational dressing in the $\hat{y}$ direction, we can consider what this operator corresponds to in the $(u,x,\bar x)$ coordinates of equation (\ref{eq:uxxbPoincare}). The  position $(y,z,\bar z)$ of this operator will be mapped to $(u(y,z,\bar z),x(y,z,\bar z),\bar x(y,z,\bar z))$, and the boundary point of the gravitational dressing $(y=0,z,\bar z)$ will be mapped to the boundary point $(u=0, x_0=f(z),\bar x_0=\bar f(\bar z)$). Therefore we can also denote the operator $\phi(y,z,\bar z)$ as follows 
\begin{equation}
	\phi(y,z,\bar z)\equiv\phi(u, x,\bar x;x_0,\bar x_0)
\end{equation}
with the understanding that  $(u,x,\bar x)$ are functions of $(y,z,\bar z)$ as given by (\ref{eq:GeneralVacuumMetrics}) and $x_0=f(z)$ and $\bar x_0=\bar f(\bar z)$. Although in the $(y,z,\bar z)$ coordinates, the gravitational dressing is in the $\hat{y}$ direction, in the $(u,x,\bar x)$ coordinates the dressing is not in the $\hat{u}$ direction, since typically $x_0\ne x$ and $\bar x_0\ne \bar x$. 

These observations imply that in the $(u,x,\bar x)$ coordinates, the bulk-boundary propagator should be given by
\begin{equation}
\label{eq:BulkBoundaryuxxb}
	\left\langle\phi\left(u, x, \bar{x} ; x_{0}, \bar{x}_{0}\right) \mathcal{O}\left(x_{1}, \bar{x}_{1}\right)\right\rangle=\left(\frac{u}{u^{2}+\left(x-x_{1}\right)\left(\bar{x}-\bar{x}_{1}\right)}\right)^{2 h}.
\end{equation}
The new operator must also satisfy the bulk-primary conditions 
\begin{equation}
\label{eq:BulkPrimaryuxxb}
[L_{n}^{\left(x_{0}\right)}, \phi\left(u, x, \bar{x} ; x_{0}, \bar{x}_{0}\right)] = 0, \quad [\bar{L}_{n}^{\left(\bar{x}_{0}\right)},  \phi\left(u, x, \bar{x} ; x_{0}, \bar{x}_{0}\right)] = 0, \quad n \geq 2
\end{equation}
The above two conditions uniquely fix $\phi\left(u, x, \bar{x} ; x_{0}, \bar{x}_{0}\right)$ to be\footnote{Performing one of the summations allows us to simplify and write $\phi$ as
\begin{equation}
\phi\left(  u, x ,\bar x ;  x_0, \bar x_0  \right)=u^{2h}\sum_{N,\bar{N}=0}^{\infty}\frac{(\Delta x)^{N}(\Delta \bar x)^{\bar{N}}}{N!\bar{N}!}\ _{2}F_{1}\left(-N,-\bar{N},2h,-\frac{u^{2}}{\Delta x \Delta \bar x}\right)\cl_{-N}\bar{\cl}_{-\bar{N}}\co\left(x_{0},\bar{x}_{0}\right).
\end{equation}
where we used $\Delta x = x - x_0$ and $ \Delta \bar x = \bar x - \bar x_0$ for concision.
}  
\begin{equation}\label{eq:Phiuxxbx0xb0}
	\phi\left(u, x, \bar{x} ; x_{0}, \bar{x}_{0}\right)=\sum_{n=0}^{\infty} \sum_{m, \bar{m}=0}^{\infty}  \frac{(-1)^n u^{2 h+2 n}}{n! (2h)_n} \frac{\left(x-x_{0}\right)^{m}\left(\bar{x}-\bar{x}_{0}\right)^{\bar{m}}}{m ! \bar{m} !} \mathcal{L}_{-n-m} \bar{\mathcal{L}}_{-n-\bar{m}} \mathcal{O}\left(x_{0}, \bar{x}_{0}\right).
\end{equation}
Note that the $\mathcal{L}_{-n-m}$ and $\bar{\mathcal{L}}_{-n-\bar{m}}$ here are defined on the boundary complex plane $(x,\bar x)$,  where $T(x), \bar T(\bar x)$ are quantized by expanding around the point $(x_0, \bar x_0)$. It's obvious that the above equation satisfies the bulk-primary conditions (\ref{eq:BulkPrimaryuxxb}), since the $\cl_{-N}$s and $\bar{\cl}_{- \bar N}$s are constructed as solutions to such conditions.
Using the property (\ref{eq:LNInQuasiPrimary}) of the $\cl_{-N}$s, it's easy to see why $\phi\left(u, x, \bar{x} ; x_{0}, \bar{x}_{0}\right)$ has the correct bulk-boundary propagator (\ref{eq:BulkBoundaryuxxb}) with $\co(x_1,\bar x_1)$, since the quasi-primary terms in $\mathcal{L}_{-n-m} \bar{\mathcal{L}}_{-n-\bar{m}} \mathcal{O}$ will not contribute in this two-point function. So when computing (\ref{eq:BulkBoundaryuxxb}), we can simply replace the $\mathcal{L}_{-n-m} \bar{\mathcal{L}}_{-n-\bar{m}} \mathcal{O}$ with $L_{-1}^{n+m}\bar{L}_{-1}^{n+\bar{m}}\co$, and the sums over $m$ and $\bar m$ becomes translation operators. We then have $\phi\left(u, x, \bar{x} ; x_{0}, \bar{x}_{0}\right)\rightarrow\sum_{n=0}^{\infty} \frac{(-1)^n u^{2 h+2 n}}{n! (2h)_n}L_{-1}^n\bar{L}_{-1}^{n}\co(x,\bar x)$, which will give us the desired bulk-boundary propagator (\ref{eq:BulkBoundaryuxxb}).

In prior work \cite{Anand:2017dav} the special case $f\left(z\right)=z,\bar{f}\left(\bar{z}\right)=z$
was studied, so that $y=u,x=z,\bar{x}=\bar{z}$, and only the terms
with $m=\bar{m}=0$ contribute in the above equation. In that case
the result reduces to 
\begin{equation}
\phi\left(y,z,\bar{z};z,\bar z\right)=\sum_{n=0}^{\infty} \frac{\left(-1\right)^{n}y^{2h+2n}}{n!\left(2h\right)_{n}}\cl_{-n}\bar{\cl}_{-n}\co\left(z,z\right)=\phi\left(y,z,\bar{z}\right)
\end{equation}
where $\phi\left(y,z,\bar{z}\right)$ here is exactly the bulk proto-field defined in section \ref{sec:ProtofieldFGGauge}. And in this case, the gravitational dressing points in the $\hat u$ direction, as it coincides with the $\hat y$ direction.

If we take the limit that $y\rightarrow0$, then  this also implies $u\to0$, and also
$\left(x-x_0\right)$, $\left(\bar{x}-\bar{x}_0\right)\to 0$, as can be seen from equation (\ref{eq:VacuumAdSDiffeomorphism}), so we
find 
\begin{equation}
\lim_{y\rightarrow0}y^{-2h}\phi\left(y,z,\bar{z}\right)=\left(f'\left(z\right)\bar{f}'\left(\bar{z}\right)\right)^{h}\co\left(f\left(z\right),\bar{f}\left(\bar{z}\right)\right)=\co\left(z,\bar{z}\right)
\end{equation}
where the Jacobian factor $\left(f'\left(z\right)\bar{f}'\left(\bar{z}\right)\right)^{h}$
comes from $\lim_{y\rightarrow0}u$ using equation (\ref{eq:VacuumAdSDiffeomorphism}). This is exactly what we expect for a bulk operator in the ($y, z,\bar z$) coordinates with a gravitational dressing in the $\hat{y}$ direction.

Up to now we have been discussing the operator $\phi(y,z,\bar z)$ in the CFT vacuum. But our definition also applies to general CFT states $|\psi\rangle$.  The semiclassical bulk metric associated with  $|\psi\rangle$ is 
\begin{equation} 
\label{eq:GeneralyzzbMetric}
	ds^2_{|\psi\rangle}=\frac{d y^{2}+d z d \bar{z}}{y^{2}}-\frac{6 T_\psi(z)}{c} d z^{2}-\frac{6 \bar{T}_\psi(\bar{z})}{c} d \bar{z}^{2}+y^{2} \frac{36 T_\psi(z) \bar{T}_\psi(\bar{z})}{c^2} d z d \bar{z}
\end{equation}
where $T_\psi(z)\equiv\langle\psi|T(z)|\psi\rangle$ and similarly for $\bar{T}_\psi(\bar z)$. Note that $T_\psi(z)$ here is the expectation value of the energy-momentum tensor in $|\psi\rangle$, where the CFT lives on a general 2d surface defined via $f, \bar f$. $T_\psi(z)$ is related to the $\langle\psi|T(x)|\psi\rangle$ (since we are focusing on the boundary here, we have $x=f(z)$) via the usual transformation rule for the energy-momentum tensor
\begin{equation}\label{eq:TforSources}
		T_\psi(z)=f'(z)^2 \langle\psi|T(x)|\psi\rangle+\frac{c}{12}S_f,
\end{equation}
where again, $S_f$ is the Schwarzian derivative. Since on the boundary, $x,\bar x$ are the coordinates on a complex plane,  we have $\langle0|T(x)|0\rangle=0$. So in the vacuum, the metric reduces to that of equation (\ref{eq:GeneralVacuumMetrics}).

An interesting example that we will consider later is a map to the cylinder via $f(z)=e^{z}$ and $\bar f(\bar z)=e^{\bar z}$, where the bulk metric dual to the CFT vacuum is  global AdS$_3$. We will  study   correlators in a heavy  state $|\psi\rangle=|\co_H\rangle$, so that the semiclassical bulk geometry is a BTZ black hole. 
Specifically, we will study  the bulk-boundary propagator $\langle\mathcal{O}_H|\phi\co|\co_H\rangle$  in such a heavy state  in section \ref{sec:HeavyLightCorrelator}.

\subsubsection*{Semiclassical Correlators in General Backgrounds}

We would like the bulk proto-field to have the property that in \emph{any} semiclassical background arising from heavy sources in a CFT state $| \psi \>$, the bulk-boundary propagator $\< \psi | \phi \CO | \psi \>$ takes the correct form.  Formally, this means that as $c \to \infty$ with heavy source dimensions $h_H \propto c$, but with the dimension of $\CO$ fixed, this correlator must obey the free bulk wave equation in the associated metric of equation (\ref{eq:GeneralyzzbMetric}).  Let us explain why this must be the case.

Given any classical fields $T_{\psi}, \bar T_\psi$ forming the metric of equation (\ref{eq:GeneralyzzbMetric}), we will assume that we can identify new diffeomorphisms $g(z), \bar g(\bar z)$ so that the metric (\ref{eq:GeneralyzzbMetric}) arises from a diffeomorphism (\ref{eq:VacuumAdSDiffeomorphism}) with $g, \bar g$ replacing $f, \bar f$.  Note that while $f, \bar f$ connect the empty Poincar\'e patch to a corresponding CFT vacuum (\ref{eq:GeneralVacuumMetrics}), the new functions $g, \bar g$ relate the empty Poincar\'e patch to a Fefferman-Graham gauge metric (\ref{eq:GeneralyzzbMetric}) where there are heavy sources\footnote{We'll give an example of this statement in section \ref{sec:HeavyLightCorrelator}.}. 

The bulk primary conditions of equation (\ref{eq:BulkPrimaryuxxb}) were chosen to guarantee that $\phi$ transforms as a scalar field in Fefferman-Graham gauge, and so it must transform as a scalar under the map from $(u,x,\bar x) \leftrightarrow (y, z, \bar z)$  induced by $g, \bar g$.  Then the condition (\ref{eq:BulkBoundaryuxxb}) guarantees that the bulk boundary propagator in $(u, x, \bar x)$ coordinates takes the correct form, and so we can conclude that it must take the correct form in the semiclassical limit in the very non-trivial metric of equation (\ref{eq:GeneralyzzbMetric}).  Thus the conditions we have used to define $\phi$ ensure that its semiclassical correlators must take the expected form in the background of any sources, as long as $\phi$ does not intersect with these sources directly.

\subsection{$\langle\phi \mathcal{O} T\rangle$ on General Vacuum AdS$_3$ Metrics and Examples}\label{sec:PhiOTinArbit}

We can compute $\left\langle \phi\mathcal{O}T\right\rangle $ with $\phi$ given by (\ref{eq:Phiuxxbx0xb0}) on general vacuum AdS$_{3}$ metrics given by (\ref{eq:GeneralVacuumMetrics}).
Specifically, we study 
\begin{equation}
\left\langle T\left(x_{1}\right)\co\left(x_{2},\bar{x}_{2}\right)\phi\left(u,x,\bar{x};x_{0},\bar{x}_{0}\right)\right\rangle \label{eq:phiOTGeneral}
\end{equation}
with 
\begin{equation}
x_{0}=f\left(z\right),\bar{x}_{0}=\bar{f}\left(\bar{z}\right),x_{1}=f\left(z_{1}\right),x_{2}=f\left(z_{2}\right),\text{ and }\bar{x}_{2}=\bar{f}\left(\bar{z}_{2}\right)
\end{equation}
Although (\ref{eq:phiOTGeneral}) is written in terms of the $\left(u,x,\bar{x}\right)$
coordinates, it should really be understood as a correlation function
in the coordinates $\left(y,z,\bar{z}\right)$ with the metric
given by (\ref{eq:GeneralVacuumMetrics}) (and $S_{f}$ given by the Schwarzian derivative of $f\left(z\right)$).
More precisely, to obtain the correct $\left\langle \phi\co T\right\rangle $
in the $\left(y,z,\bar{z}\right)$ coordinates, we would simply
transform $T\left(x_{1}\right)$ and $\co\left(x_{2},\bar{x}_{2}\right)$
to $T\left(z_{1}\right)$ and $\co\left(z_{2},\bar{z}_{2}\right)$
using the usual transformation rules for the energy-momentum tensor
and primary operators, and leave $\phi$ as it is, since it's a bulk scalar field.

We can use the OPE of $T\co$ and $T\phi$ to evaluate this correlator.
Note that when using the OPE of $T\phi$, we need to expand $T$ around
$x_{0}$ instead of $x$. Explicitly, the singular terms in the OPEs are 
\begin{align}
T\left(x_{1}\right)\co\left(x_{2},\bar{x}_{2}\right) & \sim\frac{L_{-1}\co\left(x_{2},\bar{x}_{2}\right)}{x_{1}-x_{2}}+\frac{h\co\left(x_{2},\bar{x}_{2}\right)}{\left(x_{1}-x_{2}\right)^{2}}+\cdots\\
T\left(x_{1}\right)\phi\left(u,x,\bar{x};x_{0},\bar{x}_{0}\right) & \sim\frac{L_{-1}\phi\left(u,x,\bar{x};x_{0},\bar{x}_{0}\right)}{x_{1}-x_{0}}+\frac{L_{0}\phi\left(u,x,\bar{x};x_{0},\bar{x}_{0}\right)}{\left(x_{1}-x_{0}\right)^{2}}\nonumber \\
 & \quad+\frac{L_{1}\phi\left(u,x,\bar{x};x_{0},\bar{x}_{0}\right)}{\left(x_{1}-x_{0}\right)^{3}}+\cdots
\end{align}
where we've used bulk-primary conditions (\ref{eq:BulkPrimaryuxxb}) when writing down the OPE of $T\phi$.  So when computing $\left\langle \phi\co T\right\rangle $, we simply
include all the singular terms in these two OPEs. The $L_{-1}$s will
becomes just the differential operators $\partial_{x_{2}}$ and $\partial_{x_{0}}$,
while the terms with $L_{0}\phi$ and $L_{1}\phi$ can be computed
by commuting $L_{0}$ and $L_{1}$ with $\co\left(x_{2},\bar{x}_{2}\right)$
or by writing them in terms of contour integrals. Eventually, we get
\begin{align}
 & \left\langle T\left(x_{1}\right)\co\left(x_{2},\bar{x}_{2}\right)\phi\left(u,x,\bar{x};x_{0},\bar{x}_{0}\right)\right\rangle \nonumber \\
= & h\left[\frac{1}{\left(x_{1}-x_{2}\right)^{2}}-\frac{1}{\left(x_{1}-x_{0}\right)^{2}}-2\frac{\left(x_{2}-x_{0}\right)}{\left(x_{1}-x_{0}\right)^{3}}\right]\left\langle \phi\co\right\rangle \\
 & -2h\left[-\frac{1}{x_{1}-x_{2}}+\frac{x_{2}-x_{0}}{\left(x_{1}-x_{0}\right)^{2}}+\frac{1}{x_{1}-x_{0}}+\frac{\left(x_{2}-x_{0}\right)^{2}}{\left(x_{1}-x_{0}\right)^{3}}\right]\frac{\left(\bar{x}-\bar{x}_{2}\right)}{u^{2}+\left(x-x_{2}\right)\left(\bar{x}-\bar{x}_{2}\right)}\left\langle \phi\co\right\rangle ,\nonumber 
\end{align}
where $\left\langle \phi\co\right\rangle =\left(\frac{u}{u^{2}+\left(x-x_{2}\right)\left(\bar{x}-\bar{x}_{2}\right)}\right)^{2h}$. Correlation functions of the form $\langle\phi\co T\cdots T\rangle$ can then be computed recursively using the above OPEs and also the OPE of $TT$.

Here, we give two examples of the above result.
\subsubsection{Example 1: $\langle\phi\mathcal{O}T\rangle$ on Poincare AdS$_{3}$}
To obtain the $\langle\phi\mathcal{O}T\rangle$ on the Poincare AdS$_{3}$
metric, we use $f(z)=z,\bar{f}(\bar{z})=\bar{z}$. In this case, we
have $u=y,x=z$, and $\bar{x}=\bar{z}$, and the metric (\ref{eq:GeneralVacuumMetrics}) is simply $ds^{2}=\frac{dy^{2}+dzd\bar{z}}{y^{2}}$. After simplification, $\langle\phi\mathcal{O}T\rangle$ is given by 
\begin{equation}\label{eq:PhiOTPoincareAdS}
 \left\langle T\left(z_{1}\right)\co\left(z_{2},\bar{z}_{2}\right)\phi\left(y,z,\bar{z}\right)\right\rangle =  \frac{h\left(z_{2}-z\right)^{2}}{\left(z_{1}-z\right)^{3}\left(z_{1}-z_{2}\right)^{2}}\left(z_{1}-z+\frac{2y^{2}\left(z_{1}-z_{2}\right)}{y^{2}+\left(z-z_{2}\right)\left(\bar{z}-\bar{z}_{2}\right)}\right)\langle\phi\co\rangle
	\end{equation}
This is exactly the result (\ref{eq:phiOT1708}) obtained in \cite{Anand:2017dav}.

\subsubsection{Example 2: $\langle\phi\mathcal{O}T\rangle$ on Global AdS$_{3}$
\label{sec:PhiOTGlobalAdS}}

To obtain $\left\langle \phi\co T\right\rangle $ on global AdS$_{3}$,
whose boundary is a cylinder, we use $f(z)=e^{z}$ and $\bar{f}\left(\bar{z}\right)=e^{\bar{z}}$. From equation (\ref{eq:VacuumAdSDiffeomorphism}), we have 
\begin{equation}
u=\frac{4y\sqrt{\xi\bar{\xi}}}{4+y^{2}},\quad x=\frac{4-y^{2}}{4+y^{2}}\xi,\quad\bar{x}=\frac{4-y^{2}}{4+y^{2}}\bar{\xi}.
\end{equation}
where we've defined $\xi\equiv e^{z}$ and $\bar{\xi}\equiv e^{\bar z}$ for notational convenience. The resulting metric is 
\begin{equation}
ds^{2}=\frac{dy^{2}+dzd\bar z}{y^{2}}+\frac{dz^{2}}{4}+\frac{d\bar{z}^{2}}{4}+\frac{y^{2}}{16}dzd\bar{z},\label{eq:yzzbAdS3Global}
\end{equation}
which is related to the usual global AdS$_{3}$ metric 
\begin{equation}
ds^{2}=\left(r^{2}+1\right)dt_E^{2}+\frac{dr^{2}}{r^{2}+1}+r^{2}d\theta^{2}\label{eq:GlobalAdS3}
\end{equation}
by a simple coordinate transformation
\begin{equation}\label{eq:GlobalAdS3Transformation}
z=t_E+i\theta,\quad\bar{z}=t_E-i\theta,\quad y=2\left(\sqrt{r^{2}+1}-r\right).
\end{equation}
The bulk-boundary three-point function  $\left\langle \phi\co T\right\rangle $ on this metric is given
by
\begin{align}\label{eq:PhiOTGlobalAdS3}
 & \left\langle T\left(\xi_{1}\right)\co\left(\xi_{2},\bar{\xi}_{2}\right)\phi\left(u,x,\bar{x};\xi,\bar{\xi}\right)\right\rangle \\
= & \frac{h\left(\xi-\xi_{2}\right)^{2}}{\left(\xi_{1}-\xi\right)^{3}\left(\xi_{1}-\xi_{2}\right)^{2}}\left[\xi_{1}-\xi+\frac{4y^{2}\xi\left(\xi_{1}-\xi_{2}\right)\left(\bar{\xi}+\bar{\xi}_{2}\right)}{\left(\bar{\xi}\xi+\bar{\xi}_{2}\xi_{2}\right)\left(y^{2}+4\right)+\left(\bar{\xi}\xi_{2}+\bar{\xi}_{2}\xi\right)\left(y^{2}-4\right)}\right]\nonumber \\
 & \times\left\langle \co\left(\xi_{2},\bar{\xi}_{2}\right)\phi\left(u,x,\bar{x};\xi,\bar{\xi}\right)\right\rangle .\nonumber 
\end{align}
Here, $\left\langle \co\left(\xi_{2},\bar{\xi}_{2}\right)\phi\left(u,x,\bar{x};\xi,\bar{\xi}\right)\right\rangle $
is given by 
\begin{align}\label{eq:BulkBoundaryGlobaAdSPropagator}
\left\langle \co\phi\right\rangle  & =\left(\frac{u}{u^{2}+\left(x-\xi_{2}\right)\left(\bar{x}-\bar{\xi}_{2}\right)}\right)^{2h}\\
 & =\left(\frac{4y\sqrt{\xi\bar{\xi}}}{\left(\bar{\xi}\xi+\bar{\xi}_{2}\xi_{2}\right)\left(y^{2}+4\right)+\left(\bar{\xi}\xi_{2}+\bar{\xi}_{2}\xi\right)\left(y^{2}-4\right)}\right)^{2h}.\nonumber 
\end{align}
We have also obtained the same result using the effective field theory
of gravitons developed in \cite{Cotler:2018zff} (see appendix \ref{app:phiOT}
for details of that calculation).

\subsection{General Gravitational Dressings from Singularity Structure}
\label{sec:GeneralGravitationalDressing}

In this section we will define very general gravitational dressings through a procedure analogous to that of section \ref{sec:GeneralWilsonLines}, where we studied the $U(1)$ Chern-Simons case.  To define these bulk proto-fields we will leverage the singularity structure of correlators between $\phi, \CO,$ and any number of stress tensors. We review how the singularity structure of $T(z)$ is determined by Einstein's equations in appendix \ref{app:SingularityStructureT}, generalizing the $U(1)$ case of section \ref{sec:SingularitiesGaussLaw}. 
Here the CFT will be living on a flat 2d plane, with the CFT  vacuum state corresponding to the pure Poincare metric $ds_{|0\rangle}^2=\frac{du^2+dxd\bar{x}}{u^2}$.  

The proto-field operator we will identify takes a similar form to that derived in section \ref{sec:DressingfromDiffeo}, but our interpretation here will be different, and more abstract.  Another way of motivating this section would be to ask to what extent equation (\ref{eq:Phiuxxbx0xb0}) can be given a general meaning, independent of the diffeomorphism of equation (\ref{eq:VacuumAdSDiffeomorphism}).

\subsubsection*{A General Bulk Proto-Field}

The energy associated with the bulk operator $\phi$ must be reflected in the CFT by singularities in $T(x)$ correlators.
As in the $U(1)$ Chern-Simons case of section \ref{sec:GeneralWilsonLines},  we can construct a bulk proto-field $\phi(u, x, \bar x ; x_0, \bar x_0)$ by demanding:
\begin{enumerate}
\item $\< \phi \CO \>$ must be given by $\left\langle\phi\left(u, x, \bar{x} ; x_{0}, \bar{x}_{0}\right) \mathcal{O}\left(w, \bar w\right)\right\rangle=\left(\frac{u}{u^{2}+\left(x-w\right)\left(\bar{x}-\bar{w}\right)}\right)^{2 h}
$ in the vacuum.
\item Correlators $\< \CO(w, \bar w)  T(x_1) \cdots T(x_n) \bar T(x_1) \cdots \bar T(\bar x_{\bar n})  \phi(u, x, \bar x ; x_0, \bar x_0) \>$ only have poles of up to third order in the $x_i$, which can only occur when $x_i \to x_0$ (along with up to second order poles as $x_i \to w$), and equivalently for the antiholomorphic variables.
\end{enumerate}
Note that we allow poles $\propto \frac{1}{(x_i - x_0)^3}$ in $T(x_i)  \phi(u, x, \bar x ; x_0, \bar x_0)$, whereas only second order poles occur in the OPE of $T(x_i) \CO(w)$.  This is because $\phi$ contains descendants of $\CO$, including the first descendant $L_{-1} \CO \propto \partial \CO$, and such operators necessarily induce third-order poles.  But higher order singularities are excluded by our assumptions.

The unique  operator built from Virasoro descendants of a primary $\CO$ and satisfying these conditions is
\begin{equation}
\label{eq:GeneralBulkScalarDressing}
\phi\left(u, x, \bar x ;  x_0, \bar x_0 \right) =\sum_{n=0}^{\infty} \sum_{m, \bar{m}=0}^{\infty} u^{2 h+2 n} \frac{(-1)^n}{n! (2h)_n}  \frac{(x-x_0)^{m} (\bar x - \bar x_0)^{\bar{m}}}{m ! \bar m !} \mathcal{L}_{-n-m} \bar{\mathcal{L}}_{-n-\bar{m}} \mathcal{O}\left(x_{0}, \bar{x}_{0}\right)
\end{equation}
As a formal power series expansion in CFT operators, this is identical to equation (\ref{eq:Phiuxxbx0xb0}), except that $x_0$ and $\bar x_0$ are now arbitrary, instead of given by $f(z)$ and $\bar f(\bar z)$. As we will explain below, this object can be interpreted as a bulk proto-field with a dressing that follows the geodesic path from $(u, x, \bar x)$ to $(x_0, \bar x_0)$ in any geometry.

One way to interpret this expression is as a modification of equation (\ref{eq:OriginalBulkProtoField}) (with the notation $(y,z,\bar z)$ replaced by $(u,x,\bar x)$ here), where the boundary imprint of the bulk energy has been translated to $x_0, \bar x_0$.   In appendix \ref{app:MirageTranslations} we develop a theory of non-local `mirage translation' operators that move the local energy of CFT operators (singularities in $T(x)$ correlators) without moving the apparent location of an operator $\CO$ itself (so mirage translations leave OPE singularities between local primaries fixed).  Mirage translations also provide an independent motivation for the bulk primary conditions.  

To better understand equation (\ref{eq:GeneralBulkScalarDressing}), let's consider a few simplifying limits.  If $x_{0}=x$ and $\bar x_{0}=\bar x$, then only the $m=0$ terms contribute to equation (\ref{eq:GeneralBulkScalarDressing}), and $\phi$ reduces to the bulk field of equation (\ref{eq:OriginalBulkProtoField}).  To obtain $\CO(x_0,\bar x_0)$ as we take $u \to 0$, we need to simultaneously send $x \to x_0$; otherwise we obtain a non-local operator $\tilde \CO(x,\bar x;x_0,\bar x_0)$, corresponding to $\CO(x,\bar x)$ multiplied with an additional gravitational dressing on the boundary.  The  non-local operator $\tilde \CO(x,\bar x;x_0,\bar x_0)$ can be interpreted is the mirage translation of $\CO(x,\bar x)$, as discussed in appendix \ref{app:MirageTranslations}.

The non-gravitational limit  is also easy to understand.  When $c \to \infty$ with $h$ fixed, we have $\CL_{-N} \to L_{-1}^N$.  In this limit the sum over $m, \bar m$ simplifies into a pure translation, and the sums on $m$ simply convert $\CO(x_0, \bar x_0) \to \CO(x, \bar x)$.  Only the sum on $n$ remains, and this reconstructs  the non-interacting field defined in equation (\ref{eq:FreeBulkFieldyExpansion}).

More generally, all of the terms $\CL_{-N} \bar \CL_{-\bar N} \CO$ were constructed so that they only have OPE singularities of fixed order $\leq 3$ with $T(x_1)$, and these singularities only occur when $x_1 \to x_0$.  The field $ \phi(u, x, \bar x ; x_0, \bar x_0)$ inherits this property.  However, as the $\CL_{-N}$ act as a kind of modified translation, other local CFT operators constructed from other primaries will behave as though $\phi(u, x, \bar x ; x_0, \bar x_0)$ is located at $(u, x, \bar x)$ in the bulk.

\subsubsection*{The Gravitational Dressing Naturally Follows a Geodesic}

We glossed an important issue when defining equation (\ref{eq:GeneralBulkScalarDressing}): for this expression to be meaningful, we must have some way of determining where in the bulk this field lives.  In the CFT vacuum the operator is at $(u, x, \bar x)$ in the Poincar\'e patch metric.  This observation is sufficient to determine the location of $\phi$ in perturbation theory around the Poincar\'e patch metric.  But if we turn on heavy sources, the bulk metric will change by a finite amount.  We have not fixed to Fefferman-Graham gauge in $(u, x, \bar x)$ coordinates, so these coordinates are just labels, which only have unambiguous meaning in perturbation theory or in the limit  $u \to 0$.

We can partially resolve this issue by comparing with section \ref{sec:DressingfromDiffeo}.  It is easy to see that there exist an infinite family of $f(z), \bar f(\bar z)$ functions so that $x_0, \bar x_0 = f(z), \bar f(\bar z)$ and equation (\ref{eq:VacuumAdSDiffeomorphism}) relates $(u,x, \bar x)$ to $(y, z, \bar z)$.  For each such $f, \bar f$ we implicitly define a gauge choice in $(u,x, \bar x)$ by pulling back the Fefferman-Graham gauge of the $(y, z, \bar z)$ coordinate system to $(u,x, \bar x)$ via the diffeomorphism (\ref{eq:VacuumAdSDiffeomorphism}).   So the bulk location of the proto-field could be interpreted as a bulk proto-field in any of these gauges.

However, all interpretations of the proto-field share a common feature.  The gravitational dressing in $(y, z, \bar z)$ can be chosen to be a curve with constant $(z, \bar z)$, so  that it points in the  $\hat y$-direction.  In Fefferman-Graham gauge, these curves are all geodesics.  This means that the gravitational dressing of $\phi(u, x, \bar x; x_0, \bar x_0)$ will follow a geodesic from $(u, x,\bar x)$ in the bulk to $(x_0, \bar x_0)$ on the boundary, in any dynamical metric.

Thus we conclude that $\phi(u, x, \bar x; x_0, \bar x_0)$ will behave like a scalar field in the bulk defined by fixing to any gauge where the gravitational field does not fluctuate along the geodesic connecting $(u, x, \bar x)$ and $(x_0, \bar x_0)$.   That is, if $X^\mu(\lambda)$ are coordinates on this geodesic, then $h_{\mu \nu}(X(\lambda)) \dot X^\nu(\lambda) = 0$, where $h_{\mu \nu}$ is the deviation of the bulk metric from the Poincar\'e patch form.  This leaves the gauge choice elsewhere in spacetime almost entirely undetermined.

\subsubsection*{More General Dressings}

To obtain a more general class of  dressed bulk $\phi$ we can smear $\phi(X; x_0, \bar x_0)$ over $(x_0, \bar x_0)$  via
\be
\phi[\rho](X) \equiv \int dx_0 d\bar x_0  \, \rho(x_0,\bar x_0) \phi(X; x_0,\bar x_0)
\ee
with any positive function $\rho$ that integrates to one $\int d^2x_0\rho(x_0,\bar x_0)=1$. If we work in perturbation theory around the vacuum (or any fixed semiclassical metric), then the location of the protofield in $(u, x, \bar x)$ coordinates will be unambiguous.  Then we can obtain results similar to the $U(1)$ case in section \ref{sec:Smearingz0}.

\section{Recursion Relation for Bulk-Boundary Vacuum Blocks}
\label{sec:ZRR}

Gravitational dynamics can be probed with correlation functions of the bulk proto-field (\ref{eq:Phiuxxbx0xb0}). For example, bulk locality was studied in \cite{Chen:2017dnl} by computing the bulk two point function $\langle\phi\phi\rangle$ and Euclidean black hole horizons were investigated in \cite{Chen:2018qzm} by computing the vacuum blocks of heavy-light bulk-boundary correlator $\langle\co_H\phi_L\co_L\co_H\rangle$. In those works, recursions relations were derived for computing correlators involving the proto-field of section \ref{sec:ProtofieldFGGauge} (i.e., the special case of (\ref{eq:Phiuxxbx0xb0}) with $f(z)=z$ and $\bar f(\bar z)=\bar z$). Now that we have the more general bulk proto-field (\ref{eq:Phiuxxbx0xb0}), we can also derive recursion relations for computing\footnote{Although we don't develop it in this paper, the recursion relation for computing the bulk two-point function $\langle\phi\phi\rangle$ for $\phi$ given by (\ref{eq:Phiuxxbx0xb0}) should be very similar to the one derived in \cite{Chen:2017dnl}.} its correlators.

In this section, we are going to derive a recursion relation for computing
the Virasoro vacuum block contribution to
\begin{equation}
\left\langle \co_{H}\left(\infty\right)\co_{H}\left(1\right)\phi_{L}(u,x,\bar{x};x_{0},\bar{x}_{0})\co_{L}\left(0\right)\right\rangle,
\end{equation}
where $\phi_{L}(u,x,\bar{x};x_{0},x_{0})$ (with $h=\bar{h}=h_{L}$)
is given by equation (\ref{eq:Phiuxxbx0xb0}) and the coordinates $\infty,0,1$ are on the
 complex plane with $(x, \bar x)$ coordinates.  In order to be more general and include of case of section \ref{sec:GeneralGravitationalDressing}, we are going
to assume that $x_{0},\bar{x}_{0}$ are arbitrary (i.e., we are not
assuming that they are given by $f\left(z\right)$ and $\bar{f}\left(\bar{z}\right)$),
although this will not affect the discussion in this section. Although
we use the subscripts $H$ and $L$ which usually means ``heavy''
and ``light'', the conformal dimension $h_{H}$ and $h_{L}$ in this
section are arbitrary. We will study a special case of this result in
section \ref{sec:HeavyLightCorrelator}, and compute the bulk-boundary propagator
in a black hole microstate background.

\subsection{General Structure of the Vacuum Blocks}
As usual, the vacuum block $\cv_{0}$ is obtained via the  projection
\begin{equation}
\cv_{0}=\left\langle \co_{H}\left(\infty\right)\co_{H}\left(1\right)\cp_{0}\phi_{L}\left(u,x,\bar{x};x_{0},\bar{x}_{0}\right)\co_{L}\left(0\right)\right\rangle \label{eq:VaccumBlock}
\end{equation}
where $\cp_{0}$ is the projection operator into the vacuum module\footnote{To be clear, the projection operator for a representation of the Virasoro
algebra with lowest weight state $\left|\co_{h_{1}}\right\rangle $
factorizes, that is, $\cp_{h_{1}}=\cp_{h_{1}}^{\text{holo}}\cp_{h_{1}}^{\text{anti-\text{holo}}}$
and the holomorphic part is given symbolically by 
\begin{equation}
\cp_{h_{1}}^{\text{holo}}=\sum_{\left\{ m_{i}\right\} ,\left\{ n_{j}\right\} }\frac{L_{-m_{1}}\cdots L_{-m_{i}}|\co_{h_{1}}\rangle\left\langle \co_{h_{1}}|L_{n_{j}}\cdots L_{n_{1}}\right.}{\mathcal{N}_{\left\{ m_{i}\right\} ,\left\{ n_{j}\right\} }},
\end{equation}
where $\frac{1}{\mathcal{N}_{\left\{ m_{i}\right\} ,\left\{ n_{j}\right\} }}$
is  the inverse of the inner-product matrix between the
states.}
(including holomorphic and anti-holomorphic parts).  $\phi_{L}\left(u,x,\bar{x};x_{0},\bar{x}_{0}\right)$ of equation
(\ref{eq:Phiuxxbx0xb0}) can be simplified to the following form: 
\begin{equation}
\phi_{L}\left(u,x,\bar{x};x_{0},\bar{x}_{0}\right)=u^{2h}\sum_{N,\bar{N}=0}^{\infty}\ca_{N,\bar{N}}\left(u,\Delta x,\Delta\bar{x}\right)\frac{\mathcal{L}_{-N}\bar{\mathcal{L}}_{-\bar{N}}\mathcal{O}\left(x_{0},\bar{x}_{0}\right)}{\left(2h_{L}\right)_{N}\left(2h_{L}\right)_{\bar{N}}N!\bar{N}!}
\end{equation}
where $\Delta x=x-x_{0}$ and $\Delta\bar{x}=\bar{x}-x_{0}$,
and 
\begin{equation}
\ca_{N,\bar{N}}\left(u,\Delta x,\Delta\bar{x}\right)\equiv\left(2h_{L}\right)_{N}\left(2h_{L}\right)_{\bar{N}}\left(\Delta x\right)^{N}\left(\Delta\bar{x}\right)^{\bar{N}}\ _{2}F_{1}\left(-N,-\bar{N},2h_{L},-\frac{u^{2}}{\Delta x\Delta\bar{x}}\right).
\end{equation}
The extra factors of $\left(2h_{L}\right)_{N}\left(2h_{L}\right)_{\bar{N}}$
in the  expressions above  are inserted for later convenience.

Although the Virasoro vacuum block of (\ref{eq:VaccumBlock}) doesn't
factorize into holomorphic and anti-holomorphic parts, we can make use of the fact that it does factorizes for a specific
$N$ and $\bar{N}$, since the projection operator $\cp_{0}$ factorizes.
Similar to the case in \cite{Chen:2018qzm}, we can define the holomorphic
part of $\phi_{h}$ to be 
\begin{equation}
\tilde{\phi}_{h}^{\text{holo}}\left(u,x;x_{0}\right)\equiv\sum_{N=0}^{\infty}\frac{\mathcal{L}_{-N}\mathcal{O}_{h}\left(x_{0},\bar{x}_{0}\right)}{\left(2h_{L}\right)_{N}N!},
\end{equation}
and we'll obtain a recursion relation for computing the more general holomorphic
block: 
\begin{equation}
\mathcal{V}_{\mathrm{holo}}\left(h_{1},h_{2},c\right)\equiv\left\langle \mathcal{O}_{H}(\infty)|\mathcal{O}_{H}(1)\mathcal{P}_{h_{1}}^{\mathrm{holo}}\tilde{\phi}_{h_2}^{\text{holo}}\left(u,x;x_{0}\right)|\mathcal{O}_{L}(0)\right\rangle .\label{eq:HoloBlocks}
\end{equation}
Here the holomorphic projection operator $\mathcal{P}_{h_{1}}^{\mathrm{holo}}$
only includes the holomorphic descendants of $\left|\co_{h_{1}}\right\rangle $.
We are considering this more general block for the convenience of
discussing the recursion relation in the next subsection. Eventually,
we are interested in the vacuum block $\mathcal{V}_{\mathrm{holo}}\left(0,h_{L},c\right)$,
and it will be given in the following form 
\begin{equation}
\cv_{\text{holo}}\left(0,h_{L},c\right)=\frac{1}{x_{0}^{2h_{L}}}\sum_{N=0}^{\infty}\frac{1}{x_{0}^{N}}F_{N}\left(x_{0}\right)\label{eq:V0HoloExpand}
\end{equation}
$F_{N}\left(x_{0}\right)$ is an infinite series of $x_{0}$, starting
with $F_{N}\left(x_{0}\right)=1+\cdots$ (we'll explain how to obtain $F_N(x_0)$ in next subsection). The full vacuum block $\cv_{0}$
is then obtained via the following equation 
\begin{equation}
\cv_{0}=\left(\frac{u}{x_{0}\bar{x}_{0}}\right)^{2h_{L}}\sum_{N,\bar{N}=0}^{\infty}\frac{\ca_{N,\bar{N}}\left(u,\Delta x,\Delta\bar{x}\right)}{(\Delta x_0)^{N}(\Delta \bar x_0)^{\bar{N}}}F_{N}\left(x_{0}\right)F_{\bar{N}}\left(\bar{x}_{0}\right),\label{eq:V0Expand}
\end{equation}
where $F_{\bar{N}}\left(\bar{x}_{0}\right)$ is simply $F_{\bar{N}}(x_0)$ with $x_0$ replaced by $\bar x_0$.

\subsection{Recursion Relation}

Our task now is to obtain the recursion relation for computing $\mathcal{V}_{\mathrm{holo}}\left(h_{1},h_{2},c\right)$
based on the singularity structure of $\mathcal{V}_{\mathrm{holo}}\left(h_{1},h_{2},c\right)$
as a function of the central charge $c$. Actually, the recursion
relation for computing $\cv_{\text{holo}}$ here is almost the same
as the recursion in \cite{Chen:2018qzm} for computing the $\cv_{\text{holo}}$
in that case\footnote{The recursion relation in \cite{Chen:2018qzm} is an special case
of the recursion here, with $f\left(z\right)=z$ and $\bar{f}\left(\bar{z}\right)=\bar{z}$,
but the general structures are almost the same.}, except that the seed of the recursion is different. We reproduce
the recursion relation here for convenience 
\begin{equation}
\begin{aligned}\mathcal{V}_{\text{ holo }}\left(h_{1},h_{2},c\right)= & \mathcal{V}_{\text{ holo }}\left(h_{1},h_{2},c\rightarrow\infty\right)\\
 & +\sum_{m\geq2,n\geq1}\frac{R_{m,n}\left(h_{1},h_{2}\right)}{c-c_{m,n}\left(h_{1}\right)}\mathcal{V}_{\text{ holo }}\left(h_{1}\rightarrow h_{1}+mn,h_{2},c\rightarrow c_{mn}\left(h_{1}\right)\right)\\
 & +\sum_{m\geq2,n\geq1}\frac{S_{m,n}\left(h_{1},h_{2}\right)}{c-c_{m,n}\left(h_{2}\right)}\mathcal{V}_{\text{ holo }}\left(h_{1},h_{2}\rightarrow h_{2}+mn,c\rightarrow c_{mn}\left(h_{2}\right)\right).
\end{aligned}
\end{equation}
For more details about the meaning of the symbols $R_{m,n}$, $S_{m,n}$,
$c_{m,n}$ and how to solve this recursion relation, please see section
4 and appendix C of \cite{Chen:2018qzm}.

As mentioned above, the seed of the recursion $\mathcal{V}_{\text{ holo }}\left(h_{1},h_{2},c\rightarrow\infty\right)$
is different from that of {\cite{Chen:2018qzm}}. In the $c\rightarrow\infty$
limit, only global descendants of the intermediate state $\left|\co_{h_{1}}\right\rangle $
and global descendants of $\co_{h_{2}}$ contribute, so $\mathcal{V}_{\text{ holo }}\left(h_{1},h_{2},c\rightarrow\infty\right)$
is actually the global block, i.e. 
\begin{equation}
G\left(h_{1},h_{2}\right)\equiv\mathcal{V}_{\text{ holo }}\left(h_{1},h_{2},c\rightarrow\infty\right).
\end{equation}
$G\left(h_{1},h_{2}\right)$ can be obtain by direct computation,
as follows 
\begin{align}\label{eq:GlobalBlocks}
G\left(h_{1},h_{2}\right) & =\sum_{m_{1},m_{2}=0}^{\infty}\frac{\left\langle \mathcal{O}_{H}|\mathcal{O}_{H}|L_{-1}^{m_{1}}\co_{1}\right\rangle \left\langle L_{-1}^{m_{1}}\co_{1}\left|L_{-1}^{m_{2}}\mathcal{O}_{h_{2}}\left(x_{0}\right)\right|\co_{L}\right\rangle }{\left|L_{-1}^{m_{1}}\right|\co_{1}\left.\rangle\right|^{2}\left|L_{-1}^{m_{2}}\right|\co_{2}\left.\rangle\right|^{2}}\nonumber \\
 & =\frac{1}{x_{0}^{2h_{L}}}\sum_{m_{1},m_{2=0}}^{\infty}\frac{\left(h_{1}\right)_{m_{1}}\rho_{m_{1},m_{2},0}\left(h_{1},h_{2},h_{L}\right)}{\left(2h_{1}\right)_{m_{1}}m_{1}!\left(2h_{2}\right)_{m_{2}}m_{2}!}x_{0}^{h_{1}+m_{1}}\tilde{x}_{0}^{m_{2}+h_{2}-h_{L}}
\end{align}
where $\tilde{x}_{0}\equiv\frac{1}{x_{0}}$. We are using $\tilde{x}_{0}$
here for keeping tracking the origin of each term (so that we can
obtain the $F_{N}$ in (\ref{eq:V0HoloExpand}) after we compute $\mathcal{V}_{\mathrm{holo}}$)
and for the convenience of implementation in Mathematica. Here, $\rho_{i,j,k}\left(h_{1},h_{2},h_{3}\right)$
is the three point functions of global descendants, and it's given
by \cite{Alkalaev:2015fbw}
\begin{align}
\rho_{i,j,k}\left(h_{1},h_{2},h_{3}\right) & \equiv\left\langle L_{-1}^{i}\co_{h_{1}}|L_{-1}^{j}\co_{h_{2}}\left(1\right)|L_{-1}^{k}\co_{h_{3}}\right\rangle \\
 & =\left(h_{1}+i-h_{2}-j+1-h_{3}-k\right)_{j}s_{ik}\left(h_{1},h_{2},h_{3}\right)\nonumber 
\end{align}
with 
\begin{equation}
\begin{aligned}s_{ik}\left(h_{1},h_{2},h_{3}\right)= & \sum_{p=0}^{\min(i,k)}\frac{i!}{p!(i-p)!}\left(2h_{3}+k-p\right)_{p}(i-p+1)_{p}\\
 & \times\left(h_{3}+h_{2}-h_{1}\right)_{k-p}\left(h_{1}+h_{2}-h_{3}+p-k\right)_{i-p}.
\end{aligned}
\end{equation}
And we only need the $\rho_{i,j,k}$ with $k=0$.

As in \cite{Chen:2018qzm}, solving the recursion produces  $\cv_{\text{holo}}\left(h_{1},h_{2},c\right)$
as the following sum\footnote{As shown in section 3.2.2 of \cite{Anand:2017dav}, solving the bulk primary conditions will give us $\cl_{-n}\co$ in terms of quasi-primaries and their global
descendants (see the paragraph below (\ref{eq:LNInQuasiPrimary}) for explanations of the notations here)
\begin{equation}
\cl_{-N}\co=N!\left(2h\right)_{N}\sum_{n=0}^{N}\sum_{i}\frac{L_{-1}^{N-n}\co_{h+n}^{\left(i\right)}}{\left|L_{-1}^{N-n}\co_{h+n}^{\left(i\right)}\right|^{2}}\label{eq:LNQuasiPrimaries}
\end{equation}
Similarly, $\cp_{h_{1}}^{\text{holo}}$ can be written in terms of
quasi-primaries and their global descendants. So $\cv_{\text{holo}}\left(h_{1},h_{2},c\right)$
defined in (\ref{eq:HoloBlocks}) can also be written as a sum over
quasi-primaries and their global descendants. In this way, it's easier
to see why $\cv_{\text{holo}}\left(h_{1},h_{2},c\right)$ can be decomposed
into a sum over global blocks as in (\ref{eq:SumOverGlobalBlocks}).}
\begin{equation}
\cv_{\text{holo}}\left(h_{1},h_{2},c\right)=\sum_{m,n=0}^{\infty}C_{m,n}G\left(h_{1}+m,h_{2}+n\right).\label{eq:SumOverGlobalBlocks}
\end{equation}
Here, $G\left(h_{1}+m,h_{2}+n\right)$ is the global block (\ref{eq:GlobalBlocks})
with $h_{1}\rightarrow h_{1}+m$ and $h_{2}\rightarrow h_{2}+n$.
The summand $C_{m,n}G\left(h_{1}+m,h_{2}+n\right)$ is the contribution
to $\cv_{\text{holo}}$ from all the level $m$ quasi-primaries $\co_{h_{1}+m}^{\left(i\right)}$
and level $n$ quasi-primaries $\co_{h_{2}+n}^{\left(j\right)}$ (denoted
as $\cl_{-m}^{\text{quasi},i}\co_{h_{1}}$ and $\cl_{-n}^{\text{quasi},j}\co_{h_{2}}$
in previous papers \cite{Chen:2017dnl,Chen:2018qzm}) , and their
global descendants. The coefficients $C_{m,n}$ are exactly the same
as the coefficients in \cite{Chen:2018qzm}. Basically, they encode
the three point functions of the quasi-primaries with primaries as
follows 
\begin{equation}
C_{m,n}=\sum_{i,j}\frac{\left\langle \mathcal{O}_{H}|\mathcal{O}_{H}(1)|\mathcal{O}_{h_{1}+m}^{\left(i\right)}\right\rangle \left\langle \mathcal{O}_{h_{1}+m}^{\left(i\right)}\left|\mathcal{O}_{h_{2}+n}^{\left(j\right)}\left(1\right)\right|\mathcal{O}_{L}\right\rangle }{\left|\mathcal{O}_{h_{1}+m}^{\left(i\right)}\right|^{2}\left|\mathcal{O}_{h_{2}+n}^{\left(j\right)}\right|^{2}},
\end{equation}
where we've assumed that the quasi-primaries are orthogonalized and
the sum $\sum_{i,j}$ is over all level $m$ and level $n$ quasi-primaries.
In section 4 and appendix C of \cite{Chen:2018qzm}, we discussed
in detail how to obtain the above result and how to compute them using
the recursion.

After obtaining $\cv_{\text{holo}}(0,h_L,c)$ as a polynomial of $\tilde{x}_0$, the  coefficient of $\tilde{x}_0^N$ is the  $F_N(x_0)$ in (\ref{eq:V0HoloExpand}). This is why we keep $\tilde{x}_0$ explicitly in the global blocks (\ref{eq:GlobalBlocks}), instead of using the fact that $\tilde{x}_0\equiv\frac{1}{x_0}$ to simplify the calculation of the global blocks, because that will mix $\frac{1}{x_0}$ with the $x_0$ in $F_N(x_0)$, and we will not be able to extract $F_N(x_0)$. After obtaining $F_N(x_0)$, we can simply use equation (\ref{eq:V0Expand}) to compute the full vacuum block $\cv_0$. The Mathematica code for implementing this recursion relation is attached with this paper. 

Generally, the recursion relations for computing the boundary Virasoro
blocks \cite{ZamolodchikovRecursion, Zamolodchikovq, Cho:2017oxl} and the bulk-boundary Virasoro blocks consist of two
parts. One is the computation of the coefficients $C_{m,n}$. And
the other part is the computation of the global blocks. The computation
of $C_{m,n}$ is the most complicated part of the recursion relations,
but luckily, for most observables of interest, it's universal.
The difference between observables is in the global blocks, which are the seed for the recursion relations. 

\section{Heavy-Light Bulk-Boundary Correlator on the Cylinder}
\label{sec:HeavyLightCorrelator}

Our study of bulk reconstruction was motivated by the desire
to understand near horizon dynamics and the black hole information
paradox. This program may be advanced by computing correlation functions of
bulk proto-fields in a black hole microstate background. One object
of interest is the heavy-light bulk-boundary vacuum block $\cv_{0}$
of $\left\langle \co_{H}|\co_{L}\phi_{L}|\co_{H}\right\rangle $ for
$\phi_{L}$ defined in global AdS. When $\left|\co_{H}\right\rangle $
is dual to a BTZ black hole microstate ($h_{H}>\frac{c}{24}$), $\cv_{0}$
will be dual to the bulk-boundary propagator of the light operators
in such a background. In this section, we'll compute this vacuum block
using two different methods.

Our first method utilizes the recursion relations introduced in the last section.
This method will give us an exact result for the vacuum block as an
expansion in the kinematic variables, with coefficients evaluated
exactly at finite $c$, including all the gravitational interactions
between the light probe operator and the heavy state. Our second method
is based on the idea of bulk-boundary OPE blocks \cite{Anand:2017dav} (or bulk-boundary bi-local operators)  and
effective theory for boundary gravitons in AdS$_{3}$/CFT$_{2}$ \cite{Cotler:2018zff}.
This second method will give us the vacuum block in a large $c$ expansion
with $\frac{h_{H}}{c}$ fixed. We'll carry out the calculation up
to order $\frac{1}{c}$, which corresponds to the gravitational one-loop
correction to the bulk-boundary propagator in a microstate BTZ black
hole background. We have verified that the results from these methods
agree.

We will also show analytically that the one-loop corrections are singular at the Euclidean horizon.   This effect only arises because the $1/c$ corrections to the correlators are not periodic in Euclidean time \cite{Fitzpatrick:2015dlt, Fitzpatrick:2016ive}.  When they are interpreted as correlators in the BTZ geometry, they must have a branch cut at the horizon.  

Throughout this section,
we'll assume the following limit
\begin{equation}
c\rightarrow\infty,\quad h_{H}\sim O\left(c\right),\quad h_{L}\sim O\left(1\right),
\end{equation}
although our computation using the recursion relation is valid at finite
$c$.
We'll study the bulk protofield 
\[
\phi_{L}\left(y,z,\bar{z}\right)\equiv\phi_{L}\left(u,x,\bar{x};f(z),f(z)\right)
\]
 (with $h=\bar{h}=h_{L}$) to be given by equation (\ref{eq:Phiuxxbx0xb0})
with $f\left(z\right)=e^{z}$ and $\bar{f}\left(\bar{z}\right)=e^{\bar{z}}$.
In this case, we have 
\begin{equation}
u=\frac{4y\sqrt{\xi\bar{\xi}}}{4+y^{2}},\quad x=\frac{4-y^{2}}{4+y^{2}}\xi,\quad\bar{x}=\frac{4-y^{2}}{4+y^{2}}\bar{\xi}
\end{equation}
with $\xi\equiv e^{z}\text{ and }\bar{\xi}\equiv e^{\bar{z}}$. As
discussed in section (\ref{sec:DressingfromDiffeo}), in this case, the CFT is living on the boundary
cylinder with coordinates $\left(z,\bar{z}\right)$ and the bulk
metric (\ref{eq:GeneralVacuumMetrics}) that's dual the CFT vacuum $\left|0\right\rangle $ is given
by
\begin{equation}
ds_{\left|0\right\rangle }^{2}=\frac{dy^{2}+dzd\bar{z}}{y^{2}}+\frac{dz^{2}}{4}+\frac{d\bar{z}^{2}}{4}+\frac{y^{2}}{16}dzd\bar{z},\label{eq:VacuumMetricSec5}
\end{equation}
which becomes the usual global AdS$_{3}$ metric\footnote{The reason that we consider this specific $\phi_{L}$ is because it's
easier to relate the global AdS$_{3}$ metric to the BTZ black hole
metric (since their boundaries are both cylindrical), and it enables
us to circumvent some technical (numerical) issues that we encounter
in \cite{Chen:2018qzm}.} (\ref{eq:GlobalAdS3}) via the coordinate transformation (\ref{eq:GlobalAdS3Transformation}).

In this section, we are interested in studying bulk-boundary propagator
in a heavy state background $\left|\co_{H}\right\rangle $. The semiclassical
bulk metric that's dual to this  heavy state is given by equation
(\ref{eq:GeneralyzzbMetric}), i.e. 
\begin{equation}
ds_{\left|\co_{H}\right\rangle }^{2}=\frac{dy^{2}+dzd\bar{z}}{y^{2}}+\frac{1}{4}\alpha^{2}dz^{2}+\frac{1}{4}\bar{\alpha}^{2}d\bar{z}^{2}+\frac{\alpha^{2}\bar{\alpha}^{2}y^{2}}{16}dzd\bar{z},\label{eq:HeavyMetricSec5}
\end{equation}
with $\alpha=\sqrt{1-\frac{24h_{H}}{c}}$ and $\bar{\alpha}$
its complex conjugate\footnote{In this section, we mostly consider the non-rotating BTZ black holes, but most of our formulas (especially those of section \ref{sec:OneLoopNearHorizon}, since we've kept $\alpha$ and $\bar \alpha$ independent) are easily generalized to the rotating case. In the rotating black hole case, the relations between $\alpha,\bar \alpha$, $h_H,\bar h_H$ and $r_+,r_-$ are a little bit more complicated.}. This metric is related the usual BTZ black
hole metric
\begin{equation}
ds^{2}=\left(r^{2}-r_{+}^{2}\right)dt_{E}^{2}+\frac{dr^{2}}{r^{2}-r_{+}^{2}}+r^{2}d\theta^{2}
\end{equation}
via a simple coordinate transformation,
\begin{equation}\label{eq:BTZCoordinateTransformation}
z=t_{E}+i\theta,\quad\bar{z}=t_{E}-i\theta,\quad r^{2}=\frac{\left(\alpha^{2}y^{2}-4\right)\left(\bar{\alpha}^{2}y^{2}-4\right)}{16y^{2}},
\end{equation}
where $\alpha=ir_{+}$ and $\alpha=-ir_+$ with $r_+=\sqrt{\frac{24h_H}{c}-1}$. This is why the vacuum block $\cv_{0}$ of $\left\langle \co_{H}|\co_{L}\phi_{L}|\co_{H}\right\rangle $
has the interpretation of the bulk-boundary propagator in a BTZ black
hole microstate background.

\subsection{Recursion Relation on the Cylinder}

In last section, we obtained the recursion relation for computing the bulk-boundary
vacuum block in the configuration 
\begin{equation}
\left\langle \mathcal{O}_{H}(\infty)\mathcal{O}_{H}(1)\phi_{L}\left(u,x,\bar{x};f\left(z\right),\bar{f}\left(\bar{z}\right)\right)\mathcal{O}_{L}(0)\right\rangle .
\end{equation}
We emphasize again that the coordinates $\left(0,1,\infty\right)$
here are on the boundary $\left(x,\bar{x}\right)$ complex plane
(rather than the $\left(z,\bar{z}\right)$ coordinates). In order
to study the bulk-boundary propagator in a heavy background in this
section, we'll consider the following configuration
\begin{equation}
\left\langle \mathcal{O}_{H}(\infty)\mathcal{O}_{H}(1)\phi_{L}\left(u,1-x,1-\bar{x};1-f\left(z\right),1-\bar{f}\left(\bar{z}\right)\right)\mathcal{O}_{L}(0)\right\rangle \label{eq:VOConfiguration}
\end{equation}
which is equivalent to $\left\langle \mathcal{O}_{H}|\phi_{L}\left(u,x,\bar{x};f\left(z\right),\bar{f}\left(\bar{z}\right)\right)\mathcal{O}_{L}(1)|\mathcal{O}_{H}\right\rangle $
due to translational symmetry on the complex plane\footnote{Another way of obtaining $\phi_{L}$ in (\ref{eq:VOConfiguration})
is to substitute $f\left(z\right)$ and $\bar{f}\left(\bar{z}\right)$
with $1-f\left(z\right)$ and $1-\bar{f}\left(\bar{z}\right)$
in equation (\ref{eq:Phiuxxbx0xb0}), which will not change the metrics
(\ref{eq:VacuumMetricSec5}).}.

To obtain the bulk-boundary vacuum block $\cv_{0}$ of (\ref{eq:VOConfiguration}),
we just need to adopt the result of last section to the special case
considered here. For the configuration in (\ref{eq:VOConfiguration}),
we have 
\begin{equation}
u=\frac{4y\sqrt{\xi\bar{\xi}}}{4+y^{2}},\quad\Delta x=\frac{2y^{2}}{y^{2}+4}\xi,\quad\Delta\bar{x}=\frac{2y^{2}}{y^{2}+4}\xi,\quad x_{0}=1-\xi,\quad\bar{x}_{0}=1-\bar{\xi}.
\end{equation}
So we just need to plug the above expressions for $u,\Delta x,\Delta\bar{x},x_{0},\bar{x}_{0}$
into (\ref{eq:V0HoloExpand}), (\ref{eq:V0Expand}) and (\ref{eq:GlobalBlocks})
to obtain $\cv_{0}$. The Mathematica code for computing $\cv_{0}$ using the recursion relation is attached with this paper. The first several terms of $\mathcal{V}_0$ from the recursion relation are given by 
\begin{align}
\frac{\cv_{0}}{\left(\frac{u}{x_{0}\bar{x}_{0}}\right)^{2h_{L}}} & =1+\frac{2\left(\bar{z}^{2}+z^{2}\right)h_{H}h_{L}}{c}-\frac{h_{L}}{72}\left(z^{2}\left(\bar{z}^{2}+12\right)-36z\bar{z}+12\left(\bar{z}^{2}+12\right)\right)s\\
 & \quad+\frac{h_{H}h_{L}\left(3z^{3}\bar{z}h_{L}-2z^{2}\left(\bar{z}^{2}+6\right)\left(h_{L}-1\right)+3z\bar{z}^{3}h_{L}-12\bar{z}^{2}\left(h_{L}-1\right)\right)}{3c}s+\cdots\nn
\end{align}
where we've defined $s\equiv\frac{y^{2}}{z\bar{z}}$ and expanded
the LHS in terms of small $s,z$ and $\bar{z}$ to get the RHS. We've checked that this result is consistent with the semiclassical result and $1/c$ corrections to be computed in next subsection. Since the recursion relation is valid at finite $c$, we can use it to study non-perturbative physics near the black hole horizon. Due to numerical difficulties of obtaining convergent and reliable result near the horizon, we postpone it to future work. 

\subsection{Quantum Corrections to $\left\langle \protect\co_{H}|\protect\phi_{L}\co_{L}|\protect\co_{H}\right\rangle $
on the Cylinder \label{sec:OneOverCHHLphi}}

We can use bulk-boundary bi-local operators (as a generalization of the boundary bi-local operators in \cite{Cotler:2018zff}) and the effective theory for boundary gravitons developed in \cite{Cotler:2018zff} to compute the semiclassical limit of $\cv_0$ and its $1/c$ corrections. We will briefly discuss the physical interpretation of these results at the end of this section.

\subsubsection*{Semiclassical Result}

First, we notice that the semiclassical metric (\ref{eq:HeavyMetricSec5})
can be obtained from the Poincare patch 
\begin{equation}
ds^{2}=\frac{du^{2}+dxd\bar{x}}{u^{2}}
\end{equation}
via the coordinate transformation (\ref{eq:VacuumAdSDiffeomorphism}) with 
\begin{equation}
f\left(z\right)=e^{\alpha z},\qquad\bar{f}\left(\bar{z}\right)=e^{\bar{\alpha}\bar{z}},\label{eq:BTZfz}
\end{equation}
where $\alpha=\sqrt{1-\frac{24h_{H}}{c}}$ and $\bar{\alpha}$
is the complex conjugate of $\alpha$. This means that semiclassically,
the effect of the heavy operators is trivialized via this map back
to the $\left(u,x,\bar{x}\right)$ coordinates\footnote{Note that the $\left(u,x,\bar{x}\right)$ in this subsection
are different from those of last subsection since $f\left(z\right),\bar{f}\left(z\right)$
are different now.}. So the semiclassical result of $\cv_{0}$ of $\left\langle \co_{H}|\phi_{L}\left(y,z_{1},\bar{z}_{1}\right)\co_{L}\left(z_{2},\bar{z}_{2}\right)|\co_{H}\right\rangle $
must be given by the bulk-boundary propagator $\left\langle \phi_{L}\left(u_{1},x_{1},\bar{x}_{1}\right)\co_{L}\left(x_{2},\bar{x}_{2}\right)\right\rangle$
in the $\left(u,x,\bar{x}\right)$ coordinates, that is
\begin{align}
\cv_{0} & =\left(f'\left(z_{2}\right)\bar{f}'\left(z_{2}\right)\right)^{h_{L}}\left(\frac{u_{1}}{u_{1}^{2}+\left(x_{1}-x_{2}\right)\left(\bar{x}_{1}-\bar{x}_{2}\right)}\right)^{2h_{L}}+O\left(\frac{1}{c}\right)\label{eq:V0Semiclassical1}
\end{align}
where the factor $\left(f'\left(z_{1}\right)\bar{f}'\left(z_{1}\right)\right)^{h_{L}}$
comes from the transformation rule for the primary operator $\co_{L}$
and 
\[
u_{1}=\frac{4y\sqrt{\alpha\bar{\alpha}e^{\alpha z_{1}+\bar{\alpha}\bar{z}_{1}}}}{\alpha\bar{\alpha}y^{2}+4},\quad x_{1}=\frac{e^{\alpha z_{1}}\left(4-\alpha\bar{\alpha}y^{2}\right)}{\alpha\bar{\alpha}y^{2}+4},\quad\bar{x}_{1}=\frac{e^{\bar{\alpha}\bar{z}_{1}}\left(4-\alpha\bar{\alpha}y^{2}\right)}{\alpha\bar{\alpha}y^{2}+4}
\]
and $x_{2}=e^{\alpha z_{2}}$, $\bar{x}_{2}=e^{\bar{\alpha} \bar{z}_{2}}$.
For later convenience, we'll denote the semi-classical result (the
first term in (\ref{eq:V0Semiclassical1})) as $\cv_{0}^{\text{s-c}}$,
and in terms of $\left(y,\xi,\bar{\xi}\right)$ with $\xi\equiv e^{z_{1}-z_{2}}$
and $\bar{\xi}\equiv e^{\bar{z}_{1}-\bar{z}_{2}}$,
it's given by \cite{KeskiVakkuri:1998nw,Chen:2018qzm}
\begin{equation}
\mathcal{V}_{0}^{\text{s-c}}=\left(\frac{16y^{2}\alpha^{2}\bar{\alpha}^{2}\xi^{\alpha}\bar{\xi}^{\bar{\alpha}}}{\left(4\left(1-\xi^{\alpha}\right)\left(1-\bar{\xi}^{\bar{\alpha}}\right)+y^{2}\alpha\bar{\alpha}\left(1+\xi^{\alpha}\right)\left(1+\bar{\xi}^{\bar{\alpha}}\right)\right)^{2}}\right)^{h_{L}}.\label{eq:SemiV0yzzb}
\end{equation}
The configuration of last subsection corresponds to $z_{1}=z$ and
$z_{2}=0$, so the $\xi$ and $\bar{\xi}$ defined here are the
same as those of last subsection.

\subsubsection*{$1/c$ Corrections}

In order to compute the $1/c$ corrections (gravitational
one-loop corrections) to the semiclassical result of $\cv_{0}$, we
must include perturbations in $f\left(z\right)$ and $\bar{f}\left(\bar{z}\right)$,
and be more precise about the central charge $c$. It turns out that,
as in \cite{Cotler:2018zff}, we should use the following $f\left(z\right)$ and $\bar{f}\left(\bar{z}\right)$
\begin{align}
f\left(z\right) & =e^{\alpha_{0}z+\frac{i\epsilon\left(z\right)}{\sqrt{C}}}\qquad\text{and }\bar{f}\left(\bar{z}\right)=e^{\bar{\alpha}_{0}\bar{z}+\frac{i\bar{\epsilon}\left(\bar{z}\right)}{\sqrt{C}}}.\label{eq:fzHeavyStatePerturbation}
\end{align}
where $C=c-1$, $\alpha_{0}=\sqrt{1-24\frac{h_{H}}{C}}$ and $\bar{\alpha}_{0}$
is the complex conjugate of $\alpha_{0}$. Here $\epsilon$ and $\bar{\epsilon}$
are to be understood as operators. We then obtain the large $c$ expansion
(with $\frac{h_{H}}{c}$ fixed) of $\cv_{0}$ via
\begin{align}
\cv_{0} & =\left\langle \left(f'\left(z_{2}\right)\bar{f}'\left(z_{2}\right)\right)^{h_{L}}\left(\frac{u_{1}}{u_{1}^{2}+\left(x_{1}-x_{2}\right)\left(\bar{x}_{1}-\bar{x}_{2}\right)}\right)^{2h_{L}}\right\rangle \label{eq:OPEBlockInHeavyState}
\end{align}
upon plugging in the expressions of various terms using (\ref{eq:VacuumAdSDiffeomorphism}) with $f$
and $\bar{f}$ of (\ref{eq:fzHeavyStatePerturbation}), and expanding
in large $c$ or small $\epsilon,\bar{\epsilon}$. The idea here
is roughly the same as the idea of the bulk-boundary OPE blocks used
in \cite{Anand:2017dav} to compute $\left\langle \phi\co T\right\rangle $
and in \cite{Chen:2018qzm} to compute the large $c$ expansion of vacuum block
of $\left\langle \co_{H}\co_{H}\phi_{L}\co_{L}\right\rangle $ (with
$h_{H}\sim O\left(1\right)$) in Poincare AdS$_{3}$. It's also
a generalization of the boundary bi-local operators in \cite{Cotler:2018zff} to the
bulk-boundary case.

At leading order of the large $C$ limit (with $h_{H}\sim\co\left(C\right)$),
we have 
\begin{equation}
\cv_{0}=\left(\frac{16y^{2}\alpha_{0}^{2}\bar{\alpha}_{0}^{2}\xi^{\alpha_{0}}\bar{\xi}^{\bar{\alpha}_{0}}}{\left(4\left(1-\xi^{\alpha_{0}}\right)\left(1-\bar{\xi}^{\bar{\alpha}_{0}}\right)+y^{2}\alpha_{0}\bar{\alpha}_{0}\left(1+\xi^{\alpha_{0}}\right)\left(1+\bar{\xi}^{\bar{\alpha}_{0}}\right)\right)^{2}}\right)^{h_{L}}\left(1+\co\left(\frac{1}{C}\right)\right)\label{eq:V0LeadingOrder}
\end{equation}
where $\xi\equiv e^{z_{1}-z_{2}}$
and $\bar{\xi}\equiv e^{\bar{z}_{1}-\bar{z}_{2}}$. At order $1/c$, we'll get two different kinds of contributions,
one from the $\epsilon$ and $\bar{\epsilon}$ terms, and another
from the large $c$ expansion of the leading order result (recalling
that $C=c-1$). Since the terms linear in $\epsilon$ or $\bar{\epsilon}$
will have zero expectation value, we'll only keep the terms quadratic
in $\epsilon$ or $\bar{\epsilon}$ in the large $c$ expansion.
We can then package the order $1/c$ contributions to $\cv_{0}$
in the following form: 
\begin{equation}
\cv_{0}=\mathcal{V}_{0}^{\text{s-c}}\left(1+\frac{h_{L}}{c}\cv_{h_{L}/c}+\frac{h_{L}^{2}}{c}\cv_{h_{L}^{2}/c}+\co\left(\frac{1}{c^{2}}\right)\right)\label{eq:V0Largec}
\end{equation}
where $\mathcal{V}_{0}^{\text{s-c}}$ is the semiclassical result
given in (\ref{eq:SemiV0yzzb}). The order $1/c$ terms of
the large $c$ expansion of (\ref{eq:V0LeadingOrder}) will be linear
in $h_{L}$ and will be included in $\cv_{h_{L}/c}$.

The order $h_L/c$ term $\cv_{h_{L}/c}$ of (\ref{eq:V0Largec})
is given by 
\begin{smaller}
\begin{equation}
\cv_{h_{L}/c}=\frac{96\left(1-\bar{\xi}^{\bar{\alpha}}\right)^{2}\left\langle \cv_{h_{L}/c}^{\left(0\right)}\right\rangle +4\alpha\bar{\alpha}\left(1-\bar{\xi}^{2\bar{\alpha}}\right)\left\langle \cv_{h_{L}/c}^{\left(2\right)}\right\rangle y^{2} +\bar{\alpha}^{2}\left(1+\bar{\xi}^{\bar{\alpha}}\right)^{2}\left\langle \cv_{h_{L}/c}^{\left(4\right)}\right\rangle y^{4}}{D^{2}}+\frac{\cv_{h_{L}/c}^{\text{s-c}}}{D}+c.c.\label{eq:hLOvercEpsilon}
\end{equation}
\end{smaller}
where we've defined 
\begin{equation}
D\equiv4\left(1-\xi^{\alpha}\right)\left(1-\bar{\xi}^{\bar{\alpha}}\right)+y^{2}\alpha\bar{\alpha}\left(1+\xi^{\alpha}\right)\left(1+\bar{\xi}^{\bar{\alpha}}\right)\label{eq:Denominator}
\end{equation}
for notational convenience and the various terms in the numerator
are given by
\begin{align}
\cv_{h_{L}/c}^{\left(0\right)} & =\frac{1}{12\alpha^{2}}\left(1-\xi^{\alpha}\right)^{2}\left(\left(\epsilon_{1}'\right)^{2}+\left(\epsilon_{2}'\right)^{2}\right)-\frac{1}{6}\xi^{\alpha}\left(\epsilon_{1}-\epsilon_{2}\right)^{2}\\
\cv_{h_{L}/c}^{\left(2\right)} & =\resizebox{0.9\textwidth}{!}{\text{\ensuremath{\frac{1}{\alpha^{3}}\left(4\alpha^{2}\xi^{\alpha}\left(\epsilon_{1}-\epsilon_{2}\right)\epsilon_{1}'+\left(1-\xi^{\alpha}\right)\left(2\left(\xi^{\alpha}-1\right)\epsilon_{1}'\epsilon_{1}''+\alpha\left(\xi^{\alpha}+1\right)\left(\left(\epsilon_{1}'\right){}^{2}+\left(\epsilon_{2}'\right){}^{2}\right)\right)\right)}}}\nonumber \\
\cv_{h_{L}/c}^{\left(4\right)} & =\resizebox{0.9\textwidth}{!}{\text{\ensuremath{\frac{1}{2}\left(\xi^{\alpha}+1\right)^{2}\left(\left(\epsilon_{2}'\right)^{2}-\left(\epsilon_{1}'\right)^{2}\right)}+\ensuremath{\xi^{\alpha}\left(\alpha^{2}\left(\epsilon_{1}-\epsilon_{2}\right)-4\epsilon_{1}''\right)\left(\epsilon_{1}-\epsilon_{2}\right)}+\ensuremath{\frac{\left(4\alpha\left(\xi^{2\alpha}-1\right)\epsilon_{1}'-\left(1-\xi^{\alpha}\right)^{2}\epsilon_{1}''\right)\epsilon_{1}''}{\alpha^{2}}}}}\nonumber 
\end{align}
with $\epsilon_{1}\equiv\epsilon\left(z_{1}\right)$ and $\epsilon_{2}\equiv\epsilon\left(z_{2}\right)$
and the primes means the derivatives with respect to their arguments,
respectively. The complex conjugate $c.c.$ in Equation (\ref{eq:hLOvercEpsilon})
means replacing $\xi,\bar{\xi},\alpha,\bar{\alpha}$ with
$\bar{\xi},\xi,\bar{\alpha},\alpha$, respectively, and
also changing $\epsilon$ to $\bar{\epsilon}$. The $\cv_{h_{L}/c}^{\text{s-c}}$ term plus it's complex
conjugate is from the $1/c$ expansion of (\ref{eq:V0LeadingOrder}),
and it's given by 
\begin{small}
\begin{equation}
\cv_{h_{L}/c}^{\text{s-c}}=\left(1-\alpha^{2}\right)\left(\frac{2\left(1-\bar{\xi}^{\bar{\alpha}}\right)\left(2\xi^{\alpha}-\alpha\left(1+\xi^{\alpha}\right)\log(\xi)-2\right)}{\alpha^{2}}-\frac{y^{2}}{2}\bar{\alpha}\left(1-\xi^{\alpha}\right)\left(1+\bar{\xi}^{\bar{\alpha}}\right)\log\left(\xi\right)\right)
\end{equation}
\end{small}

The order $h_L^2/c$ term $\cv_{h_{L}^{2}/c}$ of (\ref{eq:V0Largec})
is given by 
\begin{equation}
\cv_{h_{L}^{2}/c}=\frac{\left\langle \cv_{h_{L}^{2}/c}^{\left(0\right)}\right\rangle +\left\langle \cv_{h_{L}^{2}/c}^{\left(2\right)}\right\rangle y^{2} +\left\langle \cv_{h_{L}^{2}/c}^{\left(4\right)}\right\rangle y^{4} }{D^{2}}+c.c.,
\end{equation}
where $D$ is given by (\ref{eq:Denominator}). The numerator is given
by 
\begin{equation}
\left\langle \cv_{h_{L}^{2}/c}^{\left(0\right)}\right\rangle +\left\langle \cv_{h_{L}^{2}/c}^{\left(2\right)}\right\rangle y^{2}+\left\langle \cv_{h_{L}^{2}/c}^{\left(4\right)}\right\rangle y^{4}=\frac{1}{-2\alpha^{2}}\left\langle \left(V_{1}+y^{2}V_{2}\right)^{2}\right\rangle 
\end{equation}
with
\begin{align}
V_{1} & =4\left(\bar{\xi}^{\bar{\alpha}}-1\right)\left(\alpha\left(\epsilon_{1}-\epsilon_{2}\right)\left(\xi^{\alpha}+1\right)+\left(1-\xi^{\alpha}\right)\left(\epsilon_{1}'+\epsilon_{2}'\right)\right),\\
V_{2} & =\bar{\alpha}\left(\bar{\xi}^{\bar{\alpha}}+1\right)\left(\alpha^{2}\left(\xi^{\alpha}-1\right)\left(\epsilon_{1}-\epsilon_{2}\right)+\alpha\left(\xi^{\alpha}+1\right)\left(\epsilon_{1}'-\epsilon_{2}'\right)+2\left(1-\xi^{\alpha}\right)\epsilon_{1}''\right),\nonumber 
\end{align}

Now to compute $\cv_{h_{L}/c}$ and $\cv_{h_{L}^{2}/c}$, we need
the $\epsilon$ propagator. This is worked out in \cite{Cotler:2018zff} using the effective theory for boundary gravitons developed in that paper,
and it's given by\footnote{There is a subtlety about the ordering of the $\epsilon$ operators in the $\epsilon$ two-point function. When computing $\cv_0$, we  substitute $\left\langle \epsilon(z_{1})\epsilon(z_{2})\right\rangle$ with the symmetric average of the two different ordering   $\frac{1}{2}\left(\left\langle \epsilon(z_{1})\epsilon(z_{2})\right\rangle+\left\langle \epsilon(z_{2})\epsilon(z_{1})\right\rangle\right)$. This procedure gives a result matching the recursion relation.  }
\begin{equation}
\left\langle \epsilon(z_{1})\epsilon(z_{2})\right\rangle =\frac{6}{C}\left(2\ln(1-\xi)+\Phi(\xi,1,\alpha)+\Phi(\xi,1,-\alpha)\right),\quad\xi=e^{z_{1}-z_{2}}\label{eq:EpsilonPropagator}
\end{equation}
where $\Phi(z,s,a)$ is the Lerch transcendant
\footnote{For $s=1$ it is related to a certain incomplete Beta function as
\begin{equation}
B(z,a,0)=z^{a}\Phi(z,1,a).
\end{equation}
}
\begin{equation}
\Phi(z,s,a)=\sum_{n=0}^{\infty}\frac{z^{n}}{(n+a)^{s}},
\end{equation}
Using the $\epsilon$ propagator (\ref{eq:EpsilonPropagator}), we
can evaluate $\cv_{h_{L}/c}$ and $\cv_{h_{L}^{2}/c}$, but they are
logarithmically divergent and need to be renormalized. We follow the
procedure in \cite{Cotler:2018zff}, and define the renormalized expectation
value of the $\left\langle \cv_{h_{L}/c}^{\left(i\right)}\right\rangle $
as follows 
\begin{equation}
\left\langle \cv_{h_{L}/c}^{\left(i\right)}\right\rangle _{\text{R}}=\left\langle \cv_{h_{L}/c}^{\left(i\right)}\right\rangle -\left(\left\langle \cv_{h_{L}/c}^{\left(i\right)}\right\rangle _{\alpha\rightarrow1,\bar{\alpha}\rightarrow1}\right)_{w\rightarrow\alpha w,\bar{w}\rightarrow\bar{\alpha}\bar{w},y\rightarrow y\sqrt{\alpha\bar{\alpha}}}
\end{equation}
and similarly for $\left\langle \cv_{h_{L}^{2}/c}^{\left(i\right)}\right\rangle _{\text{R}}$.

Since the result of $\cv_{0}$ only depends on $z_{1}-z_{2}$, we'll
set $z_{1}=z$ and $z_{2}=0$. Eventually, the $\left\langle \cv_{h_{L}/c}^{\left(i\right)}\right\rangle _{\text{R}}$ terms are given by
\begin{small}
\begin{align}
\left\langle \cv_{h_{L}/c}^{\left(0\right)}\right\rangle _{\text{R}} & =e^{\alpha z}\left(\cf_{1}+4\log\alpha+4\log\left(2\sinh\left(\frac{z}{2}\right)\right)+2\right)+\left(1-e^{2\alpha z}\right)\alpha z\label{eq:V10}\\
 & \quad+\left(1+e^{2\alpha z}\right)\left(H_{-\alpha}+H_{\alpha}-1+i\pi-2\log\alpha\right)+2\left(1-e^{\alpha z}\right)^{2}\log\left(2\sinh\left(\frac{\alpha z}{2}\right)\right),\nonumber \\
\left\langle \cv_{h_{L}/c}^{\left(2\right)}\right\rangle _{\text{R}} & =\left(1-e^{2\alpha z}\right)\left(H_{-\alpha}+H_{\alpha}+i\pi-1-2\log\alpha+2\log\left(2\sinh\left(\frac{\alpha z}{2}\right)\right)\right)\nonumber \\
 & \quad+\left(1+e^{2\alpha z}\right)\alpha z+e^{\alpha z}\cf_{2},\\
\left\langle \cv_{h_{L}/c}^{\left(4\right)}\right\rangle _{\text{R}} & =\left(1-\xi^{\alpha}\right)^{2}+6\alpha^{3}\left(1-e^{2\alpha z}\right)z\\
 & \quad+6\alpha^{2}\left(1+e^{2\alpha z}\right)\left(H_{-\alpha}+H_{\alpha}+2\log\left(2\sinh\left(\frac{\alpha z}{2}\right)\right)-2\log\alpha-\frac{7}{6}+i\pi\right)\nonumber \\
 & \quad+6\alpha^{2}e^{\alpha z}\left(\cf_{1}-4\log\alpha-4\log\left(2\sinh\left(\frac{z}{2}\right)\right)+4\log\left(2\sinh\left(\frac{\alpha z}{2}\right)\right)+\frac{7}{3}\right),\nonumber 
\end{align}
\end{small}
and the $\left\langle \cv_{h_{L}^{2}/c}^{\left(i\right)}\right\rangle _{\text{R}}$ terms are given by
\begin{small}
\begin{align}
\left\langle \cv_{h_{L}^{2}/c}^{\left(0\right)}\right\rangle _{\text{R}} & =192\left(1-e^{\bar{\alpha}\bar{z}}\right)^{2}\left(\left(\log\left(\sinh\left(\frac{z}{2}\right)\right)-\log\left(\sinh\left(\frac{\alpha z}{2}\right)\right)+\log\alpha+1\right)\left(1-e^{\alpha z}\right)^{2}\right.\nonumber \\
 & \quad\left.+\cf_{3}+2e^{\alpha z}\left(H_{-\alpha}+H_{\alpha}+2\log\left(2\sinh\left(\frac{z}{2}\right)\right)+i\pi\right)\right),\\
\left\langle \cv_{h_{L}^{2}/c}^{\left(2\right)}\right\rangle _{\text{R}} & =48\bar{\alpha}\left(1-e^{\alpha z}\right)\left(1-e^{2\bar{\alpha}\bar{z}}\right)\left(\coth\left(\frac{z}{2}\right)\left(1-e^{\alpha z}\right)\right.\\
 & \quad\left.+2\alpha\left(e^{\alpha z}+1\right)\left(\log\alpha-\log\left(\sinh\left(\frac{\alpha z}{2}\right)\right)+\log\left(\sinh\left(\frac{z}{2}\right)\right)+\frac{1}{2}\right)\right),\nonumber \\
\left\langle \cv_{h_{L}^{2}/c}^{\left(4\right)}\right\rangle _{\text{R}} & =2\alpha^{2}\bar{\alpha}^{2}\left(1+e^{\bar{\alpha}\bar{z}}\right)^{2}\left(\frac{6}{\alpha}\coth\left(\frac{z}{2}\right)\left(1-e^{2\alpha z}\right)+6\log\alpha\left(1+e^{\alpha z}\right)^{2}-e^{2\alpha z}\right.\nonumber \\
 & \quad+6\left(1-e^{\alpha z}\right)^{2}\left(\log\left(\sinh\left(\frac{z}{2}\right)\right)-\log\left(\sinh\left(\frac{\alpha z}{2}\right)\right)+\frac{1}{6\alpha^{2}}\right)-6\cf_{3}\nonumber \\
 & \quad\left.-1-12e^{\alpha z}\left(H_{-\alpha}+H_{\alpha}+2\log\left(2\sinh\left(\frac{\alpha z}{2}\right)\right)+i\pi-\frac{13}{6}\right)\right).\label{eq:V24}
\end{align}
\end{small}
where we've define 
\begin{align*}
\cf_{1} & \equiv\Phi\left(e^{z},1,\alpha\right)+\Phi\left(e^{z},1,-\alpha\right)+\Phi\left(e^{-z},1,\alpha\right)+\Phi\left(e^{-z},1,-\alpha\right)\\
\cf_{2} & \equiv\Phi\left(e^{z},1,\alpha\right)-\Phi\left(e^{z},1,-\alpha\right)-\Phi\left(e^{-z},1,\alpha\right)+\Phi\left(e^{-z},1,-\alpha\right)\\
\cf_{3} & \equiv e^{2\alpha z}\Phi\left(e^{z},1,\alpha\right)+\Phi\left(e^{z},1,-\alpha\right)+\Phi\left(e^{-z},1,\alpha\right)+e^{2\alpha z}\Phi\left(e^{-z},1,-\alpha\right)
\end{align*}
for notational convenience.

We checked that these results agree with the  recursion relation of last subsection by expanding in small\footnote{The branch cuts in the various functions in the equations (\ref{eq:V10})-(\ref{eq:V24}) were chosen such that when  they are expanded in small $z$ assuming $z>0$ in Mathematica, the result matches that of the recursion relation. One should be careful when evaluating (\ref{eq:V10})-(\ref{eq:V24}) for $z<0$.} $s\equiv\frac{y^{2}}{z\bar{z}}$, $z$ and $\bar{z}$.
We have also verified that when we take $\phi_{L}$ to the boundary,
we recover the $1/c$ correction \cite{Fitzpatrick:2015dlt} to the
heavy-light vacuum block \cite{Fitzpatrick:2014vua}.

\subsubsection*{One-Loop Corrections Near the Black Hole Horizon \label{sec:OneLoopNearHorizon}}

We would like to see what the semiclassical result and the $1/c$
corrections computed here tell us about physics near the Euclidean
black hole horizon. For the non-rotating case considered here, the horizon is at $r=r_{+}=\sqrt{\frac{24h_H}{c}-1}$ (which corresponds to
$y=\frac{2}{r_{+}}$). In this case, the semi-classical result $\mathcal{V}_{0}^{\text{s-c}}$
of equation (\ref{eq:SemiV0yzzb}) written in terms of the $\left(r,t_{E},\theta\right)$ (using the coordinate relations (\ref{eq:BTZCoordinateTransformation}))
is given by 
\begin{equation}
\mathcal{V}_{0}^{\text{s-c}}=\left(\frac{r_{+}}{2}\right)^{2h_{L}}\frac{1}{\left(\frac{r}{r_{+}}\cosh\left(r_{+}\theta\right)-\frac{\sqrt{r^{2}-r_{+}^{2}}}{r_{+}}\cos\left(r_{+}t_{E}\right)\right)^{2h_{L}}}.
\end{equation}
which is the $n=0$ terms in the full semiclassical bulk-boundary
correltator for a free field in a BTZ black hole given by the image
sum in \cite{KeskiVakkuri:1998nw}. So one can see that the semiclassical result is periodic in $t_E$, and it's smooth
at the horizon $r=r_{+}$ (and its dependence on $t_E$ drops out there). 

In terms of $\left(y,z,\bar{z}\right)$,
the horizon is at $y=\frac{2}{r_{+}}$, and at this
value of $y$, the $\frac{1}{c}$ corrections $\cv_{h_{L}/c}$ and
$\cv_{h_{L}^{2}/c}$ to the vacuum block of $\left\langle \co_{H}|\phi_{L}\co_{L}|\co_{H}\right\rangle $
are finite, since their numerators truncate at order $y^{4}$, and
their denominators are just the same as the denominator of the semiclassical
result $\mathcal{V}_{0}^{\text{s-c}}$, which is also non-singular
at this value of $y$.  However, unlike the semiclassical vacuum block, the functions $\cv_{h_{L}/c}$ and
$\cv_{h_{L}^{2}/c}$ are not periodic in Euclidean time \cite{Fitzpatrick:2015dlt, Fitzpatrick:2016ive}.  This means that the $1/c$ correction to the 
bulk-boundary heavy-light correlator will have a branch point at the Euclidean horizon.  So the singularity of these correlators at the Euclidean horizon arises already in perturbation theory, and does not require non-perturbative effects \cite{Chen:2018qzm}.


\section{Discussion}
\label{sec:Discussion}

Our primary goal has been to develop an exact bulk reconstruction procedure with very general gravitational dressings.  The motivation was to enable future investigations into the dressing-dependence of bulk observables, as these ambiguities  present  a major caveat when drawing physical conclusions.  For example, using our results it should be possible to determine if the breakdown of bulk locality at short-distances in AdS$_3$ \cite{Chen:2017dnl}  persists with a general class of gravitational dressings.  We can also investigate BTZ black hole horizons \cite{Chen:2018qzm}, though the necessary numerics may be rather formidable.  We took the first steps in this direction in section \ref{sec:HeavyLightCorrelator}.  By exploiting the connection between the singularity structure of CFT stress-tensor correlators and gravitational dressings, it may be possible to generalize some of our results to higher dimensions.

Our reconstruction procedure only incorporates effects arising as a mandatory consequence of Virasoro symmetry.  With hard work one could add other perturbative interactions, but such methods would likely just reproduce bulk perturbation theory, without providing a deeper understanding of quantum gravity.  So our methods are limited, as they are only able to address certain universal features of quantum gravity.  Unfortunately, in the case of quantum gravity it would seem that we must either solve toy models completely, and then try to argue that they are representative, or solve a universal sector of a general class of models, and try to argue that the effects from this sector determine the relevant physics.  Given the universal nature of the gravitational force, we believe that the latter route is a more compelling way to address locality and near-horizon dynamics.

Most work on bulk reconstruction suffers from a nagging conceptual problem.  As physical observers, we  do not setup experiments by making reference to the boundary of spacetime.  And defining the bulk by reference to the boundary seems even more perverse in a  cosmological setting.  Furthermore, it has been shown that using the boundary as a reference point leads to fundamental problems, such as bulk fields that do not commute outside the lightcone  \cite{Donnelly:2015hta}, even at low-orders in gravitational perturbation theory.  Perhaps a more sensible approach defines observables relative to other objects in spacetime, just as we define our local reference frame with respect to the earth, solar system, galaxy, and galactic neighborhood.  This also seems more in keeping with interpretations of the Wheeler-DeWitt equation \cite{DeWitt:1967yk}.  We hope to formalize such a definition of local observables in future work.

\section*{Acknowledgments}
We would like to thank Ibou Bah, Liam Fitzpatrick, and David E. Kaplan for discussions.  HC, JK and US  have been supported in part by NSF grant PHY-1454083.  JK was also supported in part by the Simons Collaboration Grant on the Non-Perturbative Bootstrap. 

\appendix

\addtocontents{toc}{\protect\setcounter{tocdepth}{1}}

\section{Bulk Primary Conditions as Mirage Translations}
\label{app:MirageTranslations}

Bulk operators must leave an imprint in boundary correlators representing their conserved charges and energies.  This imprint manifests as singularities in correlators involving conserved currents $J(z)$ (see section \ref{sec:SingularitiesGaussLaw}) or the stress tensor $T(z)$.  In this section we  develop some formalism for displacing the singularities associated with local charge or energy in a CFT$_2$.  This will allow us to alter the `gravitational dressing' of bulk operators.  We will also identify an alternative explanation for the bulk primary condition \cite{Anand:2017dav}.  In appendix \ref{app:SingularityStructureT} we  provide a review of the singularity structure of $T(z)$ correlators with CFT primary operators as derived from the bulk.

\subsection{Mirage Translations}

In a translation-invariant theory such as a CFT, we use the momentum generator $P_\mu$ to move local operators around.  In a CFT$_2$ this means that 
\be
\CO(z, \bar z) = e^{z L_{-1} + \bar z \bar L_{-1}} \CO(0) e^{-z L_{-1} - \bar z \bar L_{-1}}
\ee
since $L_{-1}, \bar L_{-1}$ are the holomorphic and anti-holomorphic momentum generators.  

Now let us assume that the CFT has a holomorphic $U(1)$ current $J(z)$.  Correlators with the current such as
\be
\< J(z_1) \CO^\dag(z,\bar z) \CO(0) \> = q \frac{z}{(z-z_1)z_1}  \frac{1}{z^{2h} \bar z^{2 \bar h}}
\ee
have singularities in $z_1$ when $J$ collides with charged operators, which indicate the presence of charge localized at $0$ and $z$.
We will pose the following question:  \emph{can we find an operator that moves local charge without moving the associated primary operators?}  Or equivalently, can we move the primary operators while leaving its charge in place?  

We can sharpen these questions into precise criteria for correlators.  We would like to find a modified translation operator $G_{h, q}(z_f)$ that can appear in correlators as\footnote{We are implicitly assuming we can separate $z_i$ and $w_j$ from $0$ and $z_f$ and perform radial quantization about the non-local object $[G_{h, q} \CO]$.}
\be
\< \CO^\dag(z)    J(z_k) \cdots J(z_1) \left[ G_{h, q}(z_f) \CO(0)\right]  \> 
\ee 
We wish to choose  $G_{h, q}$ so that $\CO^\dag(z,\bar z)$ only has an OPE singularity with  $\left[ G_{h, q}(z_f) \CO(0)\right]$  when $z - z_f$ vanish, but the currents $J(z_j)$ have OPE singularities with $ \left[ G_{h, q}(z_f) \CO(0)\right]$  when $z_j \to 0$.   So the non-local object $[G_{h, q} \CO]$ behaves like a mirage, present at both $0$ and $z_f$.

Conventional translation operators automatically satisfy the first condition.  They also satisfy the second condition when the charge of $\CO$ is $q=0$, suggesting that $G_{h, 0}(z_f, 0) = e^{z_f L_{-1}}$.  So let us modify the translation generator and define
\be
\label{eq:MirageChargeTranslation}
G_{h, q}(z_f) =  \sum_{n=0}^\infty \frac{z_f^n}{n!} \CJ_{-n}
\ee
where we have $\CJ_{-n} = L_{-1}^n + O(q)$, so that $\CJ_{-n}$ implicitly depends on $h$ and $q$.  Our criterion require an OPE 
\be
J(z_1) \left[G_{h, q}(z_f)  \CO(0)\right]= \frac{q}{z_1} \left[G_{h, q}(z_f, 0)  \CO(0)\right] + \cdots
\ee  
where the ellipsis denotes finite terms, so that we are demanding that the only singularity is a simple pole at $z_1 = 0$.  This condition will automatically be satisfied if
\be
\left[ J_{m}, \CJ_{-n} \CO(0) \right] = 0 \ \ \ {\rm for \ all} \ \ \ m \geq 1
\ee
and for any $n$.  These conditions uniquely determine $\CJ_{-n}$ up to an overall factor.  These overall factors will be fixed as in equation (\ref{eq:MirageChargeTranslation}) by the requirement that $G_{h, q}(z_f)$ acts on $\co$ as a conventional translation in its two-point function with $\co^\dagger$.

We can repeat this exercise and replace charge with energy-momentum, and $J(z)$ with the CFT$_2$ energy-momentum tensor $T(z)$.  In that case, we could write 
\be
\label{eq:MirageEnergyTranslation}
G_{h}(z_f) =  \sum_{n=0}^\infty \frac{z_f^n}{n!} \CL_{-n}
\ee
where $\CL_{-n}$ implicitly depends on the dimension $h$ of the primary $\CO$ to which we are applying $G_h$.  

We must have $\CL_{-1} = L_{-1}$ simply because there are no other level-one combinations of Virasoro generators.  This means that the OPE  $T(z_1) \left[G_{h}(z_f)  \CO(0)\right]$ necessarily contains a third-order pole $\frac{1}{z_1^3}$, but it needn't have any higher order singularities.  The absence of any further singularities at $z_1 \to 0$ or anywhere else in the complex plane implies that the $\CL_{-n}$ must satisfy the bulk primary conditions
\be
\left[ L_{m}, \CL_{-n} \CO(0) \right] = 0 \ \ \ {\rm for \ all} \ \ \ m \geq 2,
\ee
which were previously discovered in the context of bulk reconstruction.  Here we see them appearing in the answer to a question  concerning the CFT alone.

\subsection{Singularity Structure of $\< T(z) \>$ from Einstein's Equations}
\label{app:SingularityStructureT}

In this section, we generalize the discussion of section \ref{sec:SingularitiesGaussLaw}
to gravity, showing how Einstein's equations in the presence of a
massive source on the boundary dictate the singularity structure of
correlators with the CFT stress tensor. This is elementary, as in
essence it amounts to Gauss's law for AdS$_{3}$ gravity \cite{Balasubramanian:1999re}.
But we review the argument to emphasize the connection between gravitational
dressing and singularities.

We wish to establish that the OPE of a scalar primary with the CFT
stress tensor has $\frac{1}{z^{2}}$ singularity by using the bulk
equations of motion. A scalar primary inserted at the origin will
create a scalar particle propagating in the bulk. We assume that the
particle is sufficiently heavy to model its wavefunction with a worldline.

In global AdS$_{3}$ with metric $ds^{2}=\left(r^{2}+1\right)dt_{E}^{2}+\frac{1}{r^{2}+1}dr^{2}+r^{2}d\theta^{2}$,
the only non-vanishing component of the bulk energy-momentum tensor of this particle is 
\begin{equation}\label{eq:BulkT}
	T_{\text{B}}^{tt}=\frac{m}{2\pi r}\delta(r),
\end{equation}
where we denote the bulk energy-momentum
tensor with a subscript ``$\text{B}$'' to avoid confusion ($T,\bar{T}$
will still be the boundary bulk energy-momentum tensor). We are interested
in describing the above particle in Poincare patch $ds^{2}=\frac{dy^{2}+dzd\bar{z}}{y^{2}}$.
The coordinate maps that connect these two metrics are 
\begin{equation}
y=\frac{e^{t_{E}}}{\sqrt{r^{2}+1}},\quad z=\frac{re^{t_{E}+i\theta}}{\sqrt{r^{2}+1}},\quad\bar{z}=\frac{re^{t_{E}-i\theta}}{\sqrt{r^{2}+1}}.
\end{equation}
The trajectory of the particle, which is simply $r=0$ in the global coordinates, is corresponding to $(y,z,\bar{z})=(e^{t_{E}},0,0)$.
Since we want to study the singularity of the boundary stress tensor
$T(z)$ as it approaches the source, we need to localize to a small
neighborhood around $r=0$, that is, we'll take the limit $z,\bar{z}\rightarrow0$
in the following calculations. The full backreacted metric will take
the form of equation (\ref{eq:FGMetrichat}), where here we will interpret
$T$ and $\bar{T}$ as components of a classical gravitational field.
The delta function in the bulk energy-momentum tensor (\ref{eq:BulkT}) can be transformed
to the Poincare patch via 
\begin{align}
\frac{1}{2\pi r}\delta(r)\rightarrow\frac{1}{y\sqrt{|g|}}\delta^{2}(z,\bar{z})
\end{align}
where the Jacobian accounts for $y=e^{t}$ at $r=0$. Thus, the covariant
bulk energy-momentum tensor in Poincare patch will be given by 
\begin{align}
T_{\text{B}}^{\mu\nu}=m\frac{v^{\mu}v^{\nu}}{v^{\alpha}v^{\beta}g_{\alpha\beta}}\frac{1}{y\sqrt{|g|}}\delta^{2}(z,\bar{z})
\end{align}
where the velocity of the particle following a geodesic is $v^{\mu}=(\dot{y},\dot{z},\dot{\bar{z}})=(y,z,\bar{z})$
with constant $r$ and $\theta$. We'll assume $T\left(z\right)$
to be more singular than $\frac{1}{z}$ and similarly for $\bar{T}\left(\bar{z}\right)$.
In the small $z,\bar{z}$ limit, we have $\sqrt{|g|}\approx\frac{18yT\bar{T}}{c^{2}}$
and $v^{\alpha}v^{\beta}g_{\alpha\beta}\approx\frac{36y^{2}z\bar{z}T\bar{T}}{c^{2}}$.
The resulting simplified form of the stress tensor is 
\begin{equation}
T_{\text{\text{B}}}^{\mu\nu}\approx\frac{mc^{4}v^{\mu}v^{\nu}}{648y^{4}z\bar{z}T^{2}\bar{T}^{2}}\delta^{2}(z,\bar{z})
\end{equation}
We apply the same limit to the LHS of Einstein equation to find the
$y\bar{z}$ component to be 
\begin{equation}
G^{y\bar{z}}-g^{y\bar{z}}\approx\frac{c^{3}\partial_{\bar{z}}T}{54y^{3}T^{2}\bar{T}^{2}}.
\end{equation}
So the $y\bar{z}$ component of the Einstein's equation $G^{\mu\nu}-g^{\mu\nu}=8\pi G_{N}T_{\text{B}}^{\mu\nu}$
is
\begin{equation}
\partial_{\bar{z}}T=\frac{\pi m}{z}\delta^{2}\left(z,\bar{z}\right),
\end{equation}
where we've used $G_{N}=\frac{3}{2c}$. So we find 
\begin{equation}
T\left(z\right)=\frac{m}{2z^{2}},
\end{equation}
where we've used $\partial_{\bar{z}}\frac{1}{z}=\pi\delta^{2}\left(z,\bar{z}\right)$.
Similarly, from the $yz$ component of the Einstein's equation, we
can get $\bar{T}\left(\bar{z}\right)=\frac{m}{2\bar{z}^{2}}$.
Other components of the Einstein's equation are trivially satisfied
once we substitute these solutions for $T\left(z\right)$ and $\bar{T}\left(\bar{z}\right)$.

So in the large mass approximation $m\approx2h$ we can conclude that
$T=\frac{h}{z^{2}}$ in the presence of a source localized at the
origin. Bulk fields must be leave a similar imprint on the boundary
$T$ correlators.

 \section{Solving the Charged Bulk Primary Conditions}
\label{app:SolvingBulkPrimary}
In this appendix we solve the charged bulk primary conditions for the operators $\CJ_{-n}$, first exactly for the first few $n$ in appendix \ref{app:ExactSolutions}, and then in appendix \ref{app:LargekSolution}, we study the large $k$ limit and obtain all the terms at order $1/k$ in $\cj_{-n}$ for all $n$.

 \subsection{Exact Solutions} \label{app:ExactSolutions}
 We'll expand the bulk charged field as
\begin{equation}
\phi\left(y,z,\bar{z}\right)=\sum_{n=0}^{\infty}y^{2h+2n}\lambda_{n}\cj_{-n}\bar{L}_{-1}^{n}\co\left(z,\bar{z}\right),\qquad\text{with }\lambda_{n}\equiv\frac{\left(-1\right)^{n}}{n!\left(2h\right)_{n}}
\end{equation}
where we've factored out $\lambda_{n}$ in $\phi$ for later convenience.
Now our task is to solve for $\cj_{-n}$s, which satisfies the following
two conditions
\begin{align}
J_{m}\cj_{-n}\co & =0,\quad\text{for }m\ge1,\nn\\
L_{1}^{n}\cj_{-n}\co & =n!\left(2h\right)_{n}\co,
\end{align}
where the first one is simply the bulk-primary condition (\ref{eq:ChargedBulkPrimaryCondition}), and the second one just is to ensure that $\phi$ has the correct bulk-boundary propagator with $\co(w,\bar w)$, i.e. $\langle\phi(y,z,\bar z)\co(w,\bar w)\rangle=\left(\frac{y}{y^2+(z-w)(\bar z-\bar w)}\right)^{2h}$. One can also understand the second one as giving a normalization condition for $\cj_{-n}$. It can be shown that the above two conditions uniquely fix $\cj_{-n}$.

At each level $n$, we simply write $\cj_{-n}\co$ as a sum over all possible
level $n$ descendant operators with unknown coefficients, and use
the above equations to fix the coefficients. There will be equal number of unknown coefficients and independent equations at each level $n$. The solutions for $n$ up to 4 are given by
\begin{align}
\cj_{-1}&=  \frac{1}{1-\frac{q^{2}}{2hk}}\left(L_{-1}-\frac{q}{k}J_{-1}\right),\nonumber \\
\cj_{-2}&=  \frac{1}{1-\frac{4h+1}{2h(2h+1)}\frac{q^{2}}{k}+\frac{1}{2h(2h+1)}\frac{q^{4}}{k^{2}}}\left(L_{-1}^{2}-\frac{q}{k}\left(J_{-2}+2J_{-1}L_{-1}\right)+\frac{q^{2}}{k^{2}}J_{-1}^{2}\right),\\
\cj_{3} &= \frac{1}{1-\frac{6h^{2}+6h+1}{2h(h+1)(2h+1)}\frac{q^{2}}{k}+\frac{3}{4h(h+1)}\frac{q^{4}}{k^{2}}-\frac{1}{4h(h+1)(2h+1)}\frac{q^{6}}{k^{3}}}\nonumber \\
 &\quad \times\left(L_{-1}^{3}-\frac{q}{k}\left(2J_{-3}+3J_{-2}L_{-1}+3J_{-1}L_{-1}^{2}\right)+\frac{3q^{2}}{k^{2}}\left(J_{-2}J_{-1}+J_{-1}J_{-1}L_{-1}\right)-\frac{q^{3}}{k^{3}}J_{-1}^{3}\right),\nonumber 
\end{align}
and
\begin{align*}
\cj_{-4}&=  \frac{1}{1-\frac{\left(16h^{3}+36h^{2}+22h+3\right)q^{2}}{2h(2h+3)\left(2h^{2}+3h+1\right)k}+\frac{\left(24h^{2}+36h+11\right)q^{4}}{4h(2h+3)\left(2h^{2}+3h+1\right)k^{2}}-\frac{(4h+3)q^{6}}{2h(2h+3)\left(2h^{2}+3h+1\right)k^{3}}+\frac{q^{8}}{4h(2h+3)\left(2h^{2}+3h+1\right)k^{4}}}\\
 & \quad \times\left[L_{-1}^{4}-\frac{q}{k}\left(6J_{-4}+8J_{-3}L_{-1}+6J_{-2}L_{-1}^{2}+4J_{-1}L_{-1}^{3}\right)\right.\\
 & \quad \qquad+\frac{q^{2}}{k^{2}}\left(8J_{-3}J_{-1}+3J_{-2}J_{-2}+12J_{-2}J_{-1}L_{-1}+6J_{-1}^{2}L_{-1}^{2}\right)\\
 & \quad \left.\qquad-\frac{6q^{3}}{k^{3}}\left(6J_{-2}J_{-1}^{2}+4J_{-1}^{3}L_{-1}\right)+\frac{q^{4}}{k^{4}}J_{-1}^{4}\right].
\end{align*}
In principle, one can continue this calculation up to arbitrarily large $n$. However, as $n$ increase, the number of descendant operators at level $n$ will increase very fast and the calculation becomes very complicated.

One useful way of organizing the terms in $\cj_{-n}\co$ is to separate
the contribution of the global descendants of $\co$ from that of
quasi-primary operators and their global descendants, i.e. 
\begin{equation}
\cj_{-n}\co=L_{-1}^{n}\co+\text{quasi-primaries plus their global descendants}.\label{eq:CJnGlobalPlusQuasiPrimary}
\end{equation}
For example, we can rewrite $\cj_{-1}\co$ as 
\begin{equation}
\cj_{-1}\co=L_{-1}\co+\frac{q^{2}}{2hk-q^{2}}\left(L_{-1}-\frac{2h}{q}J_{-1}\right)\co\label{eq:J1QuasiPrimary}
\end{equation}
where $\left(L_{-1}-\frac{2h}{q}J_{-1}\right)\co$ is a quasi-primary
operator since it satisfies $L_{1}\left(L_{-1}-\frac{2h}{q}J_{-1}\right)\co=0$.

Actually, we can be more precise about the statement (\ref{eq:CJnGlobalPlusQuasiPrimary}).
Similar to the case of gravity considered in section 3.2.2 of \cite{Anand:2017dav},
one can show that $\cj_{-n}\co$ can be written as 
\begin{equation}
\cj_{-n}\co=L_{-1}^{n}\co+n!\left(2h\right)_{n}\sum_{j=0}^{n}\sum_{i}\frac{L_{-1}^{n-j}\co_{h+j}^{\left(i\right)}}{\left|L_{-1}^{n-j}\co_{h+j}^{\left(i\right)}\right|^{2}}\label{eq:CJnInTermsOfQuasiPrimaries}
\end{equation}
where the $\co_{h+j}^{\left(i\right)}$ represent the $i$th quasi-primary
at level $j$ (so they satisfy $L_{1}\co_{h+j}^{\left(i\right)}=0$)
and we've assumed that the quasi-primaries are orthogonal to each
other. The denominator in each term is the norm of the corresponding operator. The simplest example of the above expression is given by $\cj_{-1}\co$ in (\ref{eq:J1QuasiPrimary}), which can
be written as 
\begin{equation}
\cj_{-1}\co=L_{-1}+2h\frac{\left(L_{-1}-\frac{2h}{q}J_{-1}\right)\co}{\left|\left(L_{-1}-\frac{2h}{q}J_{-1}\right)\co\right|^{2}}
\end{equation}
where we've used the fact that $\left|\left(L_{-1}-\frac{2h}{q}J_{-1}\right)\co\right|^{2}=\frac{2h\left(2hk-q^{2}\right)}{q^{2}}$.

Note that the quasi-primaries $\co_{h+j}^{\left(i\right)}$ must include
at least one $J$ generator in them (like the $J_{-1}$ in $\left(L_{-1}-\frac{2h}{q}J_{-1}\right)\co$),
so their norms must be at least order $k$ in the large $k$ limit.
This means that in the large $k$ limit, we have 
\[
\lim_{k\rightarrow\infty}\cj_{-n}=L_{-1}^{n},
\]
as expected.

The decomposition of $\cj_{-n}\co$ in (\ref{eq:CJnGlobalPlusQuasiPrimary})
is useful is because we can make use of this fact when computing some
correlation function. For example, since the two-point functions of
$\co^{\dagger}$ with quasi-primaries vanish, the terms that will
contribute to $\left\langle \phi\co^{\dagger}\right\rangle $ are
those global descendants terms $L_{-1}^{n}\co$. Therefore, we have
\begin{align}
\left\langle \phi\left(y,z,\bar{z}\right)\co\left(z_{1},\bar{z}_{1}\right)\right\rangle  & =\sum_{n=0}^{\infty}y^{2h+2n}\lambda_{n}\left\langle L_{-1}^{n}\bar{L}_{-1}^{n}\co\left(z,\bar{z}\right)\co^{\dagger}\left(z_1,\bar{z}_1\right)\right\rangle\nn \\
 & =\left(\frac{y}{y^{2}+\left(z-z_{1}\right)\left(\bar{z}-\bar{z}_{1}\right)}\right)^{2h}
\end{align}

\subsection{Large $k$ Expansion}\label{app:LargekSolution}
In this section, we are going to solve the charged bulk primary conditions
to next to leading order of the large $k$ limit. We can assume the
ansatz for $\cj_{-n}\co$ up to this order to be
\begin{equation}
\cj_{-n}=\left(1+\frac{a_{0}^{\left(n\right)}}{k}\right)L_{-1}^{n}+\sum_{i=1}^{n}\frac{a_{i}^{\left(n\right)}}{k}J_{-i}L_{-1}^{n-i}+O\left(\frac{1}{k^{2}}\right)\label{eq:JMnOrder1Overk}
\end{equation}
where $O\left(\frac{1}{k^{2}}\right)$ includes all the higher oder
terms. For the exact $\cj_{-n}\co$, we have $J_{m}\cj_{-n}\co=0$
for $m\ge1$. This means that we must have
\begin{equation}
J_{m}\left[\left(1+\frac{a_{0}^{\left(n\right)}}{k}\right)L_{-1}^{n}+\sum_{i=1}^{n}\frac{a_{i}^{\left(n\right)}}{k}J_{-i}L_{-1}^{n-i}\right]\co=O\left(\frac{1}{k}\right),
\end{equation}
since $J_{m}$ acting on some of the order $\frac{1}{k^{2}}$ terms
may give order $\frac{1}{k}$ result. So the leading order contribution
to the LHS of the above equation should vanish,
\begin{align}
J_{m}\left(\left(1+\frac{a_{0}^{\left(n\right)}}{k}\right)L_{-1}^{n}\co+\sum_{i=1}^{n}\frac{a_{i}^{\left(n\right)}}{k}J_{-i}L_{-1}^{n-i}\right)\co & \approx\left(\begin{array}{c}
n\\
m\end{array}\right)m!qL_{-1}^{n-m}+ma_{m}^{\left(n\right)}L_{-1}^{n-m}\co=0
\end{align}
where we discarded the order $\frac{1}{k}$ terms on the RHS of the
approximation symbol and we've assumed $1\le m\le n$ in the above equation. Solving the equation on the RHS for $a_{m}^{\left(n\right)}$,
we get 
\begin{equation}
a_{m}^{\left(n\right)}=-\frac{n!}{\left(n-m\right)!m}q\label{eq:ai}
\end{equation}

Now using the condition that $L_{1}^{n}\cj_{-n}\co=n!\left(2h\right)_{n}\co,$
we can solve for $a_{0}^{\left(n\right)}$. We have 
\begin{align}
L_{1}^{n}\cj_{-n}\co= & \left[\left(1+\frac{a_{0}^{\left(n\right)}}{k}\right)n!\left(2h\right)_{n}+\sum_{i=1}^{n}\frac{a_{i}^{\left(n\right)}}{k}i!\left(\begin{array}{c}
n\\
i
\end{array}\right)\left(n-i\right)!\left(2h\right)_{n-i}\right]\co+O\left(\frac{1}{k^{2}}\right).
\end{align}
Requiring the RHS to be equal to $n!\left(2h\right)_{n}\co$ up to
order $\frac{1}{k^{2}}$, we get 
\begin{equation}
a_{0}^{\left(n\right)}=-\frac{1}{\left(2h\right)_{n}}\sum_{i=1}^{n}a_{i}^{\left(n\right)}\left(2h\right)_{n-i}=-q (\psi ^{(0)}(1-2 h)-\psi ^{(0)}(-2 h-n+1))\label{eq:a0}
\end{equation}
where $\psi^{(0)}$ is the digamma function.

In summary, the order $\frac{1}{k}$ terms in $\cj_{-n}$ are given
by (\ref{eq:JMnOrder1Overk}) with $a_{i}^{\left(n\right)}$ and $a_{0}^{\left(n\right)}$
given by (\ref{eq:ai}) and (\ref{eq:a0}), respectively. In appendix \ref{app:CFTphiphiOneLoop}, we are going to use this result to compute the $\frac{1}{k}$
correction to the bulk propagator $\left\langle \phi^\dagger\phi\right\rangle $.

\section{Bulk Correlation Functions in $U(1)$ Chern-Simons Theory}
\label{app:CFTCalculationBulkCorrelationFunctions}

In this appendix we first compute the $1/k$ corrections to the bulk propagator $\left\langle \phi^\dagger\phi\right\rangle$ using CFT techniques (i.e., using the result we obtained in last section for the $1/k$ corrections in $\phi$). Then in appendix \ref{app:BulkWittenDiagrams},  we compute $\< \phi \CO^\dag J\>$ and $\< \phi^\dag \phi\>$ using Witten diagrams, and get exactly the same results as the CFT calculations. These calculations provide a non-trivial check of our definition of a bulk charged scalar field using the bulk primary condition (\ref{eq:ChargedBulkPrimaryCondition}).

\subsection{CFT Calculation of $\langle \phi^\dagger\phi\rangle$}\label{app:CFTphiphiOneLoop}
We can compute $\left\langle \phi^{\dagger}\left(y_{1},z_{1},\bar{z}_{1}\right)\phi\left(y_{2},\bar{z}_{2},\bar{z}_{2}\right)\right\rangle$,
where $\phi^{\dagger}$ is $\phi$ with $q\rightarrow-q$, using the result we obtained in last section for the $1/k$ correction terms in $\phi$. Up to order
$\frac{1}{k}$, $\phi$ is given by
\begin{equation}
\phi\left(y,z,\bar{z}\right)=y^{2h}\sum_{n=0}^{\infty}y^{2n}\lambda_{n}\cj_{-n}\bar{L}_{-1}^{n}\co\left(z,\bar{z}\right),\qquad\lambda_{n}=\frac{\left(-1\right)^{n}}{n!\left(2h\right)_{n}}
\end{equation}
with
\begin{equation}
\cj_{-n}=\left(1+\frac{a_{0}^{\left(n\right)}}{k}\right)L_{-1}^{n}+\sum_{i=1}^{n}\frac{a_{i}^{\left(n\right)}}{k}J_{-i}L_{-1}^{n-i}+O\left(\frac{1}{k^{2}}\right).
\end{equation}
and $a_{i}^{\left(n\right)}=-q\frac{n!}{\left(n-i\right)!i}$ and
$a_{0}^{\left(n\right)}=-q (\psi ^{(0)}(1-2 h)-\psi ^{(0)}(-2 h-n+1))$,
as computed in appendix \ref{app:LargekSolution}. We can compute $\left\langle \phi\left(y_{1},z_{1},\bar{z}_{1}\right)\phi\left(y_{2},\bar{z}_{2},\bar{z}_{2}\right)\right\rangle $
by directly inserting the above expansion of $\phi$ into $\left\langle \phi^\dagger\phi\right\rangle $
and summing over all the contributions. So up to order $\frac{1}{k}$,
we have
\begin{align}
\left\langle \phi^{\dagger}\left(y_{1},z_{1},\bar{z}_{1}\right)\phi\left(y_{2},\bar{z}_{2},\bar{z}_{2}\right)\right\rangle = & \left(y_{1}y_{2}\right)^{2h}\sum_{n_{1},n_{2}=0}^{\infty}y_{1}^{2n_{1}}y_{2}^{2n_{2}}\lambda_{n_{1}}\lambda_{n_{2}}\left(\ca+\cb+\cc+\cd\right)\label{eq:PhiPhiCS}
\end{align}
with
\begin{align}
\ca & \equiv\left(1+\frac{a_{0}^{\left(n_{1}\right)}+a_{0}^{\left(n_{2}\right)}}{k}\right)\left\langle L_{-1}^{n_{1}}\bar{L}_{-1}^{n_{1}}\co^{\dagger}\left(z_{1},\bar{z}_{1}\right)L_{-1}^{n_{2}}\bar{L}_{-1}^{n_{2}}\co\left(z_{2},\bar{z}_{2}\right)\right\rangle ,\\
\cb & \equiv\sum_{i_{2}=1}^{n_{2}}\frac{a_{i_{2}}^{\left(n_{2}\right)}}{k}\left\langle L_{-1}^{n_{1}}\bar{L}_{-1}^{n_{1}}\co^{\dagger}\left(z_{1},\bar{z}_{1}\right)J_{-i_{2}}L_{-1}^{n_{2}-i_{2}}\bar{L}_{-1}^{n_{2}}\co\left(z_{2},\bar{z}_{2}\right)\right\rangle ,\\
\cc & \equiv\sum_{i_{1}=1}^{n_{1}}\frac{a_{i_{1}}^{\left(n_{1}\right)}}{k}\left\langle J_{-i_{1}}L_{-1}^{n_{1}-i_{1}}\bar{L}_{-1}^{n_{1}}\co^{\dagger}\left(z_{1},\bar{z}_{1}\right)L_{-1}^{n_{2}}\bar{L}_{-1}^{n_{2}}\co\left(z_{2},\bar{z}_{2}\right)\right\rangle ,\\
\cd & \equiv\sum_{i_{1}=1}^{n_{1}}\sum_{i_{2}=1}^{n_{2}}\frac{a_{i_{1}}^{\left(n_{1}\right)}a_{i_{2}}^{\left(n_{1}\right)}}{k^{2}}\left\langle J_{-i_{1}}L_{-1}^{n_{1}-i_{1}}\bar{L}_{-1}^{n_{1}}\co^{\dagger}\left(z_{1},\bar{z}_{1}\right)J_{-i_{2}}L_{-1}^{n_{2}-i_{2}}\bar{L}_{-1}^{n_{2}}\co\left(z_{2},\bar{z}_{2}\right)\right\rangle \label{eq:CurlyD}.
\end{align}
where we've only kept the terms that may contribute to order $\frac{1}{k}$.

The $\frac{1}{k}$ term in $\ca$ will cancel the $\frac{1}{k}$ terms
in $\cb$ and $\cc$\footnote{This can be checked by direct calculation. On the other hand, we can also see that this must be true by separating $\phi$ into two parts: $\phi=\phi_0+\phi_{\text{q-p}}$, where $\phi_0$ is the free-field 
\begin{equation}\label{eq:FreePhi}
\phi_0=y^{2h}\sum_{n=0}^\infty y^{2n}\lambda_nL_{-1}^n\bar{L}_{-1}^{n}\co
\end{equation}
 and $\phi_{\text{q-p}}$ includes the contributions from quasi-primaries and their global descendants. We can do this separation because of the property of $\cj_{-n}$ in (\ref{eq:CJnInTermsOfQuasiPrimariesSec}). The point here is that $\langle\phi_0\phi_{\text{q-p}}\rangle$ must be zero, since the two-point function $\co$ with a quasi-primary is zero and this is also true for the two-points of their global descendants. So the $1/k$ corrections can only come from $\langle\phi_{\text{q-p}}^\dagger\phi_{\text{q-p}}\rangle$, that is, the terms in $\cd$ of (\ref{eq:CurlyD}).  }, so the sum
of $\ca+\cb+\cc$ will be 
\begin{equation}
\ca+\cb+\cc=\left\langle L_{-1}^{n_{1}}\bar{L}_{-1}^{n_{1}}\co^{\dagger}\left(z_{1},\bar{z}_{1}\right)L_{-1}^{n_{2}}\bar{L}_{-1}^{n_{2}}\co\left(z_{2},\bar{z}_{2}\right)\right\rangle .
\end{equation}
Since $L_{-1}$ and $\bar{L}_{-1}$ are simply derivatives $\partial_{z}$
and $\partial_{\bar{z}}$, this is simply computed to be 
\begin{equation}
\ca+\cb+\cc=\frac{\left[(2h)_{n_1+n_2}\right]^2}{z_{12}^{2h+n_{1}+n_{2}}\bar{z}_{12}^{2h+n_{1}+n_{2}}}.
\end{equation}
The sum of this term in equation (\ref{eq:PhiPhiCS}) will give the
free field limit of $\left\langle \phi\phi\right\rangle $, which
is
\begin{equation}\label{eq:FreeBulkBulk}
	\left\langle \phi_0\phi_0\right\rangle=\frac{\rho^{h}}{1-\rho}, \quad \text{ with }  \rho=\left(\frac{\xi}{1+\sqrt{1-\xi^{2}}}\right)^{2}, \quad \xi=\frac{2y_{1}y_{2}}{y_{1}^{2}+y_{2}^{2}+z_{12}\bar{z}_{12}}
\end{equation}
where $\phi_0$ is given in (\ref{eq:FreePhi}).

Now the only term left to compute is $\cd$. The anti-holomorphic
part can be computed simply, and we have
\begin{align}
 & \left\langle J_{-i_{1}}L_{-1}^{n_{1}-i_{1}}\bar{L}_{-1}^{n_{1}}\co^{\dagger}\left(z_{1},\bar{z}_{1}\right)J_{-i_{2}}L_{-1}^{n_{2}-i_{2}}\bar{L}_{-1}^{n_{2}}\co\left(z_{2},\bar{z}_{2}\right)\right\rangle \\
= & \frac{\left(2h\right)_{n_{1}+n_{2}}\left(-1\right)^{n_{1}}}{\bar{z}_{12}^{2h+n_{1}+n_{2}}}\left\langle J_{-i_{1}}L_{-1}^{n_{1}-i_{1}}\co\left(z_{1}\right)J_{-i_{2}}L_{-1}^{n_{2}-i_{2}}\co\left(z_{2}\right)\right\rangle \nonumber 
\end{align}
Since we only care about the order $\frac{1}{k}$ terms in $\cd$,
we only need to compute the order $k$ term in the two-point function on the RHS of
the above equation, which is
\begin{align}
 & \left\langle J_{-i_{1}}L_{-1}^{n_{1}-i_{1}}\co\left(z_{1}\right)J_{-i_{2}}L_{-1}^{n_{2}-i_{2}}\co\left(z_{2}\right)\right\rangle \\
= & \frac{1}{2\pi i}\oint_{z_{1}}dz\left(z-z_{1}\right)^{-i_{1}}\frac{1}{\left(z-z_{2}\right)^{i_{2}+1}}\left\langle L_{-1}^{n_{1}-i_{1}}\co\left(z_{1}\right)J_{i_{2}}J_{-i_{2}}L_{-1}^{n_{2}-i_{2}}\co\left(z_{2}\right)\right\rangle +...\nonumber \\
= & \frac{1}{2\pi i}\oint_{z_{1}}dz\frac{i_{2}k}{\left(z-z_{1}\right)^{i_{1}}\left(z-z_{2}\right)^{i_{2}+1}}\left\langle L_{-1}^{n_{1}-i_{1}}\co\left(z_{1}\right)L_{-1}^{n_{2}-i_{2}}\co\left(z_{2}\right)\right\rangle +...\nonumber \\
= & k\frac{\left(-1\right)^{n_{1}-1}\left(2h\right)_{n_{1}-i_{1}+n_{2}-i_{2}}}{z_{12}^{2h+n_{1}+n_{2}}}\frac{\left(i_{2}\right)_{i_{1}}}{\left(i_{1}-1\right)!}+...\nonumber 
\end{align}
where we've only kept the leading order terms.

Putting everything together, we have 
\begin{smaller}
\begin{align}
\left\langle \phi^{\dagger}\phi\right\rangle = & \left\langle \phi_0\phi_0\right\rangle -\frac{q^{2}}{k}\left(X_{1}X_{2}\right)^{h}\\
 & \times\sum_{n_{1},n_{2}=0}^{\infty}\frac{\left(-1\right)^{n_{1}+n_{2}}\left(2h\right)_{n_{1}+n_{2}}}{\left(2h\right)_{n_{1}}\left(2h\right)_{n_{2}}}\left(\sum_{i_{1}=1}^{n_{1}}\sum_{i_{2}=1}^{n_{2}}\frac{\left(2h\right)_{n_{1}+n_{2}-i_{1}-i_{2}}\left(i_{2}\right)_{i_{1}}}{\left(n_{1}-i_{1}\right)!\left(n_{2}-i_{2}\right)!\left(i_{1}-1\right)!i_{1}i_{2}}\right)X_{1}^{n_{1}}X_{2}^{n_{2}}+O\left(\frac{1}{k^{2}}\right)\nonumber 
\end{align}
\end{smaller}where we've defined $X_{1}\equiv\frac{y_{1}^{2}}{z_{12}\bar{z}_{12}},X_{2}\equiv\frac{y_{2}^{2}}{z_{12}\bar{z}_{12}}$.
Mathematica is not able to perform the above sums, but it can be checked
that the following expression 
\begin{equation}
\left\langle \phi^{\dagger}\phi\right\rangle =\frac{\rho^{h}}{1-\rho}\left[1-\frac{q^2}{k}\left(\frac{\rho^{2}\,_{2}F_{1}(1,2h+1;2(h+1);\rho)}{2h+1}+\frac{\rho}{2h}-\log(1-\rho)\right)\right]+O\left(\frac{1}{k^{2}}\right)
\end{equation}
gives the same expansion when expanded in small $X_{1}$ and $X_{2}$
. In appendix \ref{app:OneLoopPhiPhiWitten}, we'll show that the above expression is the result
of the bulk Witten diagram calculation. Thus this provides a non-trivial
check of our definition of a bulk charged proto-field in $U\left(1\right)$
Chern-Simons theory. 

\subsection{Bulk Witten Diagram Calculation for $\left\langle J \mathcal{O}^{\dagger} \phi\right\rangle$ and $\langle\phi^\dagger\phi\rangle$}
\label{app:BulkWittenDiagrams}
In this section, we are going to compute $\left\langle J \mathcal{O}^{\dagger} \phi\right\rangle$ and $\langle\phi^\dagger\phi\rangle$ using Witten diagrams. The calculations here will be very similar to that of $\left\langle T \mathcal{O} \phi\right\rangle$ in \cite{Anand:2017dav} and that of $\langle\phi\phi\rangle$ in \cite{Chen:2018qzm}, except that the $\left\langle A_{z}A_{z}\right\rangle $
two-point function and the bulk three-point vertex are different, but actually simpler.

The action of the $U\left(1\right)$ Chern-Simons theory in Poincare AdS$_3$ is given by
\begin{equation}
I=\int d^{3}x\frac{k}{4\pi}\epsilon^{\mu\nu\lambda}A_{\mu}\partial_{\nu}A_{\lambda}+\sqrt{\left|g\right|}\left(\nabla_{\mu}\phi\left(\nabla^{\mu}\phi\right)^{\dagger}+m^{2}\phi\phi^{\dagger}\right)\label{eq:CSaction}
\end{equation}
with $\nabla_{\mu}\phi=\left(\partial_{\mu}+iqA_{\mu}\right)\phi$
and $\left(\nabla_{\mu}\phi\right)^{\dagger}=\left(\partial_{\mu}-iqA_{\mu}\right)\phi^{\dagger}$.
The Poincare AdS$_{3}$ metric and its inverse are
\begin{align}
g_{\mu\nu} & =\left(\begin{array}{ccc}
\frac{1}{y^{2}} & 0 & 0\\
0 & 0 & \frac{1}{2y^{2}}\\
0 & \frac{1}{2y^{2}} & 0
\end{array}\right),\quad g^{\mu\nu}=\left(\begin{array}{ccc}
y^{2} & 0 & 0\\
0 & 0 & 2y^{2}\\
0 & 2y^{2} & 0
\end{array}\right)
\end{align}
in the coordinate system $\left(y,z,\bar{z}\right)$. From the
Chern-Simons action (\ref{eq:CSaction}), we can see that the photon
two point function is given by
\begin{equation}
H\left(Y_{1},Y_{2}\right)\equiv\left\langle A_{z}\left(y_{1},z_{1},\bar{z}_{1}\right)A_{z}\left(y_{2},z_{2},\bar{z}_{2}\right)\right\rangle =\frac{1}{k\left(z_{1}-z_{2}\right)^{2}}.\label{eq:PhotonPropagator}
\end{equation}
Here and in some other parts of this paper, we'll use $X$ or $Y$ to denote
$\left(y,z,\bar{z}\right)$ for convenience. The scalar two point
function is just the usual bulk-bulk propagator as in (\ref{eq:FreeBulkBulk})
\begin{equation}
G\left(Y_{1},Y_{2}\right)\equiv\left\langle \phi^{\dagger}\left(y_{1},z_{1},\bar{z}_{1}\right)\phi\left(y_{2},z_{2},\bar{z}_{2}\right)\right\rangle =\frac{\rho^{h}}{1-\rho},\label{eq:BulkBulkPropagator}
\end{equation}
with $m^{2}=4h\left(h-1\right)$. The photon-scalar-scalar three point
vertex is given by 
\begin{equation}
iqg^{\mu\nu}A_{\mu}\left(\phi\partial_{\mu}\phi^{\dagger}-\phi^{\dagger}\partial_{\mu}\phi\right).
\end{equation}
And since we'll only be interested in the $z$ component of $A_{z}$,
the above vertex becomes 
\begin{equation}
i2qy^{2}A_{z}\left(\phi\partial_{\bar{z}}\phi^{\dagger}-\phi^{\dagger}\partial_{\bar{z}}\phi\right)
\end{equation}
where we've used $g^{\bar{z}z}=2y^{2}$. We'll assume\footnote{There may be some subtleties about this, which we explain in the calculation of $\langle\phi\co J\rangle$ in next subsection.} that we are free to perform an integration by parts, so that the above
vertex can be written as $-4iqy^{2}A_{z}\phi^{\dagger}\partial_{\bar{z}}\phi.$

We'll also need to bulk-boundary propagator for the charged scalar
field and the photon field. These are given by
\begin{equation}
K\left(Y_{1},\left(z_{2},\bar{z}_{2}\right)\right)\equiv\left\langle \phi\left(y_{1},z_{1},\bar{z}_{1}\right)\co^{\dagger}\left(z_{2},\bar{z}_{2}\right)\right\rangle =\left(\frac{y_{1}}{y_{1}^{2}+\left(z_{1}-z_{2}\right)\left(\bar{z}_{1}-\bar{z}_{2}\right)^{2}}\right)^{2h}
\end{equation}
and
\begin{equation}\label{eq:BulkBoundaryPhoto}
\left\langle A_{z}\left(Y_{1}\right)J\left(z_{2}\right)\right\rangle=\frac{1}{\left(z_{1}-z_{2}\right)^{2}},
\end{equation}
where we've used the fact that $J(z)=\frac{A_z}{k}$.

\subsubsection{Witten Diagram Calculation for $\left\langle J\mathcal{O}^{\dagger}\phi\right\rangle$}

\begin{figure}[th!]
\begin{center}
\includegraphics[width=0.45\textwidth]{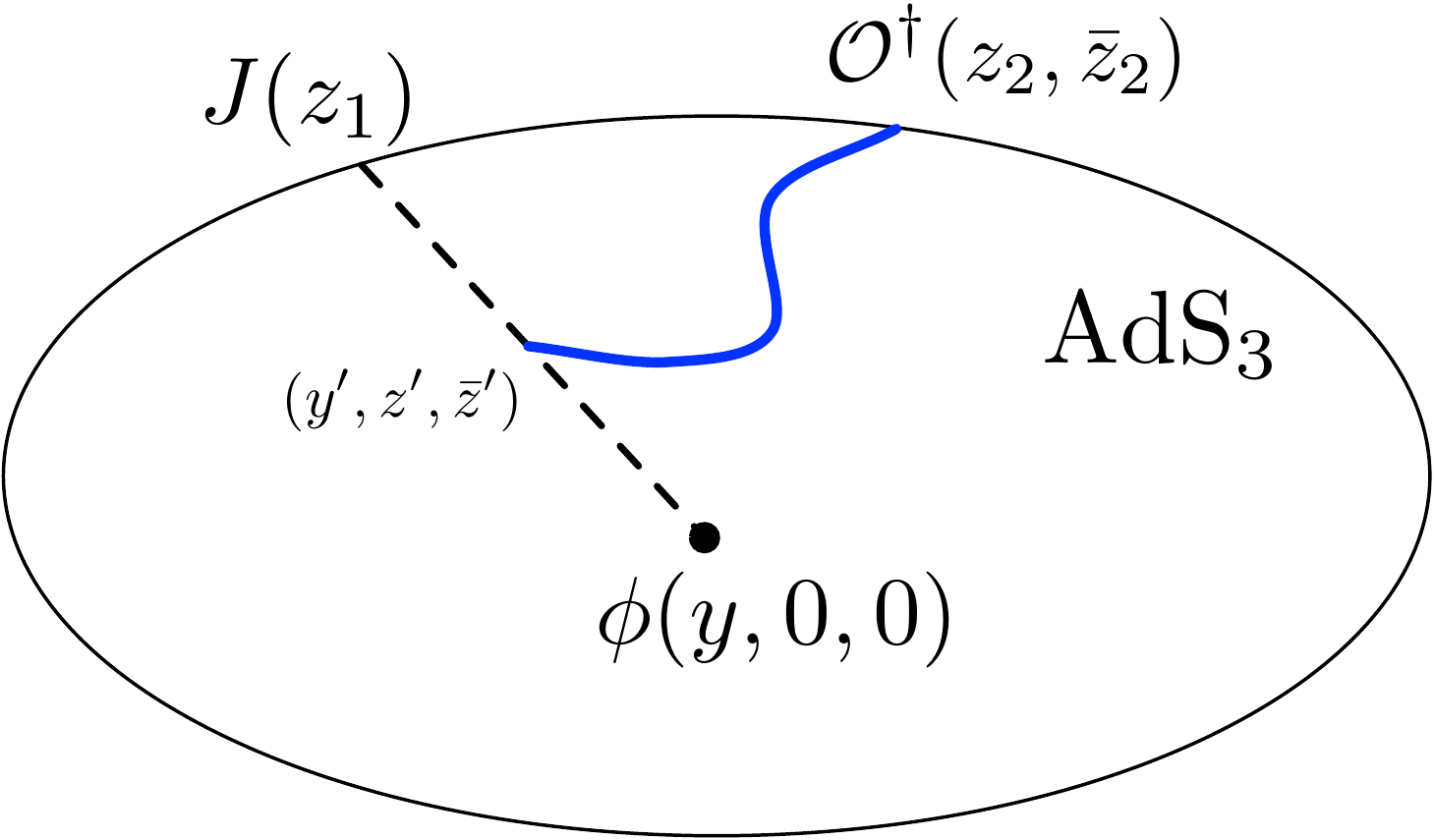}
\caption{Witten diagram for $\langle J\co^\dagger \phi\rangle$ }
\label{fig:JOPhiWitten}
\end{center}
\end{figure}

The Witten diagram for $\left\langle J\mathcal{O}^{\dagger}\phi\right\rangle $
is given by figure \ref{fig:JOPhiWitten}. We'll use the saddle point approximation to evaluate this diagram. The idea is that instead of integrating the bulk vertex point over AdS$_3$, we only integrate along the geodesic connecting $\phi$ and $\co^\dagger$. In appendix D.4 of \cite{Anand:2017dav}, we showed that the saddle point approximation for $\left\langle \phi\mathcal{O}T\right\rangle$ actually gives the exact result. We expect the same thing to happen here. 

The bulk-boundary three-point function
$\left\langle J\left(z_{1}\right)\mathcal{O}^{\dagger}\left(z_{2},\bar{z}_{2}\right)\phi\left(X_{1}\right)\right\rangle $
(with $X_{1}=\left(y,0,0\right)$) is computed as usual by
\begin{equation}
\left\langle J\mathcal{O}^{\dagger}\phi\right\rangle =\int_{\text{AdS}_{3}}\sqrt{g}d^3X'\frac{1}{\left(z'-z_{1}\right)^{2}}2iqy^{\prime2}\left[G\partial_{\bar{z}'}K-G\partial_{\bar{z}'}K\right]
\end{equation}
where where we've denoted $X^{\prime}=\left(y^{\prime},z^{\prime},\bar{z}^{\prime}\right)$,
and also used the bulk-boundary propagator for the photons (\ref{eq:BulkBoundaryPhoto}). Explicitly, the coordinate dependence of $G$
and $K$ are $G=G\left(X_{1},X^{\prime}\right)$ and $K=K\left(X^{\prime},\left(z_{2},\bar{z}_{2}\right)\right)$.
Now as mentioned above, we integrate by
parts\footnote{In fact, when performing an integration by parts, there is a delta function term coming from $\partial_{\bar{z}'}\frac{1}{z'-z_{1}}=\pi\delta^2\left(z'-z_{1},\bar{z}'-\bar{z}_1\right)$. Such terms would contaminate pure CFT correlators (as well as bulk correlators) and violate Ward identities, and so we have dropped them.  This can be viewed as a choice of regulator.  It would be very interesting to better understand regulation, and the role of these terms, in future work. } to find
\begin{equation}
\left\langle J\mathcal{O}^{\dagger}\phi\right\rangle =-\int_{\text{AdS}_{3}}\sqrt{g}d^3X^{\prime}\frac{4iqy^{\prime2}}{\left(z'-z_{1}\right)^{2}}G\partial_{\bar{z}'}K\label{eq:JOphiIntegral}
\end{equation}
Now we can do a saddle point approximation as in appendix D.4 of
\cite{Anand:2017dav}. The idea is as follows. We can write
\begin{equation}
G\left(X_{1},X^{\prime}\right)=\frac{e^{-2h\sigma_{\left(X_{1},X^{\prime}\right)}}}{1-e^{-2\sigma_{\left(X_{1},X^{\prime}\right)}}}
\end{equation}
where $\rho\equiv e^{-2\sigma}$ and $\sigma_{\left(X_{1},X^{\prime}\right)}$
is the geodesic length between $X_{1}=\left(y,0,0\right)$ and $X^{\prime}=\left(y',z',\bar{z}'\right)$, whose expression can be obtained from (\ref{eq:FreeBulkBulk}).
Similarly, the bulk boundary propagator can be written as
\begin{equation}
K\left(X^{\prime},\left(z_{2},\bar{z}_{2}\right)\right)=e^{-2h\sigma_{\left(X',\left(z_{2},\bar{z}_{2}\right)\right)}}
\end{equation}
where $\sigma_{\left(X',\left(z_{2},\bar{z}_{2}\right)\right)}=\log\frac{y^{\prime2}+\left(z^{\prime}-z_{2}\right)\left(\bar{z}^{\prime}-\bar{z}_{2}\right)}{y^{\prime}}$
is the (regularized) bulk-boundary geodesic length between $X^{\prime}=\left(y',z',\bar{z}'\right)$
and $\left(z_{2},\bar{z}_{2}\right)$. So the integral (\ref{eq:JOphiIntegral}) after simplification
can be written in the following form
\begin{equation}
\left\langle J\mathcal{O}^{\dagger}\phi\right\rangle =8ihq \int\frac{dz'd\bar{z}'dy'}{y^{'2}}e^{-2hL\left(y',z',\bar{z}'\right)}\frac{e^{-\sigma_{\left(y',z',\bar{z}'\right),\left(z_{2},\bar{z}_{2}\right)}}}{1-e^{-2\sigma_{\left(y,0,0\right),\left(y',z',\bar{z}'\right)}}}\frac{\left(z'-z_{2}\right)}{\left(z'-z_{1}\right)^{2}}
\end{equation}
where $L\left(y',z',\bar{z}'\right)\equiv\sigma_{\left(X_{1},X^{\prime}\right)}+\sigma_{\left(X',\left(z_{2},\bar{z}_{2}\right)\right)}$.
Now we can take the large $h$ limit and the integration will be dominated
by the line integral along the geodesic from $X_{1}=\left(y,0,0\right)$
to $\left(z_{2},\bar{z}_{2}\right)$. This geodesic parameterized
by $z^{\prime}$ is given by 
\begin{equation}
\bar{z}^{\prime}=\frac{\bar{z}}{z}z^{\prime},\quad y^{\prime2}=\left(1-\frac{z^{\prime}}{z}\right)\left(y^{2}+z^{\prime}\bar{z}\right),
\end{equation}
so that the saddle point approximation to equation (\ref{eq:JOphiIntegral})
is
\begin{equation}
\langle J\co^{\dagger}\phi\rangle=8\pi i qe^{-2hL(y,0,0)}\int_{0}^{z_{2}}dz^{\prime}\frac{1}{\sqrt{|\det\partial^{2}L}|}\frac{e^{-2\sigma_{\left(X',\left(z_{2},\bar{z}_{2}\right)\right)}}}{1-e^{-2\sigma_{\left(X_{1},X'\right)}}}\frac{1}{y^{\prime2}}\frac{\left(z'-z_{2}\right)}{\left(z'-z_{1}\right)^{2}},\label{eq:JOphiSaddle}
\end{equation}
where the determinant is given by 
\begin{equation}
\det\partial^{2}L=\det\left(\begin{array}{cc}
{\partial_{\bar{z}^{\prime}}^{2}L} & {\partial_{\bar{z}^{\prime}}\partial_{y^{\prime}}L}\\
{\partial_{y^{\prime}}\partial_{\bar{z}^{\prime}}L} & {\partial_{y^{\prime}}^{2}L}
\end{array}\right)=\frac{4z_{2}^{5}\left(z^{\prime}\bar{z}_{2}+y^{2}\right)}{z^{\prime2}\left(z^{\prime}-z_{2}\right)\left(z_{2}\bar{z}_{2}+y^{2}\right)^{4}}.
\end{equation}
Plugging this in (\ref{eq:JOphiSaddle}), we have 
\begin{equation}
\left\langle J(z_1)\mathcal{O}^{\dagger}(z_2,\bar{z}_2)\phi(y,0,0)\right\rangle =q\left\langle \phi\co^{\dagger}\right\rangle \int_{0}^{z_{2}}dz^{\prime}\frac{1}{\left(z_{1}-z^{\prime}\right)^{2}}=\frac{qz_{2}}{\left(z_{2}-z_{1}\right)z_{1}}\left\langle \phi\co^{\dagger}\right\rangle 
\end{equation}
where we've used the fact that $\left\langle \phi(y,0,0)\co^{\dagger}(z_2,\bar z_2)\right\rangle=e^{-2hL(y,0,0)}=\left(\frac{y}{y^2+z_2\bar z_2}\right)^{2h}$ and we neglected a numerical constant in obtaining the above result. The above result is exactly the same as the result (\ref{eq:JOPhiBoundary}) obtained via assuming $\phi$ to be a charged bulk field defined by the bulk primary condition (\ref{eq:ChargedBulkPrimaryCondition}).

\subsubsection{One-loop Correction to $\left\langle\phi^{\dagger} \phi\right\rangle$}\label{app:OneLoopPhiPhiWitten}

\begin{figure}[th!]
\begin{center}
\includegraphics[width=0.45\textwidth]{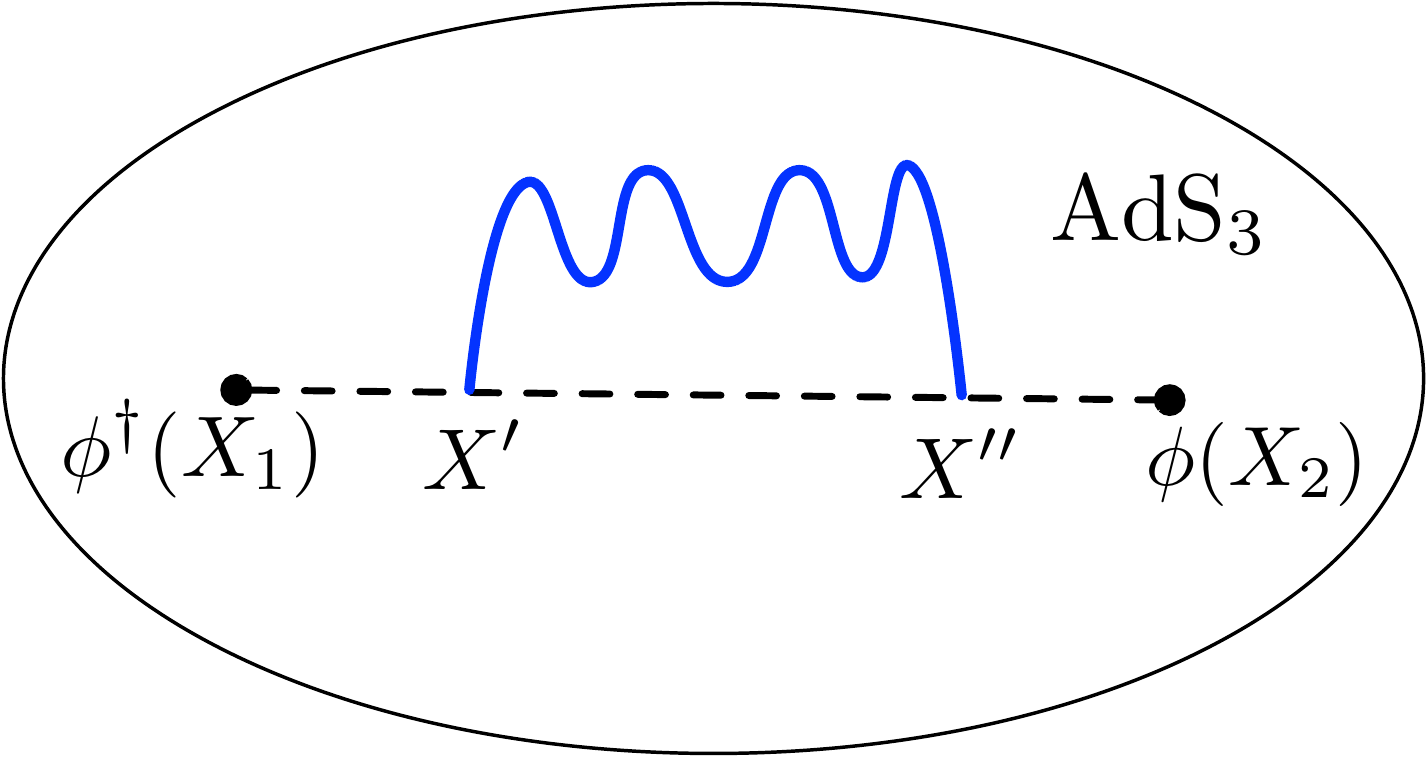}
\caption{Witten diagram for $\langle \phi^\dagger \phi\rangle$ }
\label{fig:phiphiOneLoop}
\end{center}
\end{figure}

In this subsection, we are going to compute one loop correction to
$\left\langle \phi^{\dagger}\phi\right\rangle $. The Witten diagram
is given in figure \ref{fig:phiphiOneLoop}, and it's computed as follows
\begin{align}
\left\langle \phi^{\dagger}\left(X_{1}\right)\phi\left(X_{2}\right)\right\rangle _{\text{1 loop}} & =-16q^{2}\int\sqrt{\left|g\right|}d^{3}X^{\prime}\int\sqrt{\left|g\right|}d^{3}X^{\prime\prime}y^{\prime2}y^{\prime\prime2}\\
 & \qquad\times G\left(X_{1},X^{\prime}\right)\partial_{\bar{z}^{\prime}}G\left(X^{\prime},X^{\prime\prime}\right)H\left(X^{\prime},X^{\prime\prime}\right)\partial_{\bar{z}^{\prime\prime}}G\left(X^{\prime\prime},X_{2}\right)\nonumber 
\end{align}
 Acting on this two-point function with the Klein-Gordon operator twice,
we have
\begin{equation}
\left(\nabla_{1}^{2}+m^{2}\right)\left(\nabla_{2}^{2}+m^{2}\right)\left\langle \phi^{\dagger}\left(X_{1}\right)\phi\left(X_{2}\right)\right\rangle _{\text{1 loop}}=-16q^{2}H\left(X_{1},X_{2}\right)y_{1}^{2}y_{2}^{2}\partial_{\bar{z}_{1}}\partial_{\bar{z}_{2}}G\left(X_{1},X_{2}\right)\label{eq:BoubleKGEq}
\end{equation}
By using the photon propagator (\ref{eq:PhotonPropagator}) and the
bulk-bulk propagator (\ref{eq:BulkBulkPropagator}), one can show
that the RHS of the above equation is 
\begin{equation}
-\frac{32q^{2}\rho^{h+1}\left(2h^{2}(1-\rho)^{2}+h\left(-5\rho^{2}+4\rho+1\right)+3\rho(\rho+1)\right)}{k(1-\rho)^{5}}.
\end{equation}
Assuming $P\left(\rho\right)\equiv\left\langle \phi^{\dagger}\left(X_{1}\right)\phi\left(X_{2}\right)\right\rangle _{\text{1 loop}}$
only depends on the $\rho$, equation (\ref{eq:BoubleKGEq}) becomes
\begin{align}
 & 16\left(h-1\right)^{2}h^{2}P\left(\rho\right)+\frac{64}{\rho-1}\left(-h^{2}+h+1\right)\rho^{2}P'\left(\rho\right) \\
 & -\frac{32\rho^{2}\left(\left(h-1\right)h\left(\rho-1\right)-7\rho+1\right)P''\left(\rho\right)}{\rho-1}+\frac{64\rho^{3}\left(2\rho-1\right)P^{\left(3\right)}\left(\rho\right)}{\rho-1}+16\rho^{4}P^{\left(4\right)}\left(\rho\right)\nn\\
= & -\frac{32q^{2}\rho^{h+1}\left(2h^{2}(1-\rho)^{2}+h\left(-5\rho^{2}+4\rho+1\right)+3\rho(\rho+1)\right)}{k(1-\rho)^{5}}\nonumber 
\end{align}
Luckily, Mathematica is able to solve the above fourth order differential
equation. Fixing the integration constants, we get 
\begin{equation}
P\left(\rho\right)=-\frac{q^{2}}{k}\frac{\rho^{h}}{1-\rho}\left[\frac{\rho^{2}\,_{2}F_{1}(1,2h+1;2(h+1);\rho)}{2h+1}+\frac{\rho}{2h}-\log(1-\rho)\right],
\end{equation}
which is exactly the same as the $\frac{1}{k}$ correction to the
two-point function of the bulk charged proto-field computed in appendix \ref{app:CFTphiphiOneLoop} using pure CFT technics.

\section{Computations of $\langle \phi \mathcal{O}T\rangle$ in Global AdS$_3$}
\label{app:phiOT}

In this section, we use two methods to calculate $\langle\phi\mathcal{O}T\rangle$
in global AdS$_{3}$. These two methods are essentially the same.
Both of them are based on the idea of bulk-boundary OPE blocks (as
we called them in \cite{Anand:2017dav} and \cite{Chen:2018qzm}) or the bulk-boundary bi-local operator $\phi\co$ (\ref{eq:VacuumBiLocal}) (a generalization
of the boundary bi-local operator as defined in \cite{Cotler:2018zff}). The first
method is to expand the vacuum channel of the bulk-boundary bi-local
operator in terms of the $\epsilon$ operator defined in section 6
of \cite{Cotler:2018zff}, and then use the $\epsilon$ propagator, which
was derived from the Alekseev-Shatashvilli theory of boundary gravitons
in that paper, to compute $\left\langle \phi\co T\right\rangle $. The second method is to expand the vacuum channel of
the bi-local in terms of the energy-momentum tensor $T$ and use the
two-point function of $T$ to compute $\left\langle \phi\co T\right\rangle $.
Both of them give the same result as (\ref{eq:PhiOTGlobalAdS3}) obtained using the properties of the bulk proto-field.

\subsection{$\langle\phi\mathcal{O}T\rangle$ in Global AdS$_{3}$ from Alekseev-Shatashvili
\label{PropsUsingJensen}}

The idea here very similar to that of section \ref{sec:OneLoopNearHorizon}, where we computed
the $1/c$ corrections to the bulk-boundary propagator in
a heavy background. The difference is that here we are considering
bulk-boundary bi-local operator $\phi\co$ in the vacuum.

As in section \ref{sec:PhiOTGlobalAdS}, we use $f\left(z\right)=e^{z}$ and $\bar{f}\left(\bar{z}\right)=e^{\bar{z}}$
to obtain the global AdS$_{3}$ metric (\ref{eq:yzzbAdS3Global}). But now we are going to
include perturbation as follows
\begin{equation}\label{eq:fzfbzbPerturbation}
f\left(z\right)=e^{z+i\frac{\epsilon\left(z\right)}{c}},\text{ and }\bar{f}\left(\bar{z}\right)=e^{\bar{z}-i\frac{\bar{\epsilon}\left(\bar{z}\right)}{c}}
\end{equation}
(with $\epsilon\left(z\right)$ and $\bar{\epsilon}\left(z\right)$
promoted to operators). Specifically, the vacuum contribution to the bulk-boundary bi-local
operator is given by 
\begin{equation}
\phi\left(y_{1},z_{1},\bar{z}_{1}\right)\co\left(z_{2},z_{2}\right)|_{\text{vac}}=\left(f'\left(z_{2}\right)\bar{f}'\left(\bar{z}_{2}\right)\right)^{h}\left(\frac{u_{1}}{u_{1}^{2}+\left(x_{1}-f\left(z_{2}\right)\right)\left(\bar{x}_{1}-f\left(\bar{z}_{2}\right)\right)}\right)^{2h}.\label{eq:VacuumBiLocal}
\end{equation}
The $(u_1,x_1,\bar{x}_1)$ here are functions of $(y_1,z_1,\bar{z}_1)$ given by (\ref{eq:VacuumAdSDiffeomorphism}) with $f(z)$ and $\bar{f}(\bar{z})$ given by (\ref{eq:fzfbzbPerturbation}). One can see that the above expression becomes the boundary bi-local defined in \cite{Cotler:2018zff}
when we send $\phi$ to the boundary by taking $y_{1}\rightarrow0$.
Written in terms of the coordinates $\left(y,z,\bar{z}\right)$ and
expanded in large $c$, the above vacuum block becomes 
\begin{equation}
\frac{\phi\left(y_{1},z_{1},\bar{z}_{1}\right)\co\left(z_{2},\bar{z}_{2}\right)|_{\text{vac}}}{\left\langle \phi\left(y_{1},z_{1},\bar{z}_{1}\right)\co\left(z_{2},\bar{z}_{2}\right)\right\rangle _{\text{global AdS}_{3}}}=1+\frac{1}{c}\frac{B\epsilon_{1}+C\epsilon_{1}'+D\epsilon_{1}''+E\epsilon_{2}+F\epsilon_{2}'}{A}+\co\left(\frac{1}{c^{2}}\right)\label{eq:OPEblock}
\end{equation}
where $\epsilon_{i}\equiv\epsilon\left(z_{i}\right)$ and the derivatives
are with respect to $z_{i}$. The bulk-boundary propagator $\left\langle \phi\left(y_{1},z_{1},\bar{z}_{1}\right)\co\left(z_{2},\bar{z}_{2}\right)\right\rangle _{\text{global AdS}_{3}}$
in global AdS$_{3}$ is given in (\ref{eq:BulkBoundaryGlobaAdSPropagator}) (up to a factor of $(\xi_2\bar{\xi}_2)^h$ coming from the $\left(f'\left(z_{2}\right)\bar{f}'\left(\bar{z}_{2}\right)\right)^{h}$ in (\ref{eq:VacuumBiLocal})), i.e. 
\begin{equation}
\langle\phi\mathcal{O}\rangle_{\text{global AdS\ensuremath{_{3}}}} =\left(\frac{4y_{1}\sqrt{\xi_{1}\bar{\xi}_{1}\xi_2\bar{\xi}_2}}{\left(\bar{\xi_{1}}\xi_{1}+\bar{\xi}_{2}\xi_{2}\right)\left(y_{1}^{2}+4\right)+\left(\bar{\xi}_{1}\xi_{2}+\bar{\xi}_{2}\xi_{1}\right)\left(y_{1}^{2}-4\right)}\right)^{2h},
\end{equation}
where we've defined $\xi_{i}\equiv e^{z_{i}}$ and $\bar{\xi}_{i}=e^{\bar{z}_{i}}$.
The denominator and the coefficients in the numerator of (\ref{eq:OPEblock}) are given
by
\begin{align}
A & =i\left(\left(y_{1}^{2}-4\right)\left(\xi_{2}\bar{\xi}_{1}+\xi_{1}\bar{\xi}_{2}\right)+\left(y_{1}^{2}+4\right)\left(\xi_{1}\bar{\xi}_{1}+\xi_{2}\bar{\xi}_{2}\right)\right),\nonumber \\
B & =\left(y_{1}^{2}-4\right)\left(\xi_{1}\bar{\xi}_{2}-\xi_{2}\bar{\xi}_{1}\right)+\left(y_{1}^{2}+4\right)\left(\xi_{1}\bar{\xi}_{1}-\xi_{2}\bar{\xi}_{2}\right),\nonumber \\
C & =\left(y_{1}^{2}+4\right)\left(\xi_{2}\bar{\xi}_{1}+\xi_{1}\bar{\xi}_{2}\right)+\left(y_{1}^{2}-4\right)\left(\xi_{1}\bar{\xi}_{1}+\xi_{2}\bar{\xi}_{2}\right),\\
D & =-2y_{1}^{2}\left(\xi_{1}-\xi_{2}\right)\left(\bar{\xi}_{1}+\bar{\xi}_{2}\right),\nonumber \\
E & =\left(y_{1}^{2}+4\right)\left(\xi_{2}\bar{\xi}_{2}-\xi_{1}\bar{\xi}_{1}\right)+\left(y_{1}^{2}-4\right)\left(\xi_{2}\bar{\xi}_{1}-\xi_{1}\bar{\xi}_{2}\right),\nonumber \\
F & =-\left(y_{1}^{2}-4\right)\left(\xi_{2}\bar{\xi}_{1}+\xi_{1}\bar{\xi}_{2}\right)-\left(y_{1}^{2}+4\right)\left(\xi_{1}\bar{\xi}_{1}+\xi_{2}\bar{\xi}_{2}\right).\nonumber 
\end{align}
 We've only kept the holomorphic $\epsilon$ terms in (\ref{eq:OPEblock}),
since the anti-holomorphic $\bar{\epsilon}$ will not contribute to
$\left\langle \phi\co T\right\rangle $.

To compute $\left\langle \phi\co T\right\rangle $, we also need to
write the energy-momentum tensor $T$ in terms of the $\epsilon$
operators. As explained in section 2 of \cite{Anand:2017dav}, $T$ is simply given
by the Schwarzian derivative as follows: 
\begin{equation}
T\left(z\right)=\frac{c}{12}\left[\frac{f^{\prime\prime\prime}(z)f^{\prime}(z)-\frac{3}{2}\left(f^{\prime\prime}(z)\right)^{2}}{\left(f^{\prime}(z)\right)^{2}}\right]=-\frac{c}{24}-\frac{i}{12}\left(\epsilon'(z)-\epsilon^{(3)}(z)\right)+\co\left(\frac{1}{c}\right).\label{eq:TInTermOfEpsilon}
\end{equation}
The last piece of information we need to compute $\left\langle \phi\co T\right\rangle $
is the $\epsilon$ propagator $\left\langle \epsilon\epsilon\right\rangle $.
The action obeyed by $\epsilon$ and $\bar{\epsilon}$ is the saddle
point quadratic action derived from the Alekseev-Shatashvilli action.
And the $\epsilon$ propagator is given in section 6 of \cite{Cotler:2016fpe},
\begin{equation}
\left\langle \epsilon\left(z\right)\epsilon\left(0\right)\right\rangle =6c\left(-1+\frac{3\xi}{2}-\frac{\left(1-\xi\right)^{2}}{\xi}\ln\left(1-\xi\right)\right),\quad\xi\equiv e^{z}
\end{equation}
Note that we have a different normalization for the $\epsilon$s. 

Eventually, $\left\langle \phi\left(y_{1},z_{1},\bar{z}_{1}\right)\co\left(z_{2},\bar{z}_{2}\right)T\left(z\right)\right\rangle $
is given by\footnote{There could be $1/c$ corrections to $\left\langle \phi\co T\right\rangle $
coming from the higher order terms in (\ref{eq:OPEblock}) and (\ref{eq:TInTermOfEpsilon}).
But since the result for $\left\langle \phi\co T\right\rangle $ is
exact without any $1/c$ correction (as computed using the properties of the bulk proto-field in \ref{sec:PhiOTinArbit}), such higher order
have been dropped by the regulator,  as in \cite{Fitzpatrick:2016mtp, Anand:2017dav, Chen:2018qzm}. } \begin{smaller}
\begin{equation}
\frac{\left\langle \phi\co T\right\rangle }{\left\langle \phi\co\right\rangle _{\text{global AdS\ensuremath{_{3}}}}}=\frac{h\left(\xi_{1}-\xi_{2}\right)^{2}\xi^{2}}{\left(\xi-\xi_{1}\right)^{3}\left(\xi-\xi_{2}\right)^{2}}\left[\xi-\xi_{1}+\frac{4y_{1}^{2}\xi_{1}\left(\xi-\xi_{2}\right)\left(\bar{\xi}_{1}+\bar{\xi}_{2}\right)}{\left(\bar{\xi}_{1}\xi_{1}+\bar{\xi}_{2}\xi_{2}\right)\left(y_{1}^{2}+4\right)+\left(\bar{\xi}_{1}\xi_{2}+\bar{\xi}_{2}\xi_{1}\right)\left(y_{1}^{2}-4\right)}\right]-\frac{c}{24}\label{eq:PhiOTEffectAction}
\end{equation}
\end{smaller}which agrees with (\ref{eq:PhiOTGlobalAdS3}) up to a
trivial transformation for $\co$ and $T$ from the complex plane
$\left(\xi,\bar{\xi}\right)$ to the cylinder $\left(z,\bar{z}\right)$.

\subsection{$\langle\phi\mathcal{O}T\rangle$ in Global AdS$_{3}$ from $\langle TT\rangle$ \label{PhiOTusingOPE}}

We can also compute $\left\langle \phi\co T\right\rangle $ in global
AdS$_{3}$ without using the $\epsilon$ propagator as we did in last
subsection. Instead, we'll use the two-point function of the energy-momentum
tensor $T$. This was the method that we used in \cite{Anand:2017dav} to compute
$\left\langle \phi\co T\right\rangle $ (equation (\ref{eq:PhiOTPoincareAdS})) in Poincare AdS$_{3}$,
and also in appendix B of \cite{Chen:2018qzm} to compute the $1/c$ corrections to
the vacuum block of all-light bulk-boundary correlator $\left\langle \co_{2}\co_{2}\phi_{1}\co_{1}\right\rangle $
(with $\phi_{1}$ the proto-field of section \ref{sec:ProtofieldFGGauge}) up to order $1/c^{2}$.

The idea is actually very similar to that of last section. Instead
of denoting the perturbation via $\epsilon$, we expand $f\left(z\right)$
in the large $c$ limit as
\begin{equation}
f(z)=e^{z}+\frac{1}{c}f_{1}(z)+\dots,
\end{equation}
and similarly for $\bar{f}\left(\bar{z}\right)$. Then similar
to last subsection, the energy-momentum tensor $T\left(z\right)$
is given by
\begin{align}
T(z)=\frac{c}{12}\left[\frac{f^{\prime\prime\prime}(z)f^{\prime}(z)-\frac{3}{2}\left(f^{\prime\prime}(z)\right)^{2}}{\left(f^{\prime}(z)\right)^{2}}\right] & =-\frac{c}{24}+\frac{1}{12}e^{-z}\left(2f_{1}'-3f_{1}''+f_{1}^{\left(3\right)}\right)+O\left(\frac{1}{c}\right).\label{eq:TzInTermOff1}
\end{align}
The energy-momentum tensor $T\left(\xi\right)$ on the complex plane
with coordinates $\left(\xi=e^{z},\bar{\xi}=e^{\bar{z}}\right)$
is related to $T\left(z\right)$ as usual by 
\begin{equation}
T\left(z\right)=-\frac{c}{24}+\xi^{2}T\left(\xi\right),
\end{equation}
so from (\ref{eq:TzInTermOff1}), we have 
\begin{equation}
T\left(e^{z}\right)=\frac{1}{12}e^{-3z}\left(2f_{1}'-3f_{1}''+f_{1}^{\left(3\right)}\right)
\end{equation}
We can now solve for $f_{1}\left(z\right)$ in terms of $T\left(e^{z}\right)$, and the solution is given by 
\begin{equation}
f_{1}(z)=12\int_{0}^{z}\left(\int_{0}^{z'}e^{z'+z''}(e^{z'}-e^{z''})T\left(e^{z''}\right)\,\,dz''\right)dz'.
\end{equation}

The bulk-boundary OPE block or the bulk boundary bi-local operator
(\ref{eq:VacuumBiLocal}) can then be expanded in large $c$, 
\begin{smaller}
\begin{equation}
\frac{\phi\left(y,0,0\right)\co\left(z_{2},\bar{z}_{2}\right)|_{\text{vac}}}{\left\langle \phi\left(y,0,0\right)\co\left(z_{2},\bar{z}_{2}\right)\right\rangle _{\text{global AdS}_{3}}}=1+\frac{1}{c}\left(\frac{hf_{1}'\left(z_{2}\right)}{\xi_{2}}-\frac{2h\left(\left(y^{2}+4\right)\bar{\xi}_{2}+y^{2}-4\right)f_{1}\left(z_{2}\right)}{\left(y^{2}+4\right)\left(1+\bar{\xi}_{2}\xi_{2}\right)+\left(y^{2}-4\right)\left(\xi_{2}+\bar{\xi}_{2}\right)}\right)+O\left(\frac{1}{c^{2}}\right)
\end{equation}
\end{smaller}
Using $\left\langle T\left(e^{z}\right)T\left(e^{z_{2}}\right)\right\rangle =\frac{c}{\left(e^{z}-e^{z_{2}}\right)^{4}}$
on the complex plane, we find 
\begin{align}
\left\langle f_{1}\left(z_{2}\right)T\left(e^{z}\right)\right\rangle  & =\frac{c\left(\xi_{2}-1\right)^{3}}{(\xi-1)^{3}\left(\xi-\xi_{2}\right)}\\
\left\langle f_{1}'\left(z_{2}\right)T\left(e^{z}\right)\right\rangle  & =-\frac{c\left(\xi_{2}-1\right)^{2}\xi_{2}\left(1-3\xi+2\xi_{2}\right)}{(\xi-1)^{3}\left(\xi-\xi_{2}\right)^{2}}\nonumber 
\end{align}
and $\left\langle \phi\left(y,0,0\right)\co\left(z_{2},\bar{z}_{2}\right)T\left(z\right)\right\rangle $
(after transforming $T$ to the cylinder) is given by 
\begin{small}
\begin{equation}
\frac{\left\langle \phi\co T\right\rangle }{\left\langle \phi\co\right\rangle _{\text{global AdS\ensuremath{_{3}}}}}=\frac{h\left(1-\xi_{2}\right)^{2}\xi^{2}}{\left(\xi-1\right)^{3}\left(\xi-\xi_{2}\right)^{2}}\left[\xi-1+\frac{4y_{1}^{2}\left(\xi-\xi_{2}\right)\left(1+\bar{\xi}_{2}\right)}{\left(1+\bar{\xi}_{2}\xi_{2}\right)\left(y^{2}+4\right)+\left(\xi_{2}+\bar{\xi}_{2}\right)\left(y^{2}-4\right)}\right]-\frac{c}{24}
\end{equation}
\end{small}
which is exactly the result obtained in last subsection (\ref{eq:PhiOTEffectAction})
with $y_{1}=y,z_{1}=\bar{z}_{1}=0$.

\bibliographystyle{utphys} 
\bibliography{VirasoroBib}

\end{document}